\documentclass{article}
\usepackage{arxiv}

\usepackage[utf8]{inputenc} 
\usepackage[T1]{fontenc}    
\usepackage{hyperref}       
\usepackage{url}            
\usepackage{booktabs}       
\usepackage{amsfonts}       
\usepackage{amsmath}        
\usepackage{nicefrac}       
\usepackage{microtype}      
\usepackage{lipsum}
\usepackage{graphicx}
\usepackage{transparent} 
\usepackage{tikz}        
\pdfpageattr{/Group << /S /Transparency /I false /CS /DeviceRGB >>}
\graphicspath{ {./images/} }
\usepackage{makecell}
\usepackage[toc,page]{appendix}
\usepackage{chngcntr} 
\usepackage{float}
\usepackage[labelfont=bf,labelsep=period]{caption}

\usepackage[table]{xcolor}
\usepackage{caption}

\newenvironment{bluecolorregion}{%
  \begingroup
  \color{black}

  \captionsetup{labelfont={color=black}, textfont={color=black}}

  \let\BCRtabular\tabular
  \def\tabular{\color{black}\BCRtabular}
  \ifdefined\tabularx
    \let\BCRtabularx\tabularx
    \def\tabularx{\color{black}\BCRtabularx}
  \fi
  \ifdefined\longtable
    \let\BCRlongtable\longtable
    \def\longtable{\color{black}\BCRlongtable}
  \fi

  \let\oldsection\section
  \renewcommand{\section}[1]{\oldsection{\texorpdfstring{\color{black}##1}{##1}}}
  \let\oldsubsection\subsection
  \renewcommand{\subsection}[1]{\oldsubsection{\texorpdfstring{\color{black}##1}{##1}}}
  \let\oldsubsubsection\subsubsection
  \renewcommand{\subsubsection}[1]{\oldsubsubsection{\texorpdfstring{\color{black}##1}{##1}}}
}{%
  \endgroup
}

\title{MicroLad: 2D-to-3D Microstructure Reconstruction and Generation via Latent Diffusion and Score Distillation}

\author{
 Kang-Hyun Lee \\
  Department of Mechanical Engineering\\
  Massachusetts Institute of Technology\\
  Cambridge, MA 02139 USA \\
  \texttt{kanghl@mit.edu} \\
   \And
 Faez Ahmed \\
  Department of Mechanical Engineering\\
  Massachusetts Institute of Technology\\
  Cambridge, MA 02139 USA \\
  \texttt{faez@mit.edu} \\
}

\begin{document}
\maketitle
\begin{abstract}
A major obstacle to establishing reliable structure–property (SP) linkages in materials engineering is the scarcity of diverse 3D microstructure datasets. Limited dataset availability and insufficient control over the analysis and design space restrict the variety of achievable microstructure morphologies, hindering progress in solving the inverse (property-to-structure) design problem. To address these challenges, we introduce MicroLad, a latent diffusion framework specifically designed for reconstructing 3D microstructures from 2D data. Trained on 2D images and employing multi-plane denoising diffusion sampling in the latent space, the framework reliably generates stable and coherent 3D volumes that remain statistically consistent with the original data. While this reconstruction capability enables dimensionality expansion (2D-to-3D) for generating statistically equivalent 3D samples from 2D data, effective exploration of microstructure design requires methods to guide the generation process toward specific objectives. To achieve this, MicroLad integrates score distillation sampling (SDS), which combines a differentiable score loss with microstructural descriptor-matching and property-alignment terms. This approach updates encoded 2D slices of the 3D volume in the latent space, enabling robust inverse-controlled 2D-to-3D microstructure generation. Consequently, the method facilitates exploration of an expanded 3D microstructure analysis and design space in terms of both microstructural descriptors and material properties.

\end{abstract}


\section{Introduction}

Predictive materials design relies on rigorous structure–property (SP) linkages—quantitative relationships that connect microstructural features (e.g., phase distributions, volume fractions, and morphologies) to bulk responses and their associated uncertainties \cite{bostanabad2018computational, shojaeefard2016review, bargmann2018generation}. To meet this need, Integrated Computational Materials Engineering (ICME) has significantly advanced, particularly through property simulations based on numerical methods such as the finite element method (FEM), finite difference method (FDM), finite volume method (FVM), and fast Fourier transform (FFT) \cite{ghosh2023statistically, david2020application, chen2017finite, moulinec1998numerical}. These simulations are often coupled with homogenization theories applied to representative volume elements (RVEs) of various microstructure types \cite{geers2010multi, kalamkarov2009asymptotic}, including fuel cell electrodes, battery electrodes, fiber-reinforced composites, metal matrix composites, and polycrystalline aggregates \cite{wargo2012selection, joos2012representative, pan2008numerical, jagadeesh2020review, kashkooli2017representative, benedetti2013three}. However, a comprehensive understanding of SP linkages remains elusive due to the scarcity of representative microstructure data. Material microstructures exhibit substantial stochastic variability \cite{bostanabad2018computational, holm2020overview}, so examining only a restricted region—whether with 2D imaging techniques (SEM, EBSD, TEM \cite{zaefferer2011critical, brodusch2018imaging}) or 3-D methods such as X-ray  \cite{maire2014quantitative, stock2008recent} and FIB/SEM \cite{xu2017enhanced}—offers only a narrow view of the possible morphologies and their associated properties.

\textbf{Descriptor-based microstructure characterization and reconstruction (MCR)} addresses this data gap by generating statistically representative microstructures that match prescribed statistical measures (e.g., n-point correlation functions), using methods such as the Yeong–Torquato algorithm \cite{torquato2002random, jiao2007modeling} and more recent gradient-based variants \cite{seibert2022descriptor, seibert2022microstructure, ma2024stochastic, blumer2025generative, seibert2023two}. The core idea is to generate new, statistically representative microstructures from limited data, thereby augmenting datasets for analysis and simulation. This enables researchers to (i) produce large ensembles of equivalent microstructures, (ii) explore hypothetical “interpolated” states between known morphologies, and (iii) reconstruct full 3D volumes from sparse 2D observations. A major advantage of descriptor-based MCR is its physical interpretability: the same descriptors used as loss terms can be directly compared with experimental measurements. However, the approach is limited by the information content of the chosen descriptors. For instance, relying solely on two-point correlations may not accurately characterize complex microstructures \cite{seibert2022microstructure}, as distinct microstructural patterns may share identical or nearly identical two-point statistics. Consequently, multiple physically distinct structures can serve as equally valid "solutions" to the descriptor-matching problem.

\textbf{Data-driven microstructure reconstruction} approaches, by contrast, learn microstructure distributions directly from images. Widely used architectures include deep generative models such as variational autoencoders (VAEs) \cite{jung2021exploration, zhang2022reconstruction, ji2024towards, attari2023towards, noguchi2021stochastic, white2024exploring} and generative adversarial networks (GANs) \cite{kench2022microlib, kench2021generating, hong2024mechanical, zheng2024generative, gayon2020pores, henkes2022three}.  For example, SliceGAN \cite{kench2021generating}, a GAN-based framework, has demonstrated 2D-to-3D reconstruction for a variety of anisotropic and multiphase microstructures. Instead of manually defining descriptors, these models learn salient morphological features from the data and can synthesize new instances that remain statistically close to the training set. More recently, diffusion models have emerged as state-of-the-art generative frameworks, achieving remarkable performance across diverse tasks due to their high fidelity, diversity, and stable training dynamics \cite{yang2023diffusion, ho2020denoising, dhariwal2021diffusion, song2020score, croitoru2023diffusion, cao2024survey, zhang2023text}. These models operate by progressively corrupting and then recovering data through a predefined forward and reverse diffusion process, typically formulated as a Markovian process—as in denoising diffusion probabilistic models (DDPMs) \cite{ho2020denoising}. In addition, latent diffusion models (LDMs) \cite{rombach2022high} perform the diffusion steps in a learned, lower-dimensional latent space, reducing the computational cost while preserving essential semantic and structural features. This allows them to generate samples that faithfully follow the underlying data manifold. Several studies have shown that 2D diffusion models can faithfully reproduce complex microstructures—spherical inclusions, ceramics, composites, grain structures, and more \cite{dureth2023conditional, lee2024microstructure}. Furthermore, to address the 2D-to-3D reconstruction challenge, Lee and Yun proposed a multi-plane denoising diffusion (MPDD) framework \cite{lee2024multi, lee2024denoising}. Their approach denoises multiple orthogonal 2D planes simultaneously with a pretrained 2D diffusion model, enforcing spatial coherence across slices and thereby synthesizing statistically equivalent 3D microstructures.

However, a key limitation of these approaches is their inability to generate microstructures beyond the statistical scope of the training dataset, restricting design space exploration. While useful for forward SP linkages, they struggle with inverse design—i.e., generating microstructures for user-defined properties—due to the ill-posed, non-unique nature of the SP relationship, where different structures can yield identical properties or descriptors. Thus, inverse design presents significant challenges: first, it necessitates exploration of a broader microstructure design space while preserving inherent microstructural characteristics; second, it demands methodologies capable of navigating the property design space and effectively addressing the inverse SP linkage.

\begin{figure} 
    \centering
    \includegraphics[width=0.9\linewidth]{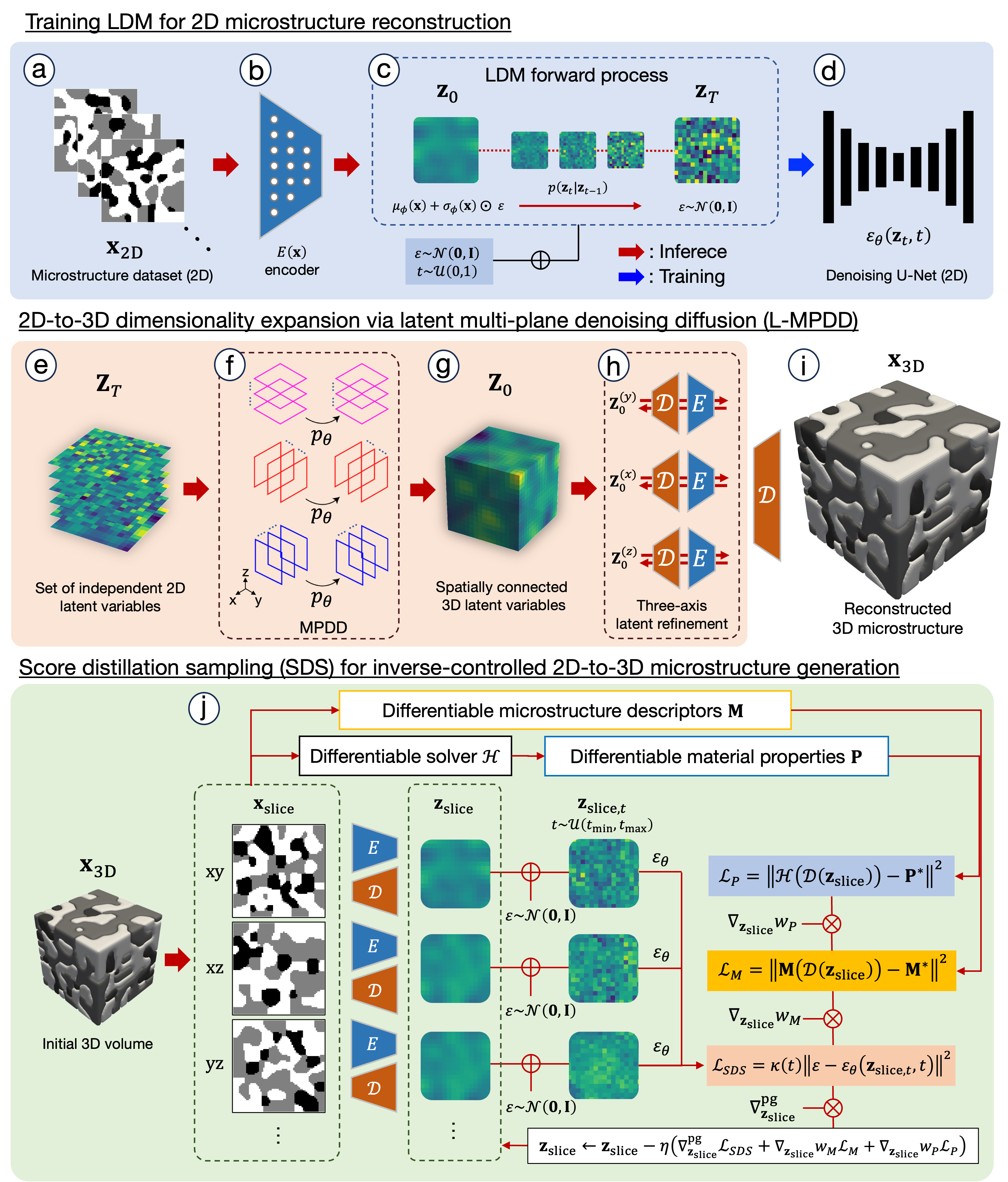}
    \caption{\textcolor{black}{MicroLad framework description: (a) acquire a 2D microstructure dataset, (b) encode the data into a latent space, (c) add noise to the latent variables, (d) train a U-Net to predict the noise, (e) initialize stacked 2D latent variables, (f) apply L-MPDD to spatially connect the 2D reverse diffusion processes, (g) generate 3D latent variables, (h) refine the latent representation to produce 3D voxels, (i) decode the final 3D microstructure, and (j) apply SDS for inverse-controlled microstructure generation with different objectives.}}
    \label{fig:framework}
\end{figure}

In this study, we introduce \textbf{MicroLad}, a latent diffusion framework that unifies 2D-to-3D microstructure reconstruction with inverse-controlled generation (Figure~\ref{fig:framework}). Leveraging a pretrained VAE on 2D microstructure data, a 2D LDM is trained to generate 2D microstructures from latent variables. The framework then performs dimensionality expansion from 2D to 3D using multi-plane denoising diffusion within the latent space (L-MPDD), decoding spatially connected 2D latent variables into coherent 3D microstructure volumes. Furthermore, using the 3D volume reconstructed via L-MPDD as initialization, the framework enables inverse-controlled 2D-to-3D microstructure generation via score distillation sampling (SDS) \cite{poole2022dreamfusion, mcallister2024rethinking}, which iteratively nudges each latent slice toward target microstructural descriptors and effective material properties. 

\paragraph{Key contributions:}

\begin{itemize}

    \item \textbf{Microstructural descriptor-guided 2D-to-3D microstructure generation}  
This work presents the first microstructure descriptor-controlled, data-driven 2D-to-3D microstructure generation framework. By combining SDS with differentiable objectives, the proposed framework enables the generation of microstructure samples guided by user-controllable microstructural descriptors, such as two-point correlation, volume fraction, and surface area. This offers great flexibility for exploring the microstructure design space. For instance, the framework can generate microstructures near the boundary of the training distribution (e.g., maximum volume fraction observed in the dataset), significantly expanding the scope of microstructure design and analysis.

    \item \textbf{Property-guided 2D-to-3D microstructure generation}  

This work also introduces the first property-guided 3D microstructure generation framework based solely on limited 2D observations, without requiring pre-labeled microstructure–property pairs. It leverages differentiable evaluations of material properties to guide generation. In other words, the framework not only facilitates exploration of the microstructure space by generating diverse samples, but also enables property-driven inverse design—addressing the ill-posed nature of inverse microstructure design problems and offering insights into how microstructures evolve under property-guided optimization.
          
    \item \textbf{2D-to-3D microstructure reconstruction within the latent space.} 
          MicroLad performs 2D-to-3D reconstruction entirely in the latent space of an LDM, exploiting a compact and relatively smooth low-dimensional manifold. This cuts wall-clock time from $\sim$30\,min for pixel-space MPDD to $<\!10$\,s for a $64^{3}$ volume using L-MPDD, while preserving full spatial coherence.          

\end{itemize}

The remainder of the paper formalizes the forward and inverse problems in the 2D-to-3D dimensionality expansion setting, describes the MicroLad architecture, and presents quantitative results on reconstruction accuracy, inverse-controlled generation, and computational cost. We validate the framework on binary carbonate \cite{prodanovic2015digital} and three-phase solid oxide fuel cell (SOFC) microstructures \cite{hsu2018mesoscale}, demonstrating accurate 3D reconstruction and inverse-controlled generation. The results are then discussed in terms of reconstruction performance and the effectiveness of inverse-controlled generation.

\begin{figure} 
    \centering
    \includegraphics[width=0.9\linewidth]{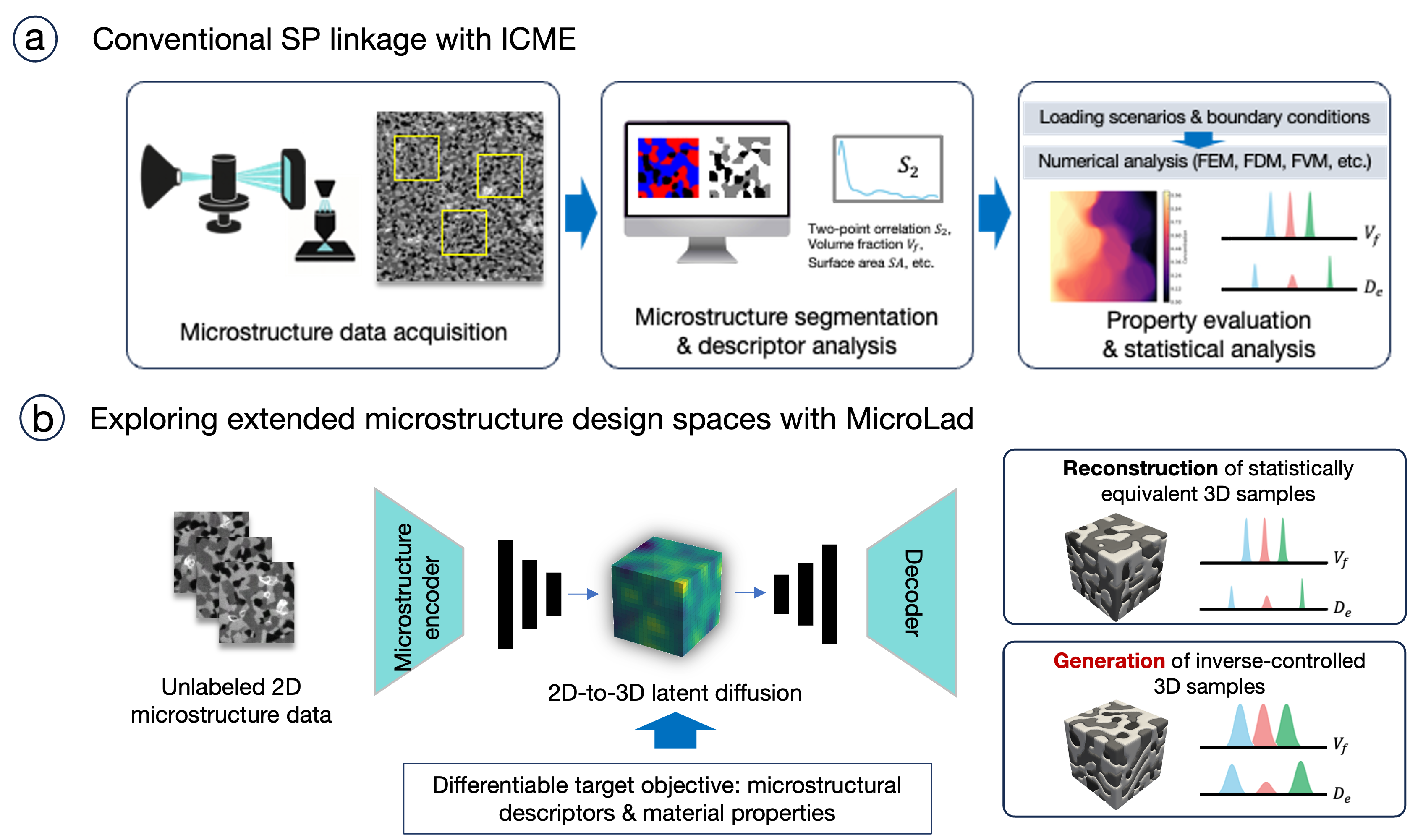}
    \caption{Comparison between (a) conventional computational materials engineering for structure–property (SP) linkage, which relies on experimental microstructure data acquisition, morphological and descriptor analysis, and numerical simulation-based property evaluation and (b) the proposed MicroLad framework, which explores 3D microstructure design spaces through reconstruction and inverse-controlled microstructure generation.}
    \label{fig:conventional_SP}
\end{figure}

\section{Problem formulation}
\label{sec:problem-formulation}

\subsection{Forward/inverse structure–property (SP) linkage}

In conventional SP linkage approaches within ICME framework (Figure~\ref{fig:conventional_SP}a), the workflow typically includes the following steps \cite{bostanabad2018computational, panchal2013key}: 1) acquisition of experimental micrograph data of material microstructures, 2) micrograph data analysis, which frequently involves filtering and segmentation of microstructure images followed by statistical analysis to extract microstructural descriptors such as n-point correlation functions and volume fractions of material phases, and 3) analysis of material responses and effective properties based on the provided microstructures, commonly using numerical methods under defined boundary conditions. In this forward linkage, one can investigate how variations in microstructural features influence material properties, enabling the evaluation of property ranges and performance predictions under specific loading scenarios, often combined with uncertainty quantification techniques \cite{chernatynskiy2013uncertainty}.

However, this workflow fundamentally supports forward SP linkage rather than inverse linkage. Since the ultimate goal of materials engineering is inverse design (i.e., developing materials that meet specific target properties) extensive forward analyses are required to thoroughly explore the property space associated with given microstructures. In principle, if a sufficiently large microstructure–property dataset were available, one could then approximate the inverse mapping—from property to structure.

In practice, however, acquiring a sufficient number of microstructure samples is often challenging, particularly for the 3D case. Additionally, obtaining extensive 3D microstructure–property data pairs can be highly time-consuming and often infeasible due to experimental or computational costs. To address this issue, we decompose the inverse problem into two components in this work. We begin by assuming the availability of unlabeled 2D data (cropped images from a single microscopy image), with the initial objective of estimating the corresponding 3D data distribution—represented as spatially connected 2D slices—through dimensionality expansion (Section~\ref{sec:dimensionality-expansion}). Subsequently, the goal is to sample 3D data guided by specified target objectives. However, because the initial data distribution is unconditional, an effective method is needed to steer the sampling process toward these objectives. In this paper, we utilize SDS combined with differentiable objectives, eliminating the need for pre-labeled microstructure–property data pairs (Section~\ref{sec:inverse-controlled-generation}). In other words, our goal is to enable both accurate 3D reconstruction and inverse-controlled 3D generation of microstructures, thereby facilitating the exploration of an expanded microstructure design space (Figure~\ref{fig:conventional_SP}b) using unlabeled 2D microstructure data.

\subsection{Dimensionality expansion (2D-to-3D) for microstructure data}\label{sec:dimensionality-expansion}
Suppose we have experimentally obtained microstructure data represented in a two-dimensional form:
\begin{equation}
\mathbf{x}_{\mathrm{2D}} \in \mathcal{X}_{\mathrm{2D}}, \quad \mathbf{x}_{\mathrm{2D}} \subseteq \mathbb{R}^{H \times W},
\end{equation}
where $\mathbf{x}_{\mathrm{2D}}$ denotes the experimental 2D microstructure data, and $H$ and $W$ represent its spatial dimensions (height and width, respectively). To model this data distribution, we train a 2D generative model $p_{\theta}$, that approximates the true but unknown distribution $p_{\text{data}}$ of experimental microstructures:
\begin{equation}
p_{\theta}(\mathbf{x}_{\mathrm{2D}}) \approx p_{\text{data}}(\mathbf{x}_{\mathrm{2D}}),
\end{equation}

\textcolor{black}{
Once trained, we generate new 2D samples from the learned distribution:
\begin{equation}\label{eq:2D_samples_depth}
\hat{\mathbf{x}}_{\mathrm{2D}}^{(i)} \sim p_{\theta}(\mathbf{x}_{\mathrm{2D}}), \quad i = 1, 2, \dots, D,
\end{equation}
where $\hat{\mathbf{x}}_{\mathrm{2D}}^{(i)}$ denotes the $i$-th generated 2D microstructure slice, 
and $D$ represents the total number of slices required to construct a 3D microstructure along the depth direction.
}In other words, we can think of 2D-to-3D microstructure reconstruction as generating multiple 2D slices that are spatially connected and collectively represent a 3D microstructure volume. In this context, the goal is now to generate multiple 2D slices $\{\hat{\mathbf{x}}_{\mathrm{2D}}^{(i)}\}_{i=1}^{D}$ that are spatially coherent and collectively represent a volume $\mathbf{x}_{\mathrm{3D}}$:
\begin{equation}
\mathbf{x}_{\mathrm{3D}} \in \mathcal{X}_{\mathrm{3D}}, \quad \mathbf{x}_{\mathrm{3D}} \subseteq \mathbb{R}^{H \times W \times D},
\end{equation}
subject to the spatial connectivity condition:
\begin{equation}
\hat{\mathbf{x}}_{\mathrm{2D}}^{(i)} \sim p_{\theta}(\mathbf{x}_{\mathrm{2D}}\mid \hat{\mathbf{x}}_{\mathrm{2D}}^{(j)}), \quad j \in \mathcal{N}(i),
\label{eq:spatial_connectivity}
\end{equation}
where $\mathcal{N}(i)$ denotes neighboring slice indices adjacent to slice $i$. This conditional distribution implies that the sampled 2D slices are not independent. Instead, each slice is generated conditioned on its adjacent slices to ensure spatial coherence in the 3D volume. In particular, it is assumed that the spatial connectivity condition must be enforced along the three orthogonal axes (height, width, and depth) to maintain structural coherence and form a cohesive 3D microstructure.

\subsection{Inverse-controlled 2D-to-3D microstructure generation}\label{sec:inverse-controlled-generation}
Common SP linkages aided by computational materials engineering typically begin with extracting descriptors from the available microstructure data (Figure~\ref{fig:conventional_SP}). This extraction step, yielding a set of microstructural descriptors \(\mathbf{M}\), can be formalized as:

\begin{equation}
\mathbf{M}: \mathcal{X} \rightarrow \mathcal{M}, \quad \mathbf{x} \mapsto \mathbf{M}(\mathbf{x}),
\end{equation}
where $\mathbf{M}(\mathbf{x})$ includes descriptors such as volume fractions, specific surface area, n-point correlation functions, grain size distributions, and other relevant morphological metrics. Subsequently, effective material properties ($\mathbf{P}$) can be evaluated through RVE-scale numerical simulations based on the microstructure data:
\begin{equation}
\mathbf{P}: \mathcal{X} \rightarrow \mathcal{P}, \quad \mathbf{x} \mapsto \mathbf{P}(\mathbf{x}),
\end{equation}
where $\mathbf{P}(\mathbf{x})$ denotes effective macroscopic (or homogenized) properties, such as effective diffusivity, thermal expansion coefficient, thermal conductivity, and modulus. The property evaluation involves solving physics-based problems defined as:
\begin{equation}
\mathbf{P}(\mathbf{x}) = \mathcal{H}\left(\mathbf{x}, \mathcal{B}\right),
\end{equation}
where $\mathcal{H}$ represents a numerical solver or homogenization operator (e.g., FEM or FFT-based solver), and $\mathcal{B}$ denotes the applied boundary conditions. Then, the effects of microstructural variations on the resultant material properties are analyzed to facilitate a comprehensive understanding of the SP relationships. 

In contrast to the forward SP linkage problem, the inverse problem aims to generate microstructure data ($\mathbf{x}$) that satisfy user-defined target microstructural descriptors ($\mathbf{M}^*$) and targeted effective material properties ($\mathbf{P}^*$). Mathematically, the inverse SP linkage can be formulated as:
\begin{equation}
\mathbf{M}^* \in \mathcal{M}, \quad \mathbf{P}^* \in \mathcal{P},
\end{equation}
we seek optimal microstructure data $\mathbf{x}^*$ that minimizes:
\begin{equation}
\mathbf{x}^* = \underset{\mathbf{x}\in\mathcal{X}}{\arg\min}\;\left[\mathcal{L}_M(\mathbf{M}(\mathbf{x}),\mathbf{M}^*) + \mathcal{L}_P(\mathbf{P}(\mathbf{x}),\mathbf{P}^*)\right],
\end{equation}
where $\mathcal{L}_M$ quantifies the discrepancy between generated descriptors $\mathbf{M}(\mathbf{x})$ and target descriptors $\mathbf{M}^*$ and $\mathcal{L}_P$ quantifies the discrepancy between computed properties $\mathbf{P}(\mathbf{x})$ and target properties $\mathbf{P}^*$. In the context of 2D-to-3D microstructure generation, the optimization problem can be formulated as 

\begin{equation}
\mathbf{x}_{\mathrm{3D}}^* = \underset{\mathbf{x}_{\mathrm{3D}}\in\mathcal{X}_{\mathrm{3D}}}{\arg\min}\;\left[\mathcal{L}_M(\mathbf{M}(\mathbf{x}_{\mathrm{3D}}),\mathbf{M}^*) + \mathcal{L}_P(\mathbf{P}(\mathbf{x}_{\mathrm{3D}}),\mathbf{P}^*)\right],
\end{equation}
subject to the critical constraint \eqref{eq:spatial_connectivity} that the 3D microstructure ($\mathbf{x}_{\mathrm{3D}}$) must be generated from spatially connected 2D slices ($\hat{\mathbf{x}}_{\mathrm{2D}}^{(i)}$) sampled from the trained $p_{\theta}$.

\section{Methodology}
\label{sec:Methodology}

\subsection{Denoising diffusion probabilistic models (DDPMs)}
DDPM \cite{ho2020denoising} defines a forward diffusion process as a Markov chain as follows:
\begin{equation}
q(\mathbf{x}_t \mid \mathbf{x}_{t-1}) = \mathcal{N}(\mathbf{x}_t; \sqrt{1 - \beta_t}\mathbf{x}_{t-1}, \beta_t \mathbf{I}), \quad t = 1, \dots, T,
\end{equation}
where $\mathbf{x}_0$ represents the original data sample, $\mathbf{x}_t$ is the noised version at timestep $t$, $\beta_t \in (0,1)$ denotes the diffusion schedule controlling the amount of noise added at each step, and $T$ is the total number of diffusion steps. After applying successive forward diffusion steps, the distribution converges to an isotropic Gaussian distribution:
\begin{equation}
q(\mathbf{x}_T \mid \mathbf{x}_0) = \mathcal{N}(\mathbf{x}_T; \mathbf{0}, \mathbf{I}).
\end{equation}

The reverse (generative) process is learned by parameterizing a neural network (typically denoted as $\mathbf{\varepsilon}_\theta$) to approximate the posterior:
\begin{equation}
p_\theta(\mathbf{x}_{t-1} \mid \mathbf{x}_t) = \mathcal{N}(\mathbf{x}_{t-1}; \mu_\theta(\mathbf{x}_t, t), \Sigma_\theta(\mathbf{x}_t, t)),
\end{equation}
where the neural network predicts the parameters of the reverse distribution, specifically trained via minimizing the following loss:
\begin{equation}
\mathcal{L}_{\text{DDPM}}(\theta) = \mathbb{E}_{t, \mathbf{x}_0, \varepsilon}[\|\varepsilon - \varepsilon_\theta(\sqrt{\bar{\alpha}_t}\mathbf{x}_0 + \sqrt{1 - \bar{\alpha}_t}\varepsilon, t)\|^2],
\end{equation}
where $\varepsilon \sim \mathcal{N}(\mathbf{0}, \mathbf{I})$ is Gaussian noise, and $\bar{\alpha}_t = \prod_{s=1}^{t}(1 - \beta_s)$. The sampling (generation) procedure involves iteratively applying the learned reverse denoising step from pure Gaussian noise:
\begin{equation}
\mathbf{x}_{t-1} \sim p_\theta(\mathbf{x}_{t-1} \mid \mathbf{x}_t), \quad \mathbf{x}_T \sim \mathcal{N}(\mathbf{0}, \mathbf{I}).
\end{equation}

\subsection{Latent diffusion models (LDMs)}
LDMs extend diffusion models by performing diffusion in a learned, lower-dimensional latent space. Specifically, LDM involves encoding the original data $\mathbf{x}$ into a latent representation $\mathbf{z}$ using an encoder $E$:
\begin{equation}\label{eq:encoder}
\mathbf{z} = E(\mathbf{x}), \quad \mathbf{z} \in \mathbb{R}^{C \times H' \times W'},
\end{equation}\label
where $C$, $H'$, and $W'$ represent latent space dimensions. The forward diffusion process in latent space is defined as:
\begin{equation}\label{eq:latent_forward}
q(\mathbf{z}_t \mid \mathbf{z}_{t-1}) = \mathcal{N}(\mathbf{z}_t; \sqrt{1 - \beta_t}\mathbf{z}_{t-1}, \beta_t \mathbf{I}), \quad t = 1, \dots, T,
\end{equation}
with latent variables $\mathbf{z}_t$ becoming increasingly noisy and converging to a Gaussian distribution:
\begin{equation}\label{eq:latent_gaussian}
q(\mathbf{z}_T \mid \mathbf{z}_0) = \mathcal{N}(\mathbf{z}_T; \mathbf{0}, \mathbf{I}).
\end{equation}

The reverse generative process in latent space, parameterized by ${\varepsilon}_\theta$, is similarly defined:
\begin{equation}\label{eq:latent_reverse}
p_\theta(\mathbf{z}_{t-1} \mid \mathbf{z}_t) = \mathcal{N}(\mathbf{z}_{t-1}; \mu_\theta(\mathbf{z}_t, t), \Sigma_\theta(\mathbf{z}_t, t)),
\end{equation}
trained via minimizing the latent diffusion loss:
\begin{equation}\label{eq:latent_loss}
\mathcal{L}_{\text{LDM}}(\theta) = \mathbb{E}_{t, \mathbf{z}_0, {\varepsilon}}[\|{\varepsilon} - {\varepsilon}_\theta(\sqrt{\bar{\alpha}_t}\mathbf{z}_0 + \sqrt{1 - \bar{\alpha}_t}{\varepsilon}, t)\|^2],
\end{equation}
where ${\varepsilon} \sim \mathcal{N}(\mathbf{0}, \mathbf{I})$. Finally, to generate new samples, latent variables are sampled through reverse diffusion and decoded back into data space using a decoder $\mathcal{D}$:
\begin{equation}\label{eq:decode}
\mathbf{x} = \mathcal{D}(\mathbf{z}_0), \quad \mathbf{z}_0 \sim p_\theta(\mathbf{z}_0).
\end{equation}

\subsection{2D-to-3D microstructure reconstruction}\label{sec:method_2dto3d_recon}
\textcolor{black}{According to Lee and Yun~\cite{lee2024denoising, lee2024multi}, MPDD (which performs iterative 2D denoising diffusion along the three Cartesian axes of a 3D volume) effectively approximates the 3D reverse diffusion process.} In this study, we adopt a similar approach but implement MPDD within the latent space (i.e., L-MPDD) using a pretrained 2D LDM. Because the pretrained 2D LDM lacks an explicit mechanism for enforcing 3D connectivity \eqref{eq:spatial_connectivity}, we leverage L-MPDD to implicitly achieve this goal by applying denoising diffusion along all three orthogonal latent dimensions: height \(H'\), width \(W'\), and depth \(D'\).

We begin by sampling a noisy latent representation of the 3D volume:
\begin{equation}
\mathbf{Z}_{T} = \{\mathbf{z}_{T}^{(i)}\}_{i=1}^{D'}, \quad \mathbf{z}_{T}^{(i)} \sim \mathcal{N}(\mathbf{0}, \mathbf{I}),
\end{equation}
where each latent slice $\mathbf{z}_{T}^{(i)}$ has dimensions $C \times H' \times W'$. The stacked latent slices are denoised through iterative reverse diffusion steps along the three Cartesian axes. For each timestep $t = T, T-1, \dots, 1$, we sequentially perform:

\textbf{Height-axis denoising ($y$-axis):}
\begin{equation}
\mathbf{z}_{t-1}^{(i)}[:, h, :] \sim p_{\theta}\left(\mathbf{z}_{t-1}^{(i)}[:, h, :] \mid \mathbf{z}_{t}^{(i)}[:, h, :]\right),\quad h = 1, \dots, H',
\end{equation}

\textbf{Width-axis denoising ($x$-axis):}
\begin{equation}
\mathbf{z}_{t-1}^{(i)}[:, :, w] \sim p_{\theta}\left(\mathbf{z}_{t-1}^{(i)}[:, :, w] \mid \mathbf{z}_{t}^{(i)}[:, :, w]\right),\quad w = 1, \dots, W',
\end{equation}

\textbf{Depth-axis denoising ($z$-axis):}
\begin{equation}
\mathbf{z}_{t-1}^{(d)} \sim p_{\theta}\left(\mathbf{z}_{t-1}^{(d)} \mid \mathbf{z}_{t}^{(d)}\right),\quad d = 1, \dots, D'.
\end{equation}

\textcolor{black}{Each denoising step applies the trained 2D diffusion model $p_\theta$ across multiple planes, thereby implicitly enforcing 3D spatial consistency. After completing the denoising process, we obtain the latent representation:
\begin{equation}\label{eq:depth_latent}
\mathbf{Z}_{0} = \{\mathbf{z}_{0}^{(i)}\}_{i=1}^{D'}, \quad \mathbf{z}_{0}^{(i)} \in \mathbb{R}^{C \times H' \times W'}.
\end{equation}}

\textcolor{black}{Due to the mismatch between the latent volume shape $\mathbf{Z} \in \mathbb{R}^{C \times D' \times H' \times W'}$ and the desired final microstructure resolution $\mathbf{x}_\mathrm{3D} \in \mathbb{R}^{D \times H \times W}$, we apply a refinement step to improve spatial coherence and fidelity. We initialize an empty volume at the final resolution:
\begin{equation}\label{eq:X_DHW}
\mathbf{X}^{(0)} \in \mathbb{R}^{D \times H \times W}.
\end{equation}}

For each refinement iteration $k = 1, \dots, K$, we sequentially refine along the three spatial axes:

\textbf{Height-axis refinement ($y$-axis):}
\begin{align}\label{eq:height_axis}
\mathbf{X}^{(k)}[:, y, :] &\leftarrow \mathcal{D}(\mathbf{Z}_{0}^{(y)}), \quad y = 1, \dots, H, \\
\mathbf{Z}_{0}^{(y)} &\leftarrow E(\mathbf{X}^{(k)}[:, y, :]).
\end{align}

\textbf{Width-axis refinement ($x$-axis):}
\begin{align}\label{eq:width_axis}
\mathbf{X}^{(k)}[:, :, x] &\leftarrow \mathcal{D}(\mathbf{Z}_{0}^{(x)}), \quad x = 1, \dots, W, \\
\mathbf{Z}_{0}^{(x)} &\leftarrow E(\mathbf{X}^{(k)}[:, :, x]).
\end{align}

\textbf{Depth-axis refinement ($z$-axis):}
\begin{align}\label{eq:depth_axis}
\mathbf{X}^{(k)}[z, :, :] &\leftarrow \mathcal{D}(\mathbf{Z}_{0}^{(z)}), \quad z = 1, \dots, D, \\
\mathbf{Z}_{0}^{(z)} &\leftarrow E(\mathbf{X}^{(k)}[z, :, :]).
\end{align}

After completing all three directional refinements, we average the results to obtain the refined volume for the $k$-th iteration:
\begin{equation}\label{eq:latent_average}
\mathbf{X}^{(k)} \leftarrow \frac{1}{3} \left( \mathbf{X}^{(k)}[z,:,:] + \mathbf{X}^{(k)}[:,y,:] + \mathbf{X}^{(k)}[:,:,x] \right).
\end{equation}

After $K$ refinement iterations, the final 3D microstructure is obtained (additional schematic in Appendix~\ref{sec:Refinement of 2D slices on orthogonal planes}):

\begin{equation}\label{eq:X_DHW_Krefine}
\mathbf{x}_{\mathrm{3D}} = \mathbf{X}^{(K)} \in \mathbb{R}^{D \times H \times W}.
\end{equation}

\subsection{Inverse-controlled 2D-to-3D microstructure generation with score-distillation sampling}\label{Method inverse-controlled 2D-to-3D microstructure generation}

After reconstructing the 3D microstructure volume \(\mathbf{x}_{\mathrm{3D}} \in \mathbb{R}^{D \times H \times W}\), we apply SDS to refine randomly selected 2D slices within the volume, guiding the 3D microstructure toward user-specified target microstructural descriptors (\(\mathbf{M}^*\)) and target effective material properties (\(\mathbf{P}^*\)). At each SDS iteration, a 2D slice \(\mathbf{x}_{\text{slice}}\) is randomly selected from the 3D volume along one of the three Cartesian axes (depth, height, or width). The selected slice is then encoded into the latent space:
\begin{equation}
\mathbf{z}_{\text{slice}} = E(\mathbf{x}_{\text{slice}}).
\end{equation}

To ensure consistency with the distribution learned by the pretrained 2D diffusion model, we compute the diffusion score loss \(\mathcal{L}_{\text{SDS}}\) in the latent space. A timestep \(t\) is randomly sampled from a predefined range \([t_{\min}, t_{\max}]\), and Gaussian noise \(\varepsilon \sim \mathcal{N}(\mathbf{0}, \mathbf{I})\) is added to form the noised latent slice:
\begin{equation}
\mathbf{z}_{\text{slice}, t} = \sqrt{\bar{\alpha}_t} \, \mathbf{z}_{\text{slice}} + \sqrt{1 - \bar{\alpha}_t} \, \varepsilon.
\end{equation}

The SDS loss is then computed by comparing the predicted noise with the true noise:
\begin{equation}
\mathcal{L}_{\text{SDS}} = \kappa(t)\,\|\varepsilon - \varepsilon_\theta(\mathbf{z}_{\text{slice},t},t)\|^2,
\end{equation}
\begin{equation}
\kappa(t) = \frac{1 - \bar{\alpha}_t}{\bar{\alpha}_t}.
\end{equation}

Next, the latent slice is decoded, and the descriptor matching loss is computed:
\begin{equation}
\hat{\mathbf{x}}_{\text{slice}} = \mathcal{D}(\mathbf{z}_{\text{slice}}),
\end{equation}
\begin{equation}
\mathcal{L}_{M} = \left\| \mathbf{M}(\hat{\mathbf{x}}_{\text{slice}}) - \mathbf{M}^* \right\|^2.
\end{equation}

\textcolor{black}{To define the objective function for property-guided generation, a differentiable physics solver (e.g., differentiable FEM, \(\mathcal{H}\)) is employed, and the corresponding loss is formulated as follows:
\begin{equation}\label{eq:property_loss}
\mathcal{L}_{P} = \left\| \mathcal{H}(\hat{\mathbf{x}}_{\text{slice}}) - \mathbf{P}^* \right\|^2.
\end{equation}}

\textcolor{black}{
Then, we update the latent slice as follows:
\begin{align}
\mathbf z_{\text{slice}}
\;\leftarrow\;
\mathbf z_{\text{slice}}
- \eta \Big(
\nabla^{\mathrm{pg}}_{\mathbf z_{\text{slice}}}\mathcal L_{\text{SDS}}
+ w_{M}\,\nabla_{\mathbf z_{\text{slice}}}\mathcal L_{M}
+ w_{P}\,\nabla_{\mathbf z_{\text{slice}}}\mathcal L_{P}
\Big),
\end{align}}

\textcolor{black}{
where $w_M$ and $w_P$ are weighting terms and $\eta$ is the learning rate. Here $\nabla^{\mathrm{pg}}$ denotes a pseudo-gradient update. In the SDS term, we treat $\varepsilon_\theta(\cdot)$ as a constant target and do not back-propagate through the frozen diffusion network (i.e., we drop the Jacobian $\partial \varepsilon_\theta / \partial \mathbf{z}_{\text{slice}}$). This mirrors the original SDS setup \cite{poole2022dreamfusion}, where the diffusion prior is frozen and only the external generator is optimized. In our case, the “external generator” is the latent slice itself.} The updated latent slice is decoded and replaces the original:
\begin{equation}
\mathbf{x}_{\text{slice}}^{\text{new}} = \mathcal{D}(\mathbf{z}_{\text{slice}}).
\end{equation}

Although SDS is applied to individual 2D slices of the 3D volume, these slices are spatially connected through the reconstruction process. Therefore, we assume that the iterative optimization of local 2D slices—each guided by target descriptors and material properties—propagates improvements throughout the volume. As the SDS steps are repeated across random slices and along different Cartesian directions, this process gradually leads to a global refinement of the entire 3D microstructure. In addition, we assume the following conditions to justify that iterative application of SDS on spatially connected 2D slices in a LDM setting leads to global optimization of the full 3D microstructure volume:
\begin{enumerate}
    \item \textbf{Local consistency}: The optimization of each individual 2D latent slice via SDS is locally consistent, meaning improvements in descriptors and properties in one latent slice coherently propagate to adjacent slices due to the spatial connectivity ~\eqref{eq:spatial_connectivity}.
    
    \item \textbf{Spatial connectivity and overlap}: Each latent slice shares overlapping latent-space coordinates with its adjacent slices, allowing coherent spatial information to propagate effectively through iterative updates.
    
    \item \textbf{Smoothness and Lipschitz continuity of latent space}: Due to the continuous and smooth nature of the pretrained latent space representation learned by the LDM, the total loss function \(\mathcal{L}_{\text{total}}\) is assumed to have Lipschitz-continuous gradients. Formally, there exists a constant \(L > 0\) such that for all latent representations \(\mathbf{z}_a, \mathbf{z}_b\):
    \begin{equation}
    \|\nabla \mathcal{L}_{\text{total}}(\mathbf{z}_a) - \nabla \mathcal{L}_{\text{total}}(\mathbf{z}_b)\| \leq L \|\mathbf{z}_a - \mathbf{z}_b\|.
    \label{eq:lipschitz_total_loss}
    \end{equation}

    \item \textbf{Coherent diffusion prior from LDM}: The pretrained 2D LDM effectively captures coherent microstructural features, constraining optimized slices to remain within realistic microstructural latent manifolds, thus implicitly enforcing global coherence and realistic microstructural patterns.
\end{enumerate}

\subsection{Implementation details}

All models in the MicroLad framework were implemented using PyTorch and trained on a single NVIDIA RTX 4090 GPU with 24GB VRAM. Differentiable property evaluations, including effective diffusivity, were performed using a custom finite element solver implemented in PyTorch. 

The VAE was pretrained on a variety of microstructure types—including Voronoi grains, generic grain microstructures, spherical inclusions, aligned elliptical inclusions, graphite, nickel-based superalloys (NBSA), Gaussian random fields (GRF), and random elliptical inclusions—using the open-source microstructure dataset from Robertson et al.\cite{robertson2024micro2d}. It was then fine-tuned on the target microstructures of interest: binary carbonate and three-phase SOFC microstructures. Latent representations were encoded as $4 \times 16 \times 16$ feature maps. The LDM operates on this latent space and was trained for 1000 diffusion steps using a linear noise schedule $\beta \in [1\times10^{-4}, 2\times10^{-2}]$. Full architectural specifications and hyperparameters for both the VAE and LDM are provided in Appendix \ref{sec:configuration parameters for the models}.

At inference time, 3D microstructures are generated by applying L-MPDD along depth, height, and width axes to sample a spatially coherent 3D latent volume. This latent volume is then decoded slice-by-slice and refined via a three-axis latent update procedure to improve structural coherence (Section~\ref{sec:method_2dto3d_recon}).

SDS is then applied iteratively to optimize randomly selected 2D slices from the reconstructed 3D volume (Section~\ref{Method inverse-controlled 2D-to-3D microstructure generation}). Each latent slice is updated using a gradient-based loss that combines (i) a diffusion score loss to maintain realism, (ii) descriptor-based losses for matching volume fraction or surface area, and (iii) a differentiable physical loss based on effective diffusivity computed via FEM.

\subsection{Dataset preparation}

Within the MicroLad framework, we assume that the sole microstructure datum is a single 2D image.  From this image, small cropped patches \((64 \times 64)\) are extracted and used to train the VAE and LDM models, which subsequently generate new 2D samples. This setup is essential for verifying whether the proposed framework can reconstruct full 3D volumes from independently sampled 2D patches, without any prior knowledge of their 3D spatial connectivity.

For the binary microstructure case, patches were randomly sampled from a $318\times337$ binary image of a carbonate structure \cite{prodanovic2015digital} to demonstrate 2D-to-3D reconstruction using MicroLad. For the multiphase case, a SOFC anode microstructure \cite{hsu2018mesoscale} was selected, which consists of three phases: pores, nickel (Ni), and yttria-stabilized zirconia (YSZ). In SOFC anodes, each phase plays a distinct role in electrochemical performance: Ni serves as the electron-conducting phase and catalyzes hydrogen oxidation, YSZ is an oxygen-ion conductor that enables ion transport from the electrolyte, and the pore phase provides gas channels for fuel diffusion. A single $950\times843$ microstructure image was taken from a reference SOFC dataset and manually segmented into the three constituent phases before being used for training. For each case, a total of 300 (binary) and 600 (three-phase) cropped images were randomly sampled, and 8-fold data augmentation was applied to enhance training diversity.

\subsection{Metrics}

\paragraph{Two-point correlation function.}
To evaluate the spatial distribution of each material phase, we compute the two-point correlation function, denoted as \( S_2(r) \). This function measures the probability that two points separated by a distance \( r \) both belong to the same phase and is widely used to characterize spatial patterns such as clustering, dispersion, and periodicity in heterogeneous microstructures.

Let \( \chi(\mathbf{r}) \in \{0,1\} \) denote the binary phase indicator function defined over spatial coordinates \( \mathbf{r} \in \Omega \), where \( \chi(\mathbf{r}) = 1 \) if \( \mathbf{r} \) belongs to the target phase and 0 otherwise. The isotropic two-point correlation function is then defined as the ensemble (or spatial) average of the autocorrelation of the phase field:
\begin{equation}
S_2(r) = \left\langle \chi(\mathbf{r}) \cdot \chi(\mathbf{r} + \mathbf{d}) \right\rangle_{\|\mathbf{d}\| = r},
\end{equation}
where \( \mathbf{d} \) is a displacement vector of fixed magnitude \( r \), and \( \langle \cdot \rangle \) denotes spatial averaging over all \( \mathbf{r} \in \Omega \). In practice, we compute \( S_2(r) \) by taking the autocorrelation of the binary phase mask using a fast Fourier transform. To ensure directional invariance, the function is computed slice-wise across all three Cartesian directions and averaged to yield a single representative curve per 3D microstructure. To quantify how closely a generated microstructure reproduces the spatial statistics of the original data, we compute the relative error between the areas under their respective \(S_2(r)\) curves:
\[
\varepsilon_{\text{rel}} = \left|1 - \frac{A_{\text{gen}}}{A_{\text{ori}}}\right| \times 100,
\]
where \(A_{\text{gen}}\) and \(A_{\text{ori}}\) are the integrated areas under the \(S_2(r)\) curves for the generated and original samples, respectively.  
This metric captures the overall difference in spatial correlation structure, independent of fine-scale fluctuations.

\paragraph{Volume fraction (\(V_f\)).}
Volume fraction represents the proportion of each material phase present in a microstructure. It is one of the most fundamental and interpretable descriptors, particularly important for systems where phase balance determines performance—such as porous electrodes or composite materials. 

\paragraph{Relative surface area ($SA$).}
Relative surface area is a commonly used microstructural descriptor that quantifies the interfacial complexity of each phase, which is particularly important in multiphase materials where phase boundaries influence functional performance (e.g., ion/electron transport pathways in SOFCs) \cite{kench2021generating, hsu2018mesoscale}. In our implementation, surface area is computed by applying Gaussian smoothing to the phase masks and computing their total variation (i.e., the gradient magnitude across the spatial domain) to enable differentiability. The surface area of each phase is then normalized to obtain its relative contribution.

\paragraph{Relative diffusivity (\(D_e\)).}
Relative diffusivity provides a physics-based measure of transport capability within each material phase. In this study, the 2D relative diffusivity for each material phase is computed using a custom differentiable FEM solver, which interprets the binary mask of each phase as a spatially varying diffusivity field. The governing equation is the steady-state diffusion equation:

\begin{equation}\label{eq:rel_diff_governing}
\nabla \cdot \left( \kappa(\mathbf{r}) \nabla u(\mathbf{r}) \right) = 0, \quad \mathbf{r} \in \Omega,
\end{equation}

where \( \kappa(\mathbf{r}) \) is the local diffusivity (or conductivity) defined from the microstructure mask, \( u(\mathbf{r}) \) is the potential field (e.g., concentration), and \( \Omega \) is the spatial domain. Dirichlet boundary conditions are imposed to drive diffusion across the domain:

\begin{equation}\label{eq:rel_diff_bc}
u(\mathbf{r_1}) = 1 , \qquad u(\mathbf{r_0}) = 0.
\end{equation}

The effective diffusivity \( D_e \) is computed from the total energy dissipation:

\begin{equation}\label{eq:rel_diff_homogenization}
D_e = \frac{1}{|\Omega|} \int_{\Omega} \kappa(\mathbf{r}) \left\| \nabla u(\mathbf{r}) \right\|^2 \, d\mathbf{r}.
\end{equation}

This formulation is sensitive to both phase connectivity and tortuosity~\cite{holzer2013redox, zhang2023analysing}, making \( D_e \) an indicator of microstructural functionality. \textcolor{black}{Additional details on the FEM implementation are provided in Section~\ref{sec:Implementation of differentiable FEM}.}

\section{Results}
\label{sec:results}

\subsection{2D-to-3D reconstructed microstructure samples}\label{sec:results_2dto3d_recon}
\begin{figure} 
    \centering
    \includegraphics[width=1.0\linewidth]{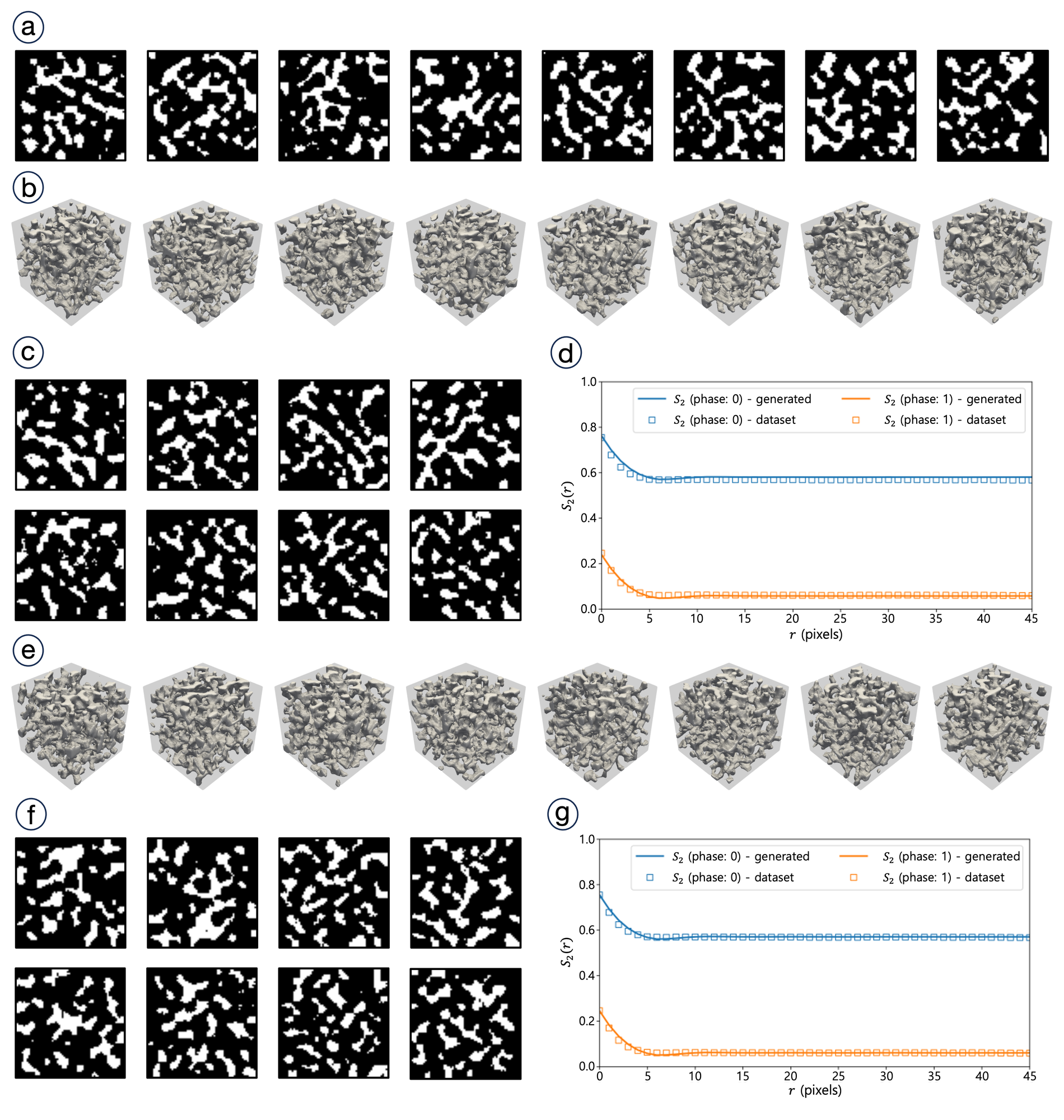}
    \caption{2D-to-3D microstructure reconstruction results for binary microstructure: (a) original 2D microstructure images, (b) reconstructed 3D microstructure, (c) random cross-sectional slices of (b), (d) two-point correlation functions of 2D slices from the reconstructed 3D samples in (b) compared to those from the original 2D dataset in (a), (e) reconstructed 3D microstructure guided by target two-point correlation functions, (f) random cross-sectional slices of (e), and (g) two-point correlation functions of 2D slices from the guided reconstruction in (e) compared to the original 2D dataset.}
    \label{fig:Recon_carbonates}
\end{figure}

To demonstrate the 2D-to-3D microstructure reconstruction performance on binary microstructures, a 3D carbonate microstructure was reconstructed using MicroLad, as shown in Figure~\ref{fig:Recon_carbonates}. The original 2D carbonate microstructure exhibits spatially distributed binary phases, where black pixels (0) represent the matrix phase and white pixels (1) represent the pore phase, as illustrated in Figure~\ref{fig:Recon_carbonates}(a). By training the LDM on this dataset and performing 2D-to-3D dimensionality expansion, MicroLad generates multiple spatially connected 3D samples (Figure~\ref{fig:Recon_carbonates}(b)). As seen in the randomly selected cross-sectional slices (Figure~\ref{fig:Recon_carbonates}(c)) from the reconstructed 3D volumes, the volumes display coherent and well-connected binary structures that visually resemble the original 2D micrographs. The two-point correlation functions of the generated slices (along all three orthogonal directions) also closely match those of the original 2D dataset (Figure~\ref{fig:Recon_carbonates}(d)), indicating that MicroLad effectively preserves the spatial statistics of the microstructure.

Furthermore, to enhance microstructure reconstruction with respect to spatial correlations, SDS is applied to guide the generation process by minimizing the loss between the two-point correlation function of the generated microstructure and that of the original dataset. This results in reconstructed 3D microstructures that more closely match the spatial correlation characteristics of the training data, as shown in Figure~\ref{fig:Recon_carbonates}(e). Although the initial reconstructions already exhibit high visual and statistical fidelity, the incorporation of \( S_2 \)-guided SDS further improves the alignment between the \( S_2 \) curve of the generated slices (solid lines) and that of the training dataset (square markers). As summarized in Table~\ref{tab:2d-to-3d reconstruction accuracy}, both the mean absolute error (MAE) and the relative error in \( S_2 \) decrease when \( S_2 \) guidance is applied, achieving relative errors below 1\%. These results demonstrate that the proposed MicroLad framework effectively combines generative sampling with conventional microstructural descriptors to achieve descriptor-informed 2D-to-3D microstructure reconstruction.

\begin{figure} 
    \centering
    \includegraphics[width=1.0\linewidth]{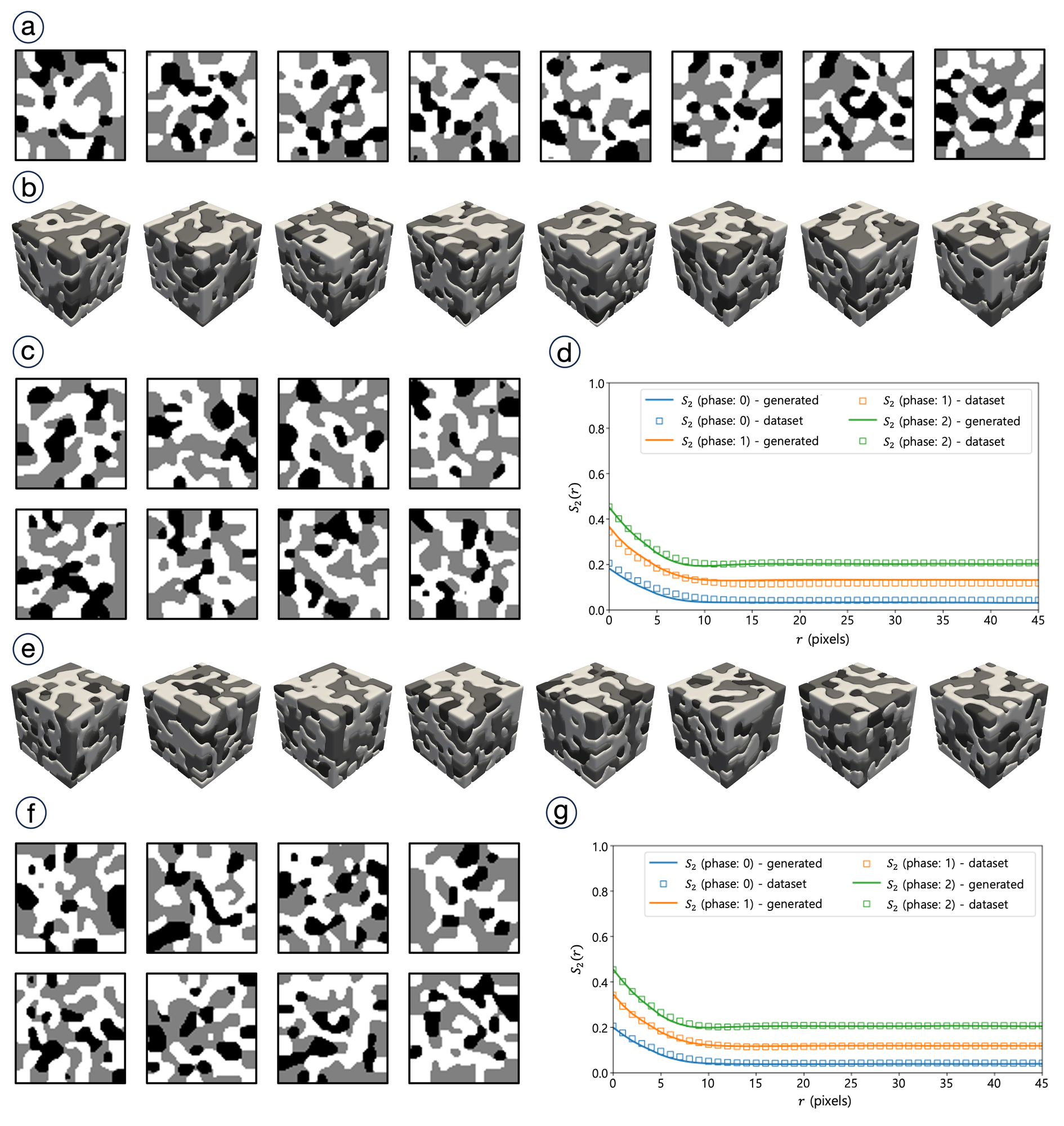}
    \caption{2D-to-3D microstructure reconstruction results for three-phase microstructure: (a) original 2D microstructure images, (b) reconstructed 3D microstructure, (c) random cross-sectional slices of (b), (d) two-point correlation functions of 2D slices from the reconstructed 3D samples in (b) compared to those from the original 2D dataset in (a), (e) reconstructed 3D microstructure guided by target two-point correlation functions, (f) random cross-sectional slices of (e), and (g) two-point correlation functions of 2D slices from the guided reconstruction in (e) compared to the original 2D dataset.}
    \label{fig:Recon_SOFC}
\end{figure}

Figure~\ref{fig:Recon_SOFC} presents the 2D-to-3D microstructure reconstruction results for the three-phase SOFC microstructures. In the original 2D micrograph (Figure~\ref{fig:Recon_SOFC}(a)), the darkest regions represent the pore phase (0), the intermediate gray regions denote the YSZ phase (1), and the brightest regions correspond to the Ni phase (2). The reconstructed 3D microstructure volumes are shown in Figure~\ref{fig:Recon_SOFC}(b), and randomly selected 2D cross-sectional slices from these volumes are illustrated in Figure~\ref{fig:Recon_SOFC}(c). Similar to the binary reconstruction case, the generated samples exhibit strong visual similarity to the original 2D micrographs, indicating that MicroLad effectively reconstructs complex multiphase microstructures. Furthermore, the two-point correlation functions for each material phase in the generated samples closely match those of the original dataset (Figure~\ref{fig:Recon_SOFC}(d)). However, as shown in Table~\ref{tab:2d-to-3d reconstruction accuracy}, the relative error in \( S_2 \) is slightly higher than in the binary case. This increase is primarily attributed to the overall reduction in \( S_2 \) values as the number of phases increases, which results in a smaller denominator when computing the relative error.

To mitigate discrepancies in spatial statistics, \( S_2 \)-guided optimization was applied using SDS. The resulting 3D microstructure samples and their corresponding randomly selected 2D slices are shown in Figure~\ref{fig:Recon_SOFC}(e) and Figure~\ref{fig:Recon_SOFC}(f), respectively. As illustrated in Figure~\ref{fig:Recon_SOFC}(g), the discrepancy between the \( S_2 \) curves of the generated data and those of the original dataset is reduced. This improvement is also reflected in Table~\ref{tab:2d-to-3d reconstruction accuracy}, where the maximum relative error is 6.13\%. It is also worth noting that the relatively higher error in the pore phase arises from its lower volume fraction (approximately 0.2), as seen in the \( S_2 \) curve near \( r = 0 \). The smaller volume fraction results in lower \( S_2 \) values and, consequently, a smaller denominator when computing relative errors, making the metric more sensitive to small discrepancies compared to the Ni and YSZ phases.

To further validate the reconstruction performance of the proposed MicroLad framework, we compare its accuracy on three-phase SOFC microstructures with two existing methods: MPDD~\cite{lee2024denoising, lee2024multi} and SliceGAN~\cite{kench2021generating}. As shown in Appendix~\ref{sec:comparison of microstructure reconstruction performance}, the maximum relative error in terms of two-point correlation for MPDD is 12.44\%, and for SliceGAN it is 11.01\%. In contrast, MicroLad achieves lower errors when \( S_2 \) guidance is applied, demonstrating strong performance in 2D-to-3D microstructure reconstruction with respect to spatial correlation.

\begin{table}[ht]
  \caption{2D-to-3D reconstruction accuracy based on two-point correlation functions ($S_2$).
           Metrics are computed by comparing 2D slices of reconstructed 3D
           microstructures with the original 2D micrographs.}
  \centering
  \renewcommand{\arraystretch}{1.2}
  \begin{tabular}{l l c c c}
    \toprule
    \textbf{Reconstruction method} & \textbf{Microstructure type} & \textbf{Phase} & \textbf{MAE} & \makecell{\textbf{Relative error} (\%) \\ $\varepsilon_{\text{rel}}$} \\
    \midrule
    Without $S_2$ guidance & Binary & 0 & 0.0164 & 2.03 \\
                           & (matrix / pore)       & 1 & 0.0043 & 4.62 \\
    \cmidrule(lr){2-5}
                           & Three-phase & 0 & 0.0129 & 10.81 \\
                           & (pore / YSZ / Ni)            & 1 & 0.0141 & 8.31  \\
                           &             & 2 & 0.0180 & 2.80  \\
    \midrule
    With $S_2$ guidance    & Binary & 0 & 0.0032 & 0.36 \\
                           &        & 1 & 0.0020 & 0.41 \\
    \cmidrule(lr){2-5}
                           & Three-phase & 0 & 0.0059 & 6.13 \\
                           &             & 1 & 0.0028 & 0.59 \\
                           &             & 2 & 0.0031 & 0.24 \\
    \bottomrule
  \end{tabular}
  \label{tab:2d-to-3d reconstruction accuracy}
\end{table}

\subsection{Inverse-controlled 2D-to-3D generated microstructure samples}\label{Results: Inverse-controlled 2D-to-3D generated microstruture samples}

To expand the microstructure analysis and design space and enrich the dataset by capturing microstructural variability, we demonstrate the inverse-controlled 2D-to-3D microstructure generation capability of MicroLad. Specifically, we control \(V_f\), \(SA\), and \(D_e\) for each microstructural phase (pore, Ni, and YSZ) in the SOFC anode microstructure during 2D-to-3D generation. To validate the model’s ability to generate samples near the boundary of the data distribution, the target values for each objective were set to the maximum observed in the original 2D training dataset. For each case, 50 microstructure volumes were generated to evaluate the statistical distribution of the generated samples.

\begin{figure} 
    \centering    \includegraphics[width=1.0\linewidth]{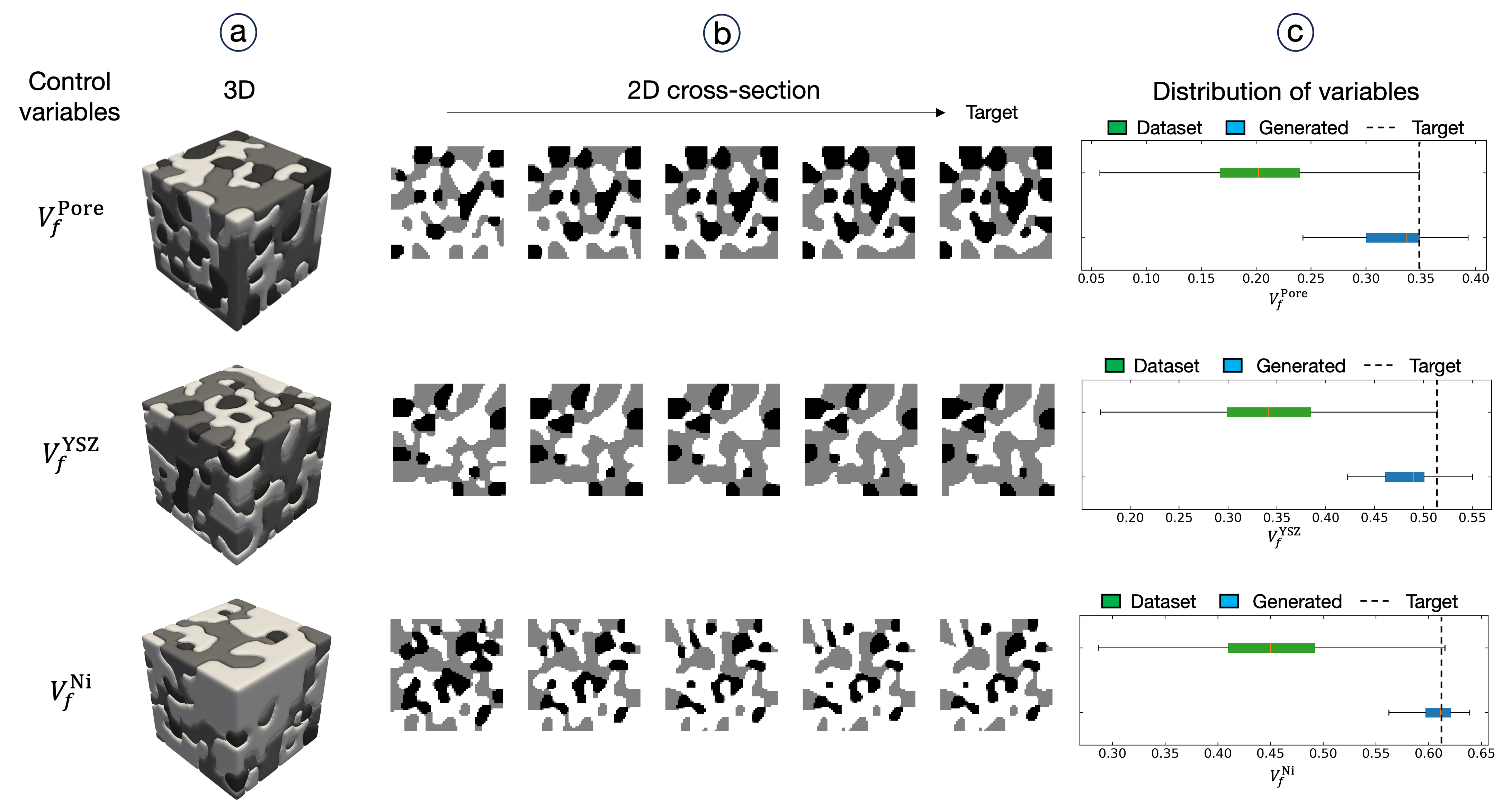}
    \caption{Inverse 2D-to-3D generation of three-phase microstructures with controlled volume fractions for each material phase: (a) generated 3D volumes, (b) variation in 2D cross-sectional images (randomly selected slices) as SDS steps increase, and (c) distribution of variables in the training dataset compared to generated samples with target objectives.}
    \label{fig:Inverse 2D-to-3D geration Vf}
\end{figure}

Figure~\ref{fig:Inverse 2D-to-3D geration Vf} illustrates the generated 3D microstructure volumes targeting the maximum $V_f$ observed in the original 2D training dataset for each phase (Figure~\ref{fig:Inverse 2D-to-3D geration Vf}(a)), along with the evolution of 2D cross-sectional slices as SDS progresses (Figure~\ref{fig:Inverse 2D-to-3D geration Vf}(b)). As shown, the volume fraction of each target phase clearly increases during the SDS process. Eventually, the distribution of 2D slices within the 3D volume shifts toward the desired targets, as depicted in Figure~\ref{fig:Inverse 2D-to-3D geration Vf}(c), leading to an overall increase in the volume fraction of the corresponding phase in 3D space. It is also worth noting that the generated samples retain their visual realism and do not exhibit structural artifacts (e.g., pixel noise), underscoring the robustness of the inverse-controlled generation process.

\begin{figure} 
    \centering    \includegraphics[width=1.0\linewidth]{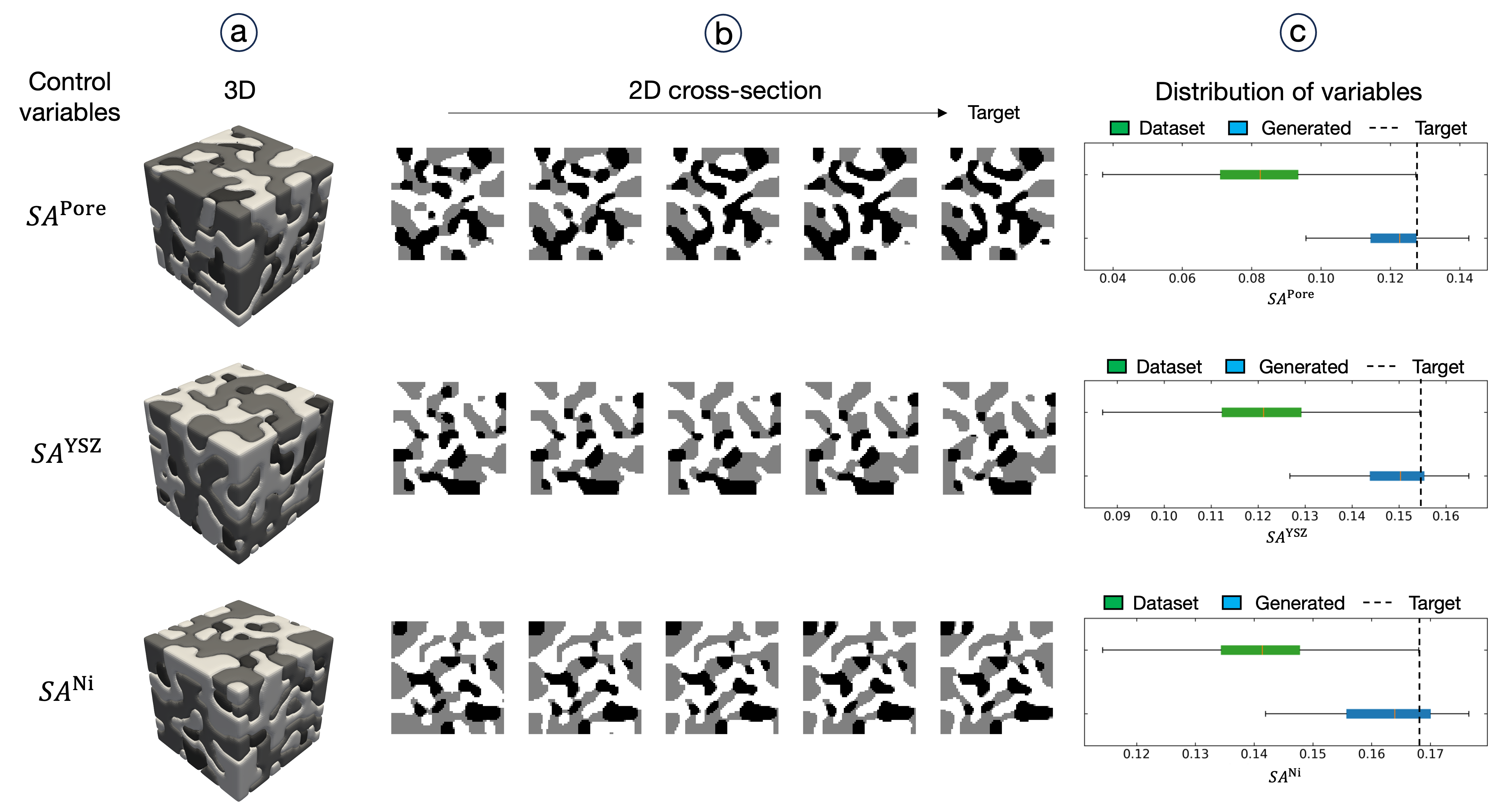}
    \caption{Inverse 2D-to-3D generation of three-phase microstructures with controlled relative surface area for each material phase: (a) generated 3D volumes, (b) variation in 2D cross-sectional images (randomly selected slices) as SDS steps increase, and (c) distribution of variables in the training dataset compared to generated samples with target objectives.}
    \label{fig:Inverse 2D-to-3D geration SA}
\end{figure}

Figure~\ref{fig:Inverse 2D-to-3D geration SA} presents the generated 3D microstructure volumes targeting the maximum relative surface area observed in the original dataset for each phase. In contrast to the $V_f$-guided generation, the surface area-guided results reveal a tendency for each material phase to form elongated, thin, and interconnected structures that maximize interfacial area with adjacent phases. This behavior also demonstrates that the proposed inverse-controlled generation framework can capture and reflect physically meaningful trends, as the surface area between phases plays a critical role in the electrochemical performance of SOFC anodes by influencing phase boundary reactions~\cite{hsu2018mesoscale}. These results highlight MicroLad’s ability to control and analyze phase-specific morphological features and to generate realistic 3D structures suitable for studying interfacial complexity.

\begin{figure} 
    \centering    \includegraphics[width=1.0\linewidth]{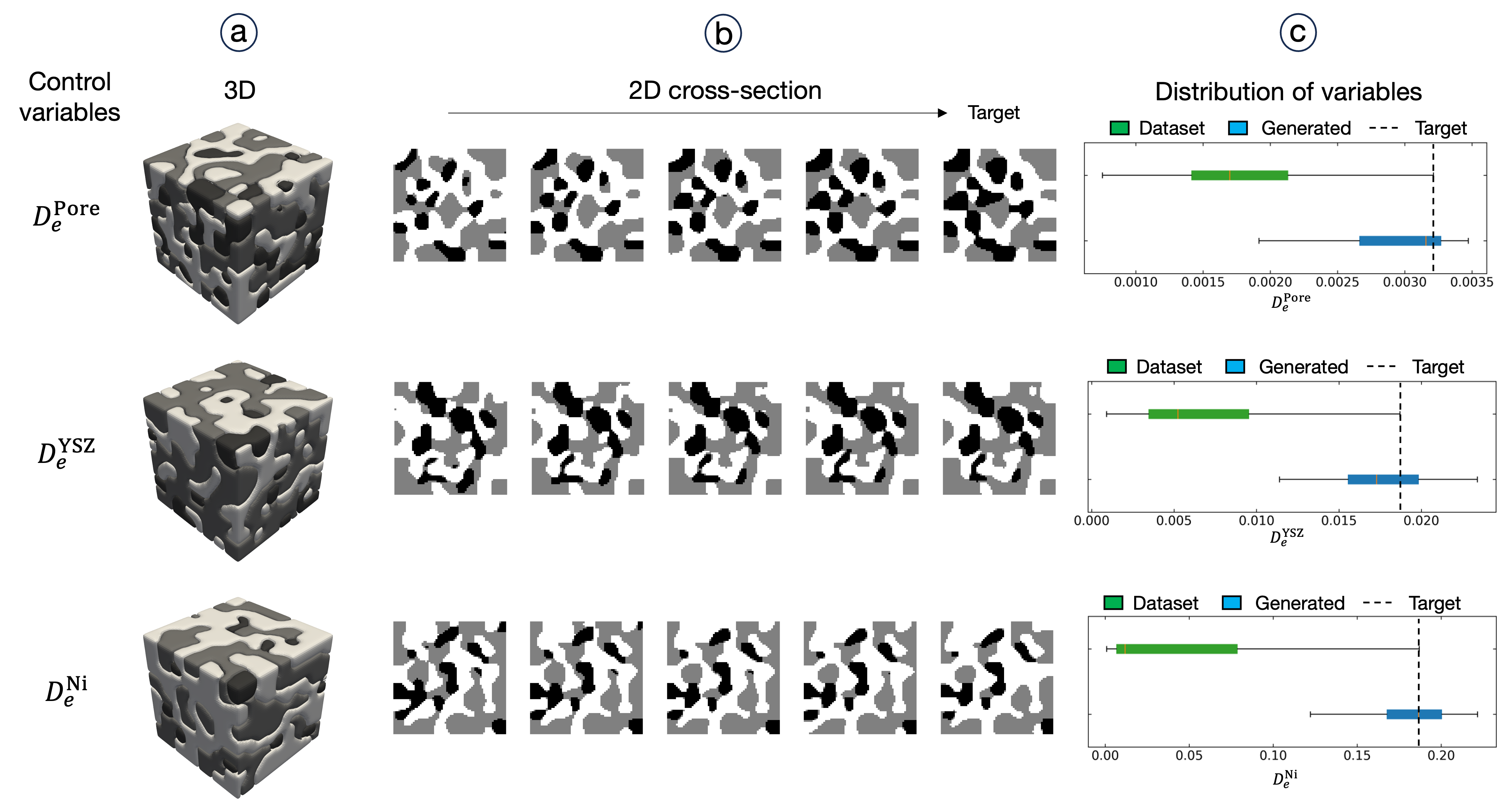}
    \caption{Inverse 2D-to-3D generation of three-phase SOFC microstructures with controlled relative diffusivity for each material phase: (a) generated 3D volumes, (b) variation in 2D cross-sectional images (randomly selected slices) as SDS steps increase, and (c) distribution of variables in the training dataset compared to generated samples with target objectives.}
    \label{fig:Inverse 2D-to-3D geration De}
\end{figure}

Next, property-guided 2D-to-3D generation is performed using the target \(D_e\), as shown in Figure~\ref{fig:Inverse 2D-to-3D geration De}. As the generation progresses toward the specified target, the phase of interest begins to grow and interconnect, forming an effective pathway for diffusion (or conduction)—not merely increasing its volume fraction (Figure~\ref{fig:Inverse 2D-to-3D geration Vf}) or interfacial area (Figure~\ref{fig:Inverse 2D-to-3D geration SA}). This trend is particularly evident in the case targeting increased \(D_e^{\text{Pore}}\), as shown in Figure~\ref{fig:Inverse 2D-to-3D geration De}(b), where initially disconnected pores (due to their relatively low volume fraction compared to the YSZ and Ni phases) gradually become interconnected, improving the overall diffusivity. The effect becomes even more pronounced when the target diffusivity is set to twice the maximum value observed in the training dataset, as illustrated in Figure~\ref{fig app: Inverse Deff x2}. In this case, the pores form a more continuous and interconnected network, enabling more efficient gas diffusion pathways throughout the volume. These results show that the proposed MicroLad framework can not only generate realistic 3D microstructures from 2D inputs, but also embed physics-based property objectives directly into the generative process. This capability enables a deeper analysis of how microstructures may evolve to meet specific functional requirements, thereby advancing our understanding of the complex relationship between morphological features and effective material properties. Additional generated samples for each case, along with the progression of 2D slices during optimization, are provided in Appendix~\ref{sec:additional inverse 2D-to-3D generation results}.

\section{Discussion}
\label{sec:Discussion}

\subsection{Distributional shifts and objective trends in latent space}

To assess whether generation with the proposed MicroLad framework genuinely expands the microstructure analysis space, we can analyze the property distribution as done in Section~\ref{Results: Inverse-controlled 2D-to-3D generated microstruture samples}. Additionally, we can examine the data distribution in the latent space of the LDM. To investigate the behavior of the inverse-controlled generation and the distance between the 2D-to-3D generated data (i.e., 2D slices within the 3D volume) and the original 2D data, we examine their latent embeddings using uniform manifold approximation and projection (UMAP) \cite{mcinnes2018umap}. A detailed explanation of the latent embedding procedure and UMAP projection is provided in Appendix~\ref{sec:umap}. 

Figure~\ref{fig:UMAP clustering}(a) visualizes the distribution of data points from both the inverse-controlled generation and the original dataset, with the latter outlined by a red dotted boundary. Black dots indicate the centroid of each group of generated samples corresponding to different target objectives, where “+” denotes generation aimed at increasing the associated quantity. As shown, the generated samples extend beyond the original data distribution, demonstrating that the framework is capable of producing novel data points in accordance with the specified target objectives during 2D-to-3D microstructure generation.

As shown in Figure~\ref{fig:UMAP clustering}(b)–(j), the target quantities exhibit distinct trends in the learned latent space. In particular, Figure~\ref{fig:UMAP clustering}(b)–(d) illustrates that the volume fractions follow relatively clear directional patterns in the UMAP projection. For example, \(V_f^{\text{Pore}}\) tends to increase as the data points move toward the lower-left region, \(V_f^{\text{YSZ}}\) increases toward the upper region, and \(V_f^{\text{Ni}}\) increases toward the lower-right. This observation is consistent with expectations, as volume fractions are simple global statistics that can be easily learned and are nearly linearly separable within the latent space. The trend for increasing \(SA^{\text{Pore}}\) in Figure~\ref{fig:UMAP clustering}(e) closely resembles that of \(V_f^{\text{Pore}}\). This is likely because the pore phase has a relatively low volume fraction compared to the YSZ and Ni phases. Thus, increasing its surface area also requires increasing its volume fraction. In contrast, the trends for \(SA^{\text{YSZ}}\) and \(SA^{\text{Ni}}\), shown in Figure~\ref{fig:UMAP clustering}(f)–(g), differ more noticeably from their volume fraction counterparts. This is likely due to the interfacial nature of surface area—an increase in the surface area of YSZ or Ni inherently increases the interface area shared with neighboring phases. While increasing volume fraction generally contributes to an increase in relative surface area, the more irregular and scattered distribution of high surface area regions (i.e., brighter points) compared to those of volume fraction suggests that optimizing surface area is a more complex process. It involves not only expanding a single phase but also modifying the interfacial relationships and spatial arrangements between multiple phases.

Figure~\ref{fig:UMAP clustering}(h) shows that the trend for increasing the effective diffusivity of the pore phase closely resembles those of \(V_f^{\text{Pore}}\) and \(SA^{\text{Pore}}\). This similarity is expected, as the volume fraction of the pore phase is relatively low compared to the YSZ and Ni phases, and the relative diffusivity of the pore phase in 2D slices is also quite small (with a maximum value around 0.0039). Consequently, achieving a higher \(D_e^{\text{Pore}}\) effectively requires increasing both the pore volume fraction and surface area. In contrast, Figures~\ref{fig:UMAP clustering}(i) and \ref{fig:UMAP clustering}(j) show more abrupt, threshold-like behavior for \(D_e^{\text{YSZ}}\) and \(D_e^{\text{Ni}}\), respectively. In these cases, the effective diffusivity remains low until a certain structural threshold is reached, beyond which the diffusivity rapidly increases. This is physically reasonable: once a given phase becomes sufficiently connected across the domain, forming a continuous pathway from one side to the other, its effective diffusivity improves significantly. This on/off behavior is particularly evident in \(D_e^{\text{Ni}}\), where a cluster of high-diffusivity samples appears in the lower-left region of the plot in Figure~\ref{fig:UMAP clustering}(j). Although the diffusivity of YSZ remains lower than that of Ni—primarily due to its granular structure and presence of sintered necks \cite{pecho20153d, hsu2018mesoscale}, as opposed to the more continuous network formed by Ni—a similar trend can still be observed in \(D_e^{\text{YSZ}}\), as shown in Figure~\ref{fig:UMAP clustering}(i).

{\color{black}
In addition, while the proposed inverse-controlled microstructure generation formulates microstructural descriptors and properties as objectives to facilitate expansion of the microstructure design and analysis space, the descriptors can also be incorporated as constraints within the same framework for practical inverse design scenarios. For example, in manufacturing-aware design problems, descriptors such as particle size, aspect ratio, or phase connectivity may be restricted by processing limitations or functional requirements. This can be expressed as
\begin{equation}\label{eq:opt_constraints}
\mathbf{x}^* = \underset{\mathbf{x}\in\mathcal{X}}{\arg\min}\;\left[\mathcal{L}_P(\mathbf{P}(\mathbf{x}),\mathbf{P}^*)\right] \quad 
\text{subject to } \mathbf{M}(\mathbf{x}) \in \mathcal{C},
\end{equation}
where $\mathcal{C}$ denotes the feasible set of microstructural descriptors. In such cases, the framework can accommodate these requirements through penalty or barrier terms, ensuring that the generated microstructures not only explore new regimes but also satisfy practical design constraints. It is suggested that future work place greater emphasis on manufacturability constraints and expand the framework toward manufacturability-aware process–structure–property linkages.
}

\begin{figure} 
    \centering    \includegraphics[width=0.8\linewidth]{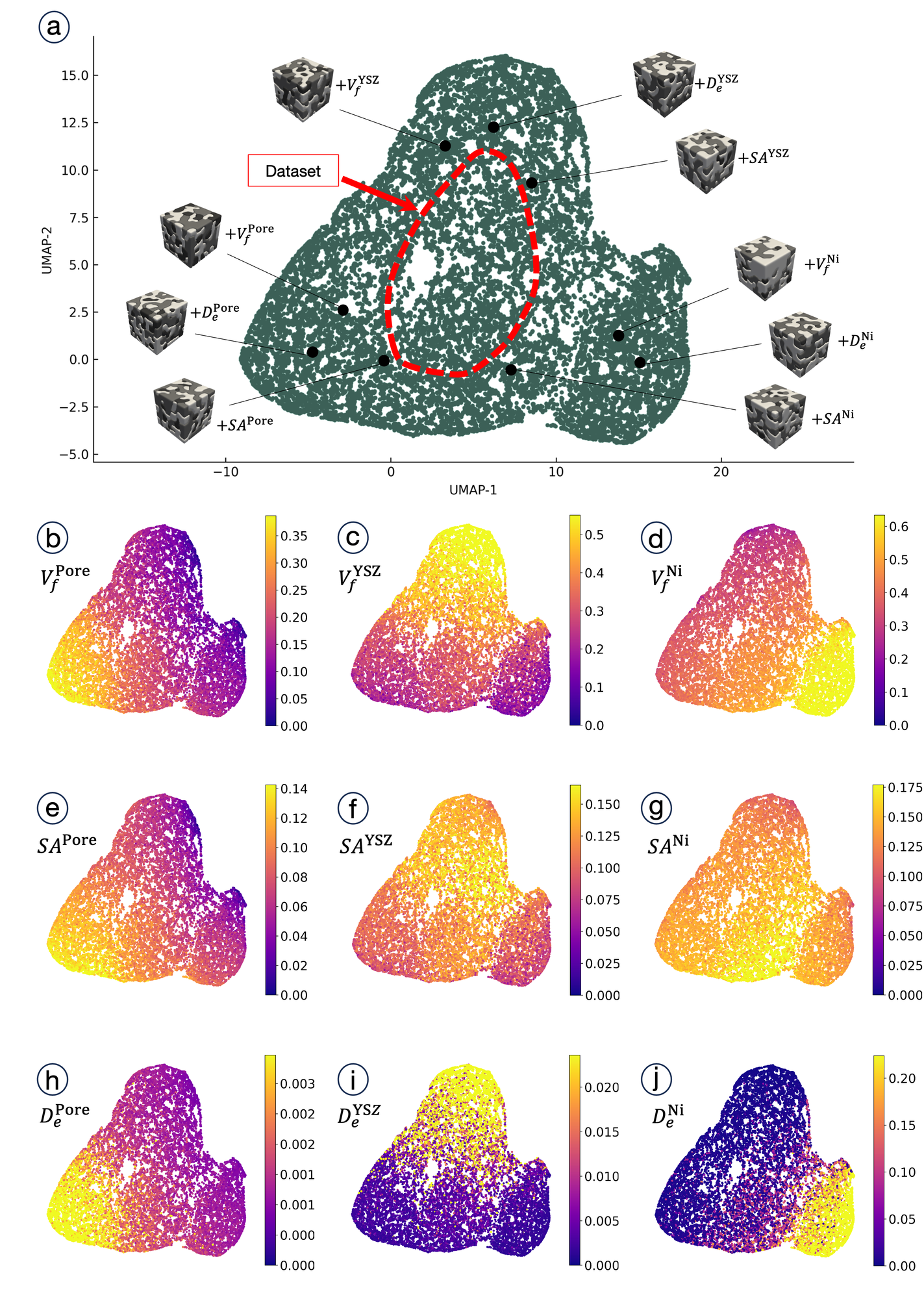}
    \caption{UMAP clustering results for VAE latent variables of generated and training microstructures: (a) data points projected onto two-dimensional UMAP axes, (b)–(d) volume fraction variation of each material phase, (e)–(g) surface area fraction variation of each material phase, (h)–(j) effective relative diffusivity variation for each material phase in the clustering results.}
    \label{fig:UMAP clustering}
\end{figure}

\begin{figure} 
    \centering    \includegraphics[width=0.9\linewidth]{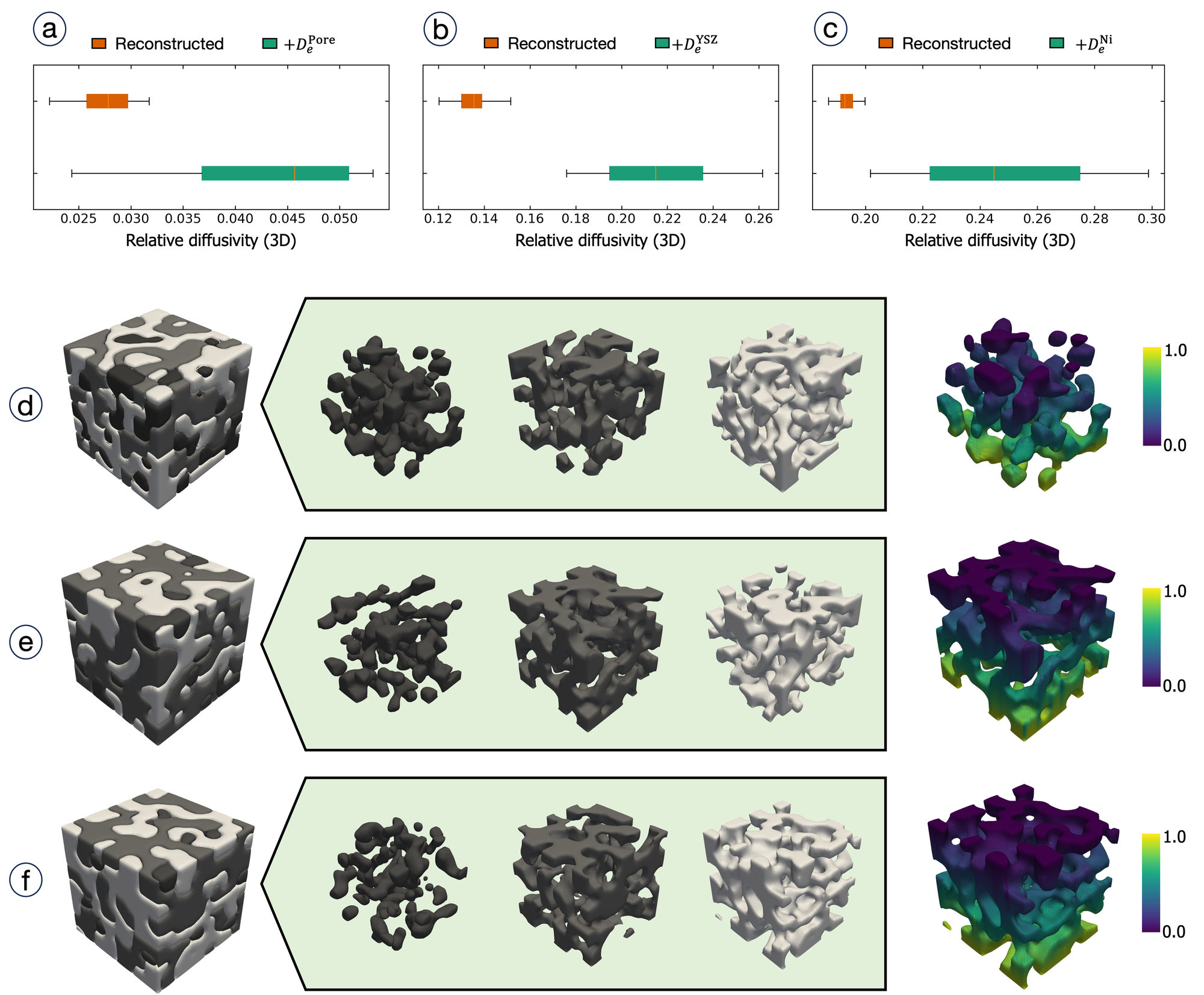}
    \caption{The evaluated relative diffusivity distributions in 3D for each phase of interest are shown in (a) +\(D_e^{\text{Pore}}\), (b) +\(D_e^{\text{YSZ}}\), and (c) +\(D_e^{\text{Ni}}\). (d) presents an example generated 3D microstructure for the +\(D_e^{\text{Pore}}\) case, with the full volume shown on the left, the individual material phases displayed separately in the center, and the steady-state concentration field for the phase of interest (pore) visualized on the right during evaluation of its relative diffusivity. The same visualization format is applied in (e) for +\(D_e^{\text{YSZ}}\) and in (f) for +\(D_e^{\text{Ni}}\).}
    \label{fig:Prop3D}
\end{figure}

\subsection{2D/3D material property behavior}

It is worth noting that material properties in 2D and 3D can differ significantly, as microstructural variability in 3D is inherently more complex and allows for a greater range of structural configurations. For instance, while the 2D-to-3D inverse-controlled microstructure generation for improving relative diffusivity for each material phase behaves as intended (i.e., successfully increasing the relative diffusivity for each material phase as shown in Figure~\ref{fig:Inverse 2D-to-3D geration De}), the absolute property values can still be limited in 2D. For example, the relative diffusivity of the pore phase in 2D is on the order of \(10^{-3}\), which is too low for it to function effectively as a gas diffusion pathway. This is because a single 2D slice often fails to form a fully connected pore network across the domain, resulting in a broken diffusion path and, consequently, low diffusivity. 

In the proposed MicroLad framework, 2D-to-3D generation is guided by optimizing individual 2D slices embedded within a 3D volume. While this approach allows for efficient optimization using only 2D computations, it is important to evaluate how well these guided generations translate to improved material properties in full 3D space. To assess this, the relative diffusivity in 3D is evaluated for each material phase, as shown in Figure~\ref{fig:Prop3D}. As seen in Figure~\ref{fig:Prop3D}(a), the relative diffusivity of the pore phase increases to the order of \(10^{-2}\), which is significantly higher than its 2D counterpart, owing to the greater likelihood of forming connected pathways in 3D. Similarly, the diffusivity of the YSZ phase also improves in 3D (Figure~\ref{fig:Prop3D}(b)). In 2D, its diffusivity is limited by the granular structure and the obstructive presence of pores, but in 3D, the enhanced connectivity enables more effective transport. 

To compare the diffusive behavior of each phase in the inverse-controlled generated 3D microstructures with those reconstructed (Recon.) without property guidance, the 3D tortuosity \cite{ghanbarian2013tortuosity, cooper2016taufactor} of both sets of volumes was computed along the three Cartesian axes (x, y, and z), as shown in Table~\ref{tab:tortuosity_axis_phase}. As observed, the reconstructed 3D volumes generally exhibit higher tortuosity (i.e., lower diffusivity), while the inverse-controlled samples show reduced tortuosity as the target +\(D_e\) is optimized for each material phase of interest. This observation suggests that enforcing target diffusivity at the slice level during 3D microstructure generation can effectively promote globally connected diffusive networks within the volume. Future work should explore more advanced control strategies, such as directionally weighted SDS loss or hierarchical optimization across multi-scale representations, to further enhance the spatial tailoring of 3D transport properties. More sophisticated control over 3D material behavior also warrants further investigation.

\begin{table}[ht]
  \centering
  \caption{Tortuosity comparison along each axis between inverse-controlled generated 3D microstructure volumes and reconstructed ones without property guidance.}
  \renewcommand{\arraystretch}{1.3}
  \setlength{\tabcolsep}{6pt}
  \begin{tabular}{lcc|cc|cc}
    \toprule
    \textbf{Axis} & \multicolumn{2}{c|}{\textbf{Pore}} & \multicolumn{2}{c|}{\textbf{YSZ}} & \multicolumn{2}{c}{\textbf{Ni}} \\
    & +$D_e^{\text{Pore}}$ & Recon. & +$D_e^{\text{YSZ}}$ & Recon. & +$D_e^{\text{Ni}}$ & Recon. \\
    \midrule
    $x$ & 5.118 $\pm$ 0.870 & 7.300 $\pm$ 1.192 & 1.582 $\pm$ 0.091 & 2.548 $\pm$ 0.201 & 1.579 $\pm$ 0.067 & 2.381 $\pm$ 0.161 \\
    $y$ & 5.101 $\pm$ 0.834 & 7.417 $\pm$ 1.217 & 1.581 $\pm$ 0.090 & 2.553 $\pm$ 0.208 & 1.498 $\pm$ 0.053 & 2.301 $\pm$ 0.154 \\
    $z$ & 5.143 $\pm$ 0.817 & 7.271 $\pm$ 1.182 & 1.574 $\pm$ 0.079 & 2.541 $\pm$ 0.221 & 1.580 $\pm$ 0.071 & 2.378 $\pm$ 0.172 \\
    \bottomrule
  \end{tabular}
  \label{tab:tortuosity_axis_phase}
\end{table}

\subsection{Slice connectivity and spatial coherence}

The underlying assumption in 2D-to-3D microstructure reconstruction and generation is that consecutive and adjacent 2D slices in a 3D volume should be closely related to preserve local consistency, spatial overlap, and facilitate smooth optimization (see Section~\ref{Method inverse-controlled 2D-to-3D microstructure generation}). To evaluate this connectivity, we compute the mean distance between each pair of consecutive slices (along the x, y, and z directions) in both reconstructed and inverse-controlled generated 3D volumes, under different target objectives. This analysis is performed using two metrics: the distance between the mean latent embeddings (\(d_u\)) and the distance in the UMAP-projected space (\(d_{\mathrm{UMAP}}\)), as summarized in Table~\ref{tab:evaluation of pairwise distances}. Compared to the original dataset—where 2D slices are independently cropped from a single micrograph and are not spatially connected—the slices in the reconstructed 3D volumes exhibit latent distances that are approximately an order of magnitude smaller. This indicates that the 2D slices in the reconstructed volumes are closely clustered in latent space, indicating their spatial coherence. Similarly, the inverse-controlled generated volumes also show consistently small distances between consecutive slices, demonstrating that optimization was conducted while preserving spatial connectivity across slices.

\begin{figure}[ht]
    \centering    \includegraphics[width=1.0\linewidth]{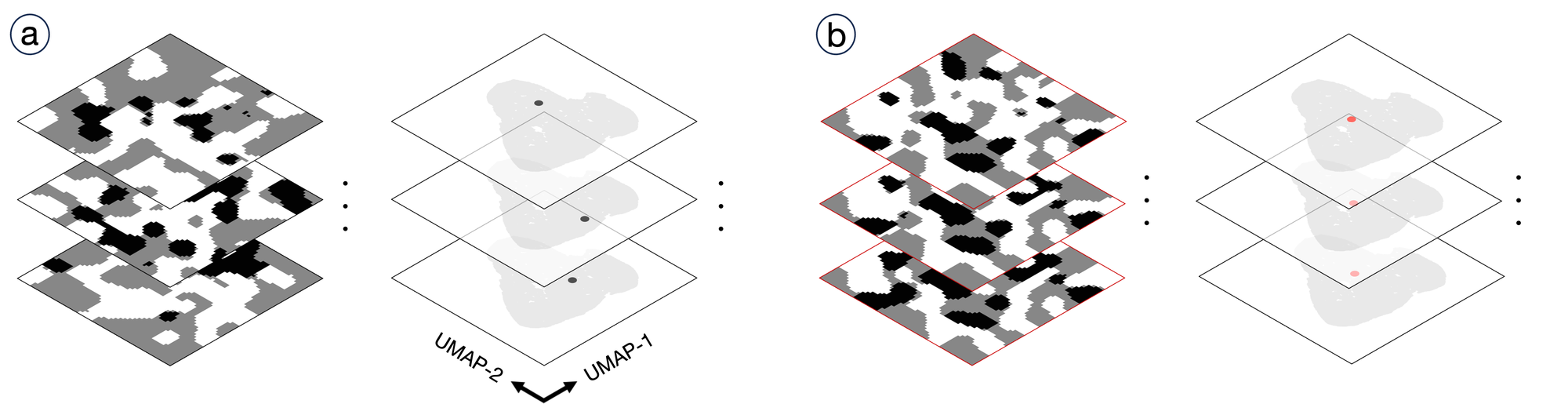}
    \caption{Visualization of slice connectivity in each 3D volume using latent representations (i.e., UMAP projections): (a) independent 2D slices from the 2D dataset, and (b) spatially connected 2D slices from a reconstructed 3D volume.}
    \label{fig:slice_connectivity}
\end{figure}

The UMAP-based distances exhibit a similar trend: both the reconstructed and generated volumes show significantly lower inter-slice distances compared to the unconnected 2D slices from the dataset. This observation is visually supported in Figure~\ref{fig:slice_connectivity}. In Figure~\ref{fig:slice_connectivity}(a), the independent 2D slices from the dataset appear scattered in the UMAP projection, with no close spatial grouping between consecutive pairs. In contrast, Figure~\ref{fig:slice_connectivity}(b) illustrates that consecutive slices from MicroLad-generated 3D volumes remain closely clustered in the latent space, suggesting that spatial coherence is effectively preserved across slices. These findings indicate that the proposed MicroLad framework is capable of generating spatially consistent 2D slices within 3D volumes, thereby maintaining the local continuity necessary for both accurate 2D-to-3D microstructure reconstruction and inverse-controlled generation.

\textcolor{black}{
In the meantime, an important point of discussion is that the present 2D-to-3D microstructure reconstruction/generation framework relies on the assumption that spatial connectivity in 3D space can be represented by connectivity along three orthogonal directions (Eq.~\eqref{eq:spatial_connectivity}). The local consistency assumption (Section~\ref{Method inverse-controlled 2D-to-3D microstructure generation}) was likewise built on the optimization of latent slices along these three directions. Although LDM with a microstructure encoder (i.e., VAE) was adopted to enable the computationally efficient 2D-to-3D expansion and to generate 2D slices that remain close to one another in the latent embedding space (Table~\ref{tab:evaluation of pairwise distances} and Figure~\ref{fig:slice_connectivity}), consistency in other directions in 3D space (e.g., diagonal) \cite{Bostanabad2020Reconstruction} is not explicitly guaranteed. While the proposed approach and its underlying assumptions are practical, given that most experimentally observed 2D micrographs are obtained along one axis or a few orthogonal planes, statistical consistency in non-orthogonal directions or planes in 3D space also needs to be evaluated and addressed in future work. Nevertheless, the proposed approach performed well for both binary and three-phase microstructures, and additional examples, including anisotropic cases presented in Appendix~\ref{sec:Reconstruction of various types of microstructures}, further demonstrate that the proposed framework can effectively handle a various types of microstructures.
}

\begin{table}[ht]
  \centering
  \caption{
Evaluation of pairwise distances between consecutive slices in reconstructed and generated 3D volumes under different target objectives.}
  \renewcommand{\arraystretch}{1.3}
  \setlength{\tabcolsep}{3pt}
  \begin{tabular}{lllccccccccc}
    \toprule
    \textbf{Metric} & \textbf{Dataset} & \textbf{Recon.} & \multicolumn{9}{c}{\textbf{Target objective}} \\
    \cmidrule(lr){4-12}
    & & & +$V_f^{\text{Pore}}$ & +$V_f^{\text{YSZ}}$ & +$V_f^{\text{Ni}}$
    & +$SA^{\text{Pore}}$ & +$SA^{\text{YSZ}}$ & +$SA^{\text{Ni}}$
    & +$D_e^{\text{Pore}}$ & +$D_e^{\text{YSZ}}$ & +$D_e^{\text{Ni}}$ \\
    \midrule
    $d_u$ & 0.121 & 0.0137 &
    0.0140 & 0.0117 & 0.0127 &
    0.0198 & 0.0166 & 0.0153 &
    0.0172 & 0.0168 & 0.0159 \\
    
    $d_{\mathrm{UMAP}}$ & 14.708 & 2.253 &
    2.422 & 2.334 & 2.165 &
    3.296 & 2.891 & 2.303 &
    2.884 & 2.880 & 2.805 \\
    \bottomrule
  \end{tabular}
  \label{tab:evaluation of pairwise distances}
\end{table}

\subsection{Computational cost}

Since the proposed MicroLad framework employs latent diffusion in a lower-dimensional space, the L-MPDD procedure for generating initial 3D microstructures is highly efficient. Generating a \(64 \times 64 \times 64\) volume takes less than 10 seconds—comparable to generating a single \(64 \times 64\) image with a batch size of 64 (wall-clock times were measured on a single NVIDIA RTX 4090 GPU). In contrast, the original MPDD framework required over 30 minutes to produce a single 3D sample with comparable accuracy \cite{lee2024multi}. While dimensionality expansion via L-MPDD is fast, the most computationally intensive component is SDS applied to enforce target objectives. For each objective, the wall-clock time increases approximately linearly with the number of SDS steps. Appendix~\ref{sec:parameteric study for SDS} presents a parametric study evaluating the effects of SDS learning rate and objective weights on optimization performance. For volume fraction and surface area targets, 500 SDS steps typically require about one minute to generate a single 3D sample. In contrast, SDS targeting effective diffusivity requires approximately two and a half minutes per sample, due to the overhead of computing gradients through the differentiable FEM solver. Across all objectives, using 3000 SDS steps consistently results in low error between generated samples and their respective targets. Thus, the number of SDS steps can be chosen by balancing computational cost with the desired accuracy of inverse generation. 

\textcolor{black}{It is also worth noting that the main purpose of adopting a differentiable model for guided generation is to eliminate the need for pre-labeling data and training conditional models by directly integrating differentiable simulations into the guided generation framework. This formulation can be adapted depending on the application. If a sufficiently large dataset is available (or can be affordably generated) and reducing inference time is critical—perhaps even more important than minimizing training cost—a conditional formulation (e.g., conditional diffusion models with classifier-free guidance~\cite{ho2022classifier}) may be more suitable. In such cases, differentiable model–based guidance may no longer be necessary. The limitation of this approach, however, is that the achievable design or property range tends to be restricted by the data-conditioned generation process, whereas guided generation with differentiable models emphasizes optimization while leveraging the diffusion prior to ensure that generation remains within the unconditional data manifold~\cite{maze2023diffusion,poole2022dreamfusion}. Therefore, the appropriate formulation should be selected based on the user’s specific objectives and design requirements.}  

\textcolor{black}{In addition, to further illustrate the computational cost and scalability of the proposed approach, we provided a 2D-to-3D reconstruction and generation case with a larger microstructure size (\(256\times256\times256\)) in Appendix~\ref{sec:Experiments with larger microstructure size}. The reconstruction of a 3D volume of this size required only 71\,s, with a maximum two-point correlation function relative error of 5.23\%, demonstrating both the efficiency and accuracy of the proposed framework for larger microstructure size. Meanwhile, the \(D_e\)-guided generation successfully shifted the property distribution toward the target but required 156.1\,minutes, primarily because the FEM computation substantially increased the overall computational cost. This efficiency can be further improved by employing more efficient FEM solvers and differentiable neural network–based surrogate modeling approaches~\cite{xue2023jax,kudela2022recent}, which would enable faster and more efficient gradient computation. Incorporating such surrogates represents a promising direction for future work.}

\subsection{Future work}
The proposed MicroLad framework presents considerable potential for extension to a broader class of multiphase materials, enabling the study of microstructural variability and its influence on resulting material properties. One promising direction is to leverage the framework for uncertainty quantification by incorporating user-defined microstructural variability, thereby enhancing its utility in data-driven material design. Furthermore, MicroLad can be extended to a wider range of physics-informed generative modeling tasks, including 2D-to-3D microstructure generation under multiphysics loading conditions and complex material behavior. A particularly promising avenue for future research is the integration of advanced differentiable solvers~\cite{xue2023jax} capable of modeling nonlinear constitutive laws. Such an extension would allow the framework to capture a richer spectrum of microstructural phenomena governed by diverse physical mechanisms. \textcolor{black}{The incorporation of a wider variety of microstructures, including those with more phases and even continuous-valued features (e.g., crystallographic orientations), is also suggested, as the proposed approach does not intrinsically restrict the input to discrete phases. This is supported by the fact that latent diffusion models have been successfully applied to colored images as well \cite{rombach2022high}}

On the generative modeling side, the current implementation of MicroLad applies SDS to enable property- and descriptor-guided generation without requiring pre-labeled microstructure–descriptor or microstructure–property pairs. This approach is particularly advantageous in scenarios where labeled data are scarce due to computational or experimental constraints. However, a natural extension would be to train conditional diffusion models that accept target objectives as direct inputs through conditioning mechanisms~\cite{ho2022classifier, yang2023diffusion}. This would eliminate the need for iterative SDS optimization and enable more efficient and scalable inverse-controlled generation.

\bibliographystyle{unsrt}
\bibliography{references}  

\newpage
\appendix
\renewcommand{\thesection}{\Alph{section}} 
\renewcommand{\thetable}{\Alph{section}\arabic{table}} 
\renewcommand{\thefigure}{\Alph{section}\arabic{figure}} 
\counterwithin{table}{section}
\counterwithin{figure}{section}

\begin{bluecolorregion}
\section{Refinement of 2D slices on orthogonal planes}\label{sec:Refinement of 2D slices on orthogonal planes}

To handle unexpected noise or deviations that may occur during the 2D-to-3D dimensionality expansion, a refinement step is applied to the reconstructed 3D volume. This process does not involve training a new denoising model. Instead, it utilizes the originally trained VAE to refine the 2D slices. The key idea is that a VAE trained exclusively on clean data implicitly projects inputs back onto the learned manifold. Because the latent space captures only the distribution of clean samples, details inconsistent with the training distribution are not faithfully reconstructed~\cite{hassanaly2024evaluation, meng2017magnet, chen2018unsupervised}. Therefore, for the generated 3D volumes, this serves as an additional refinement step applied to the slices along the three orthogonal directions, as illustrated in Figure~\ref{fig app: latent refinement}.

\begin{figure}[H]
    \centering    \includegraphics[width=0.6\linewidth]{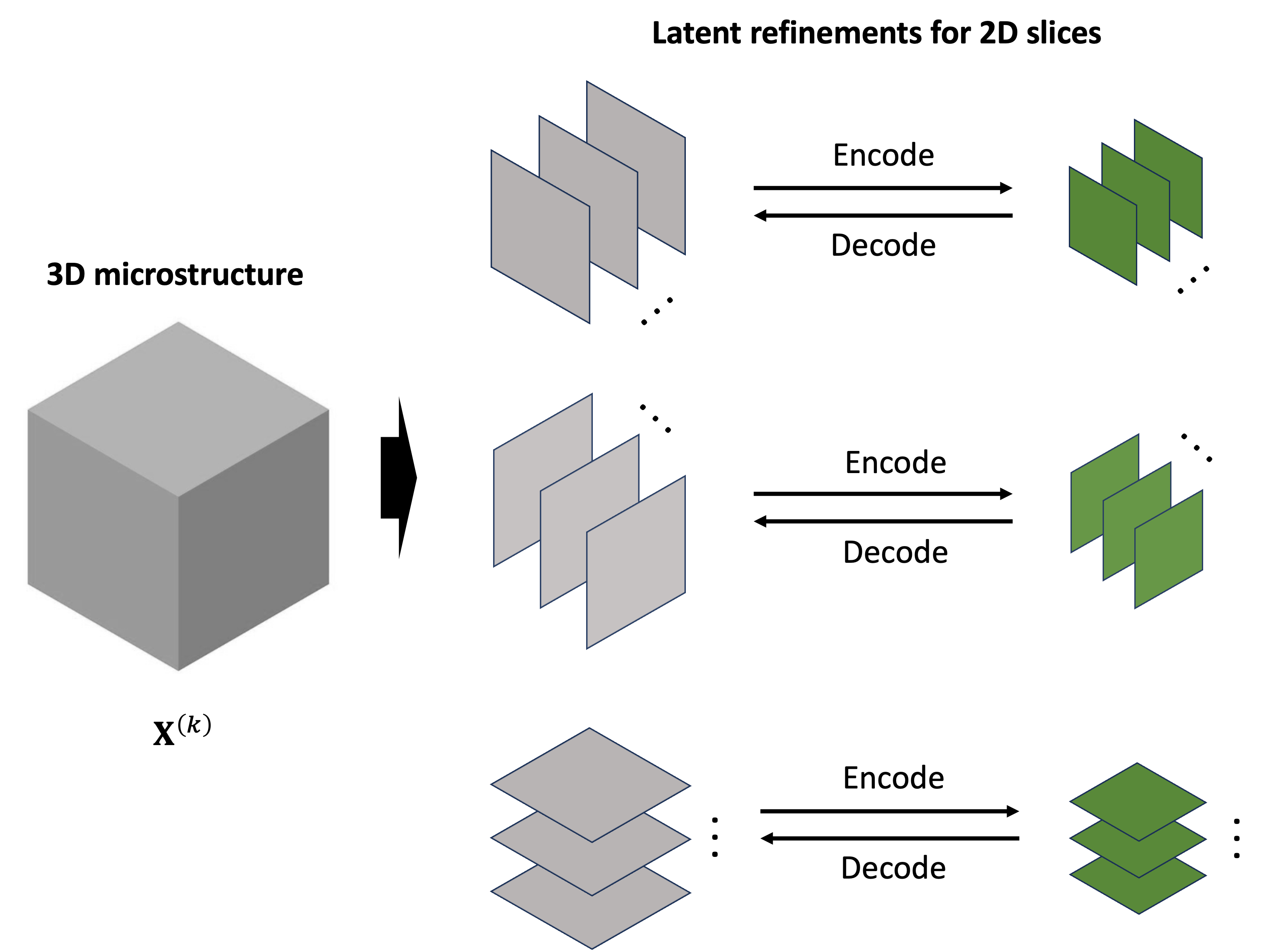}
    \caption{Schematic of the refinement process for 2D slices of a 3D volume using the trained encoder and decoder.}
    \label{fig app: latent refinement}
\end{figure}
\end{bluecolorregion}

\section{Configuration parameters for the models}\label{sec:configuration parameters for the models}

To encode 2D microstructure images into a compact and structured latent space, we use a VAE. The VAE consists of an encoder $E(\cdot)$ that maps an input image $\mathbf{x}$ to a latent distribution $\mathcal{N}(\boldsymbol{\mu}, \boldsymbol{\sigma}^2)$, and a decoder $\mathcal{D}(\cdot)$ that reconstructs the image $\hat{\mathbf{x}} = \mathcal{D}(\mathbf{z})$ from a latent sample $\mathbf{z} \sim \mathcal{N}(\boldsymbol{\mu}, \boldsymbol{\sigma}^2)$. The model is trained by minimizing the total VAE loss:
\begin{equation}
\mathcal{L}_{\text{VAE}} = \mathbb{E}_{\mathbf{x} \sim p_{\text{data}}} \left[ 
\underbrace{ \| \hat{\mathbf{x}} - \mathbf{x} \|_1 }_{\text{reconstruction loss}} +
\lambda_{\text{KL}} \underbrace{ \text{KL}\left[ \mathcal{N}(\boldsymbol{\mu}, \boldsymbol{\sigma}^2) \| \mathcal{N}(0, I) \right] }_{\text{KL divergence}} 
\right],
\end{equation}
where $\lambda_{\text{KL}}$ is a weighting factor set to 0.5 in our implementation. The reconstruction quality of the VAE was evaluated on binary and three-phase microstructure datasets using three metrics: mean absolute error (MAE), peak signal-to-noise ratio (PSNR), and structural similarity index (SSIM). Table~\ref{tab:autoencoder_recon} reports the results for multiple latent space sizes, showing that the VAE achieves high-fidelity reconstructions across both binary and multiphase datasets, especially when using a $16 \times 16$ latent feature map.

After training the VAE, we train a LDM that operates on the compressed latent space to enable generative modeling. Following the DDPM formulation \cite{ho2020denoising}, the LDM defines a forward Markov diffusion process that gradually adds Gaussian noise to a clean latent code. The LDM is trained with 1000 diffusion steps using a linear noise schedule $\beta \in [1 \times 10^{-4}, 2 \times 10^{-2}]$. During inference, new samples are generated by reversing the diffusion process starting from pure noise in latent space. These are decoded back to pixel space using the pretrained VAE decoder $\mathcal{D}(\cdot)$. The training hyperparameters for both the VAE and LDM are summarized in Tables~\ref{tab:vae_hyperparams} and \ref{tab:diffusion_hyperparams}. The encoder–decoder architecture and full latent U-Net pipeline are detailed in Table~\ref{tab:latent_pipeline},

\begin{table}[H]
  \centering
  \caption{VAE reconstruction performance across different latent space dimensions and microstructure datasets.}
  \renewcommand{\arraystretch}{1.2}
  \begin{tabular}{llccc}
    \toprule
    \textbf{Microstructure Dataset} & \textbf{Latent Space} &  \textbf{MAE ↓} & \textbf{PSNR ↑} & \textbf{SSIM ↑}  \\
    \midrule
    {Binary}     
        & $4 \times 4$   & 0.0236  & 20.39  & 0.9344 \\
        & $8 \times 8$   & 0.0121  & 21.66 & 0.9824 \\
        & $16 \times 16$ & 0.0085  & 25.77 & 0.9849 \\
    \midrule
    {Three-phase} 
        & $4 \times 4$   & 0.0281  & 19.27 & 0.9182 \\
        & $8 \times 8$   & 0.0242  & 21.92 & 0.9477 \\
        & $16 \times 16$ & 0.0076  & 27.30 & 0.9886 \\
    \bottomrule
  \end{tabular}
  \label{tab:autoencoder_recon}
\end{table}

\begin{table}[H]
  \centering
  \caption{Configuration parameters for the VAE.}
  \renewcommand{\arraystretch}{1.2}
  \begin{tabular}{p{5cm} p{7.5cm}}
    \toprule
    \textbf{Parameter} & \textbf{Value / Description} \\
    \midrule
    Latent space feature map & $16 \times 16$ \\
    Latent channels & 4  \\
    KL divergence weight & 0.5  \\
    Optimizer & Adam \\
    Learning rate & $2 \times 10^{-4}$ \\
    \bottomrule
  \end{tabular}
  \label{tab:vae_hyperparams}
\end{table}

\begin{table}[H]
  \centering
  \caption{Configuration parameters for the LDM.}
  \renewcommand{\arraystretch}{1.2}
  \begin{tabular}{p{5cm} p{7.5cm}}
    \toprule
    \textbf{Parameter} & \textbf{Value / Description} \\
    \midrule
    Diffusion steps & 1000 \\
    Noise schedule & Linear, $\beta \in [1\times10^{-4}, 2\times10^{-2}]$ \\
    Base channel width & 128 \\
    Time embedding dimension & 64 \\
    Optimizer & Adam \\
    Learning rate & $2 \times 10^{-4}$ \\
    \bottomrule
  \end{tabular}
  \label{tab:diffusion_hyperparams}
\end{table}

\begin{table}[H]
\centering
\caption{Encoder–latent-space U-Net–decoder pipeline for MicroLad (latent map $4\times16\times16$).}
\label{tab:latent_pipeline}
\begin{tabular}{@{}lll@{}}
\toprule
\textbf{Layer} & \textbf{Operation} & \textbf{Output size} \\ \midrule
\multicolumn{3}{l}{\itshape VAE encoder}\\
Input             & $3\times3$ Conv                                  & $128\times64\times64$ \\
Downsample 1       & DownBlock ($\times2$ Res)                         & $128\times32\times32$ \\
Downsample 2       & DownBlock ($\times2$ Res)                         & $256\times16\times16$ \\
Encoder head       & Res $+$ Attention $+$ Res                         & $256\times16\times16$ \\
Latent params      & $1\times1$ Conv ($\mu,\sigma$)                    & $4\times16\times16$ \\[4pt]

\multicolumn{3}{l}{\itshape U-Net in latent space}\\
Processing block   & ResidualConv $+$ Attention                        & $128\times16\times16$ \\
Hierarchical path  & Two down–up pairs (scales $8,\;4$)                & symmetric \\
Output             & $1\times1$ Conv                                   & $4\times16\times16$ \\[4pt]

\multicolumn{3}{l}{\itshape VAE decoder}\\
Latent input       & $3\times3$ Conv                                   & $256\times16\times16$ \\
Decoder core       & Res $+$ Attention $+$ Res                         & $256\times16\times16$ \\
Upsample 1          & UpBlock (factor 2)                               & $128\times32\times32$ \\
Upsample 2          & UpBlock (factor 2)                               & $64\times64\times64$ \\
Output             & $3\times3$ Conv $+$ Sigmoid                       & $1\times64\times64$ \\ \bottomrule
\end{tabular}
\end{table}

\begin{bluecolorregion}
\section{Implementation of FEM}\label{sec:Implementation of differentiable FEM}

In this study, a differentiable FEM solver was implemented in PyTorch to compute the effective diffusivity of SOFC microstructures for each material phase. The solver discretizes the microstructure into linear rectangular elements (e.g., \(64\times64\) or \(256\times256\)) and solves the steady-state diffusion equation. Dirichlet boundary conditions are applied to simulate transport through each microstructure phase and obtain the relative diffusivity (Figure~\ref{fig app: FEM}). The solver assembles a sparse stiffness matrix based on element-wise contributions and solves the resulting linear system using a PyTorch-based linear solver, enabling automatic gradient computation for property-guided microstructure generation.

\begin{figure}[H]
    \centering    \includegraphics[width=0.6\linewidth]{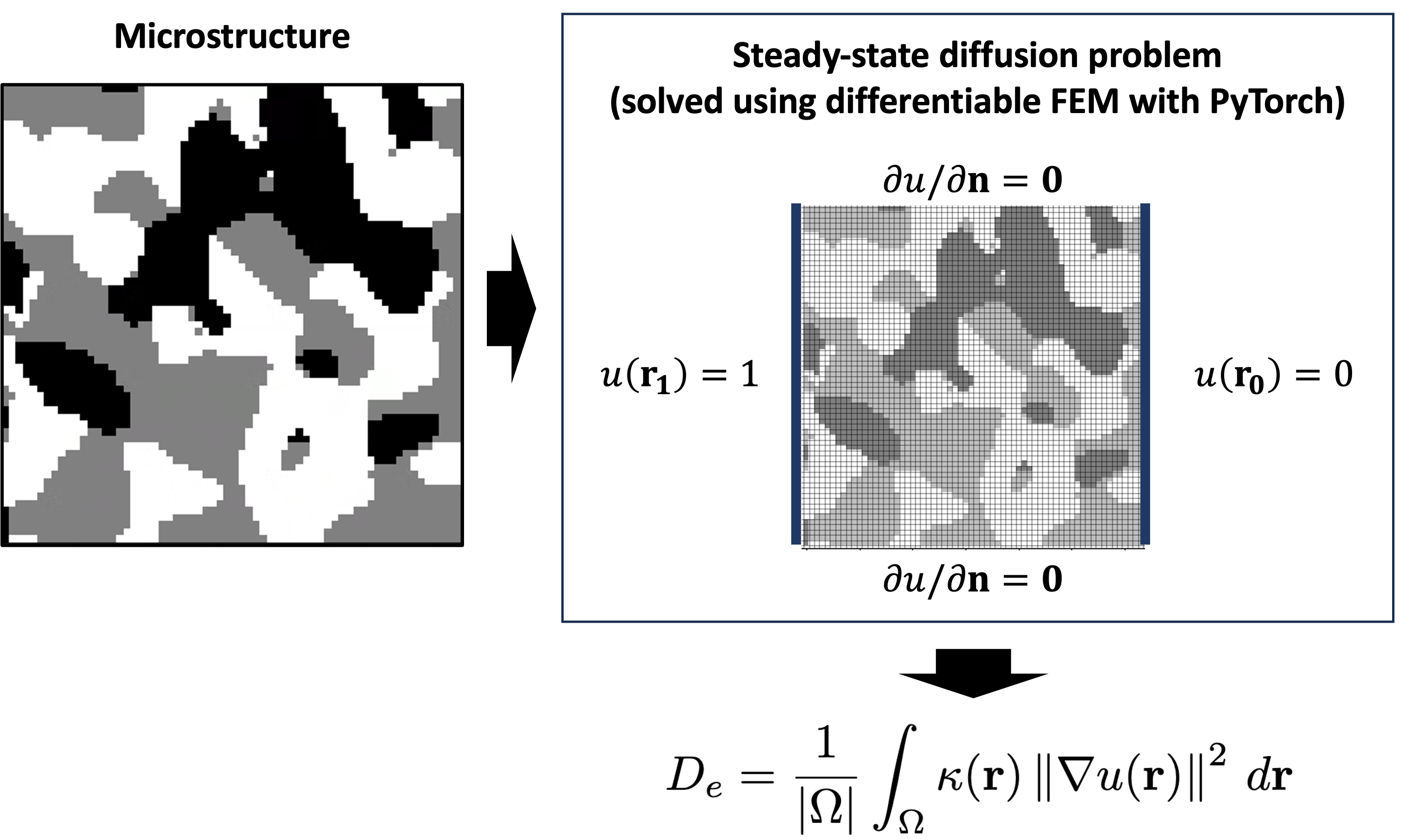}
    \caption{Schematic of the FEM simulation for a given microstructure used to obtain the relative diffusivity.}
    \label{fig app: FEM}
\end{figure}
\end{bluecolorregion}

\section{Comparison of microstructure reconstruction performance with previous methods}\label{sec:comparison of microstructure reconstruction performance}

To benchmark the reconstruction accuracy of the proposed MicroLad framework, we compare its performance against two established methods: the MPDD approach and SliceGAN. All methods are evaluated on three-phase microstructures using MAE and relative error metrics for two-point correlation function of each material phase. Table~\ref{tab:recon_accuracy_threephase} reports the phase-wise reconstruction errors, showing that MicroLad (with $S_2$ guidance) achieves superior fidelity across all phases. For completeness, we also include architectural details of the MPDD and SliceGAN baselines in Tables~\ref{tab:pixel_unet} and \ref{tab:slicegan}, respectively.

\begin{table}[H]
  \caption{Reconstruction accuracy (three-phase case).}
  \centering
  \renewcommand{\arraystretch}{1.15}
  \begin{tabular}{l c c c}
    \toprule
    \textbf{Reconstruction method} & \textbf{Phase} &
    \textbf{MAE} & \makecell{\textbf{Relative error (\%)}\\$\varepsilon_{\text{rel}}$} \\
    \midrule
    MicroLad                      & 0 & 0.0059 & 6.13 \\
                                  & 1 & 0.0028 & 0.59 \\
                                  & 2 & 0.0031 & 0.24 \\
    \cmidrule(lr){1-4}    MPDD                         & 0 & 0.0159 & 12.44 \\
                                  & 1 & 0.0117 & 6.60  \\
                                  & 2 & 0.0171 & 2.69  \\
    \cmidrule(lr){1-4}
    SliceGAN                     & 0 & 0.0134 & 11.01 \\
                                  & 1 & 0.0147 & 8.52  \\
                                  & 2 & 0.0143 & 1.99  \\
    \bottomrule
  \end{tabular}
  \label{tab:recon_accuracy_threephase}
\end{table}

\begin{table}[H]
\centering
\caption{U-Net backbone used for MPDD on $64{\times}64$ images.}
\label{tab:pixel_unet}
\begin{tabular}{@{}lll@{}}
\toprule
\textbf{Layer} & \textbf{Operation} & \textbf{Output size} \\ \midrule
Input            & noisy image                                           & $1\times64\times64$ \\
Encoding 1        & ResidualConv $+$ Attention                             & $128\times64\times64$ \\
Downsample 1      & $2\times2$ MaxPool                                    & $128\times32\times32$ \\
Encoding 2        & ResidualConv $+$ Attention                             & $256\times32\times32$ \\
Downsample 2      & $2\times2$ MaxPool                                    & $256\times16\times16$ \\
Bottleneck        & ResidualConv $+$ Attention                             & $512\times16\times16$ \\
Upsample 1        & $2\times2$ ConvTranspose                               & $256\times32\times32$ \\
Decoding 1        & ResidualConv (skip-fusion)                             & $256\times32\times32$ \\
Upsample 2        & $2\times2$ ConvTranspose                               & $128\times64\times64$ \\
Decoding 2        & ResidualConv (skip-fusion)                             & $128\times64\times64$ \\
Output            & $1\times1$ Conv                                        & $1\times64\times64$ \\ \bottomrule
\end{tabular}
\end{table}

\begin{table}[H]
\centering
\caption{SliceGAN generator and discriminator for three-phase volumes ($64^3$ voxels).}
\label{tab:slicegan}
\begin{tabular}{@{}lll@{}}
\toprule
\textbf{Layer} & \textbf{Operation} & \textbf{Output size} \\ \midrule
\multicolumn{3}{l}{\itshape Generator (3D)}\\
Latent vector   & reshape (32)                                           & $32\times1\times1\times1$ \\
Block 1         & Deconv3D $+$ BN $+$ ReLU (1024)                         & $1024\times2\times2\times2$ \\
Block 2         & Deconv3D $+$ BN $+$ ReLU (512)                          & $512\times4\times4\times4$ \\
Block 3         & Deconv3D $+$ BN $+$ ReLU (128)                          & $128\times8\times8\times8$ \\
Block 4         & Deconv3D $+$ BN $+$ ReLU (32)                           & $32\times16\times16\times16$ \\
Block 5         & Deconv3D (3 channels)                                   & $3\times32\times32\times32$ \\
Upsample        & Trilinear $\times2$ $+$ Conv3D (3)                      & $3\times64\times64\times64$ \\[4pt]

\multicolumn{3}{l}{\itshape Discriminator (2D)}\\
Input slice     & —                                                      & $3\times64\times64$ \\
Block 1         & Conv2D $+$ ReLU (64)                                    & $64\times32\times32$ \\
Block 2         & Conv2D $+$ ReLU (128)                                   & $128\times16\times16$ \\
Block 3         & Conv2D $+$ ReLU (256)                                   & $256\times8\times8$ \\
Block 4         & Conv2D $+$ ReLU (512)                                   & $512\times4\times4$ \\
Output          & Conv2D (1 channel)                                      & $1\times2\times2$ \\ \bottomrule
\end{tabular}
\end{table}

\section{Additional inverse 2D-to-3D generation results}\label{sec:additional inverse 2D-to-3D generation results}

\begin{figure}[H]
    \centering    \includegraphics[width=0.8\linewidth]{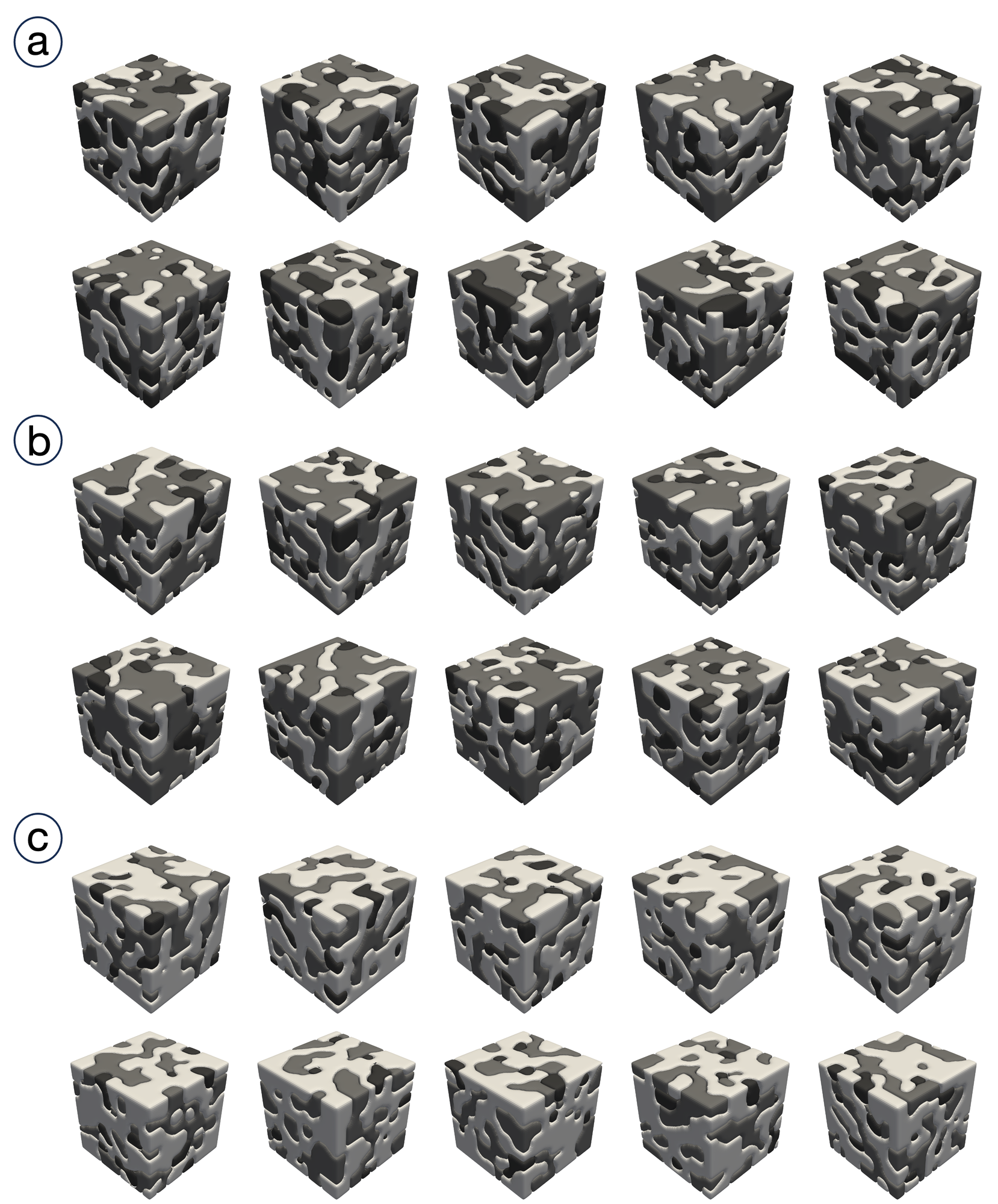}
    \caption{Visualization of the 2D-to-3D generated samples using MicroLad, along with their target objectives: (a) +$V_f^{\text{Pore}}$, (b) +$V_f^{\text{YSZ}}$ and (c) +$V_f^{\text{Ni}}$}
    \label{fig app: visualization Vf}
\end{figure}

\begin{figure}[H]
    \centering    \includegraphics[width=0.8\linewidth]{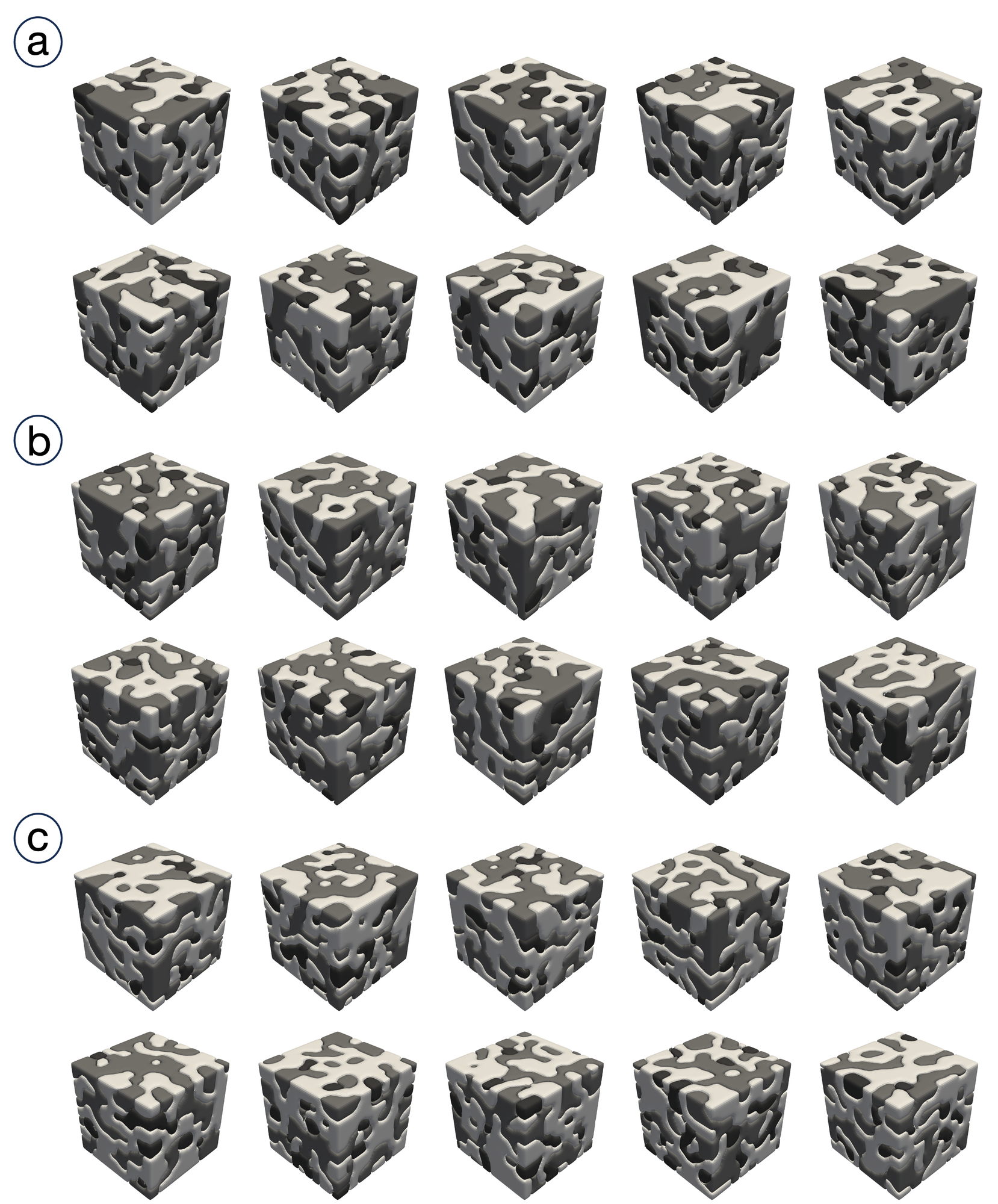}
    \caption{Visualization of the 2D-to-3D generated samples using MicroLad, along with their target objectives: (a) +$SA^{\text{Pore}}$, (b) +$SA^{\text{YSZ}}$ and (c) +$SA^{\text{Ni}}$}
    \label{fig app: visualization SA}
\end{figure}

\begin{figure}[H]
    \centering    \includegraphics[width=0.8\linewidth]{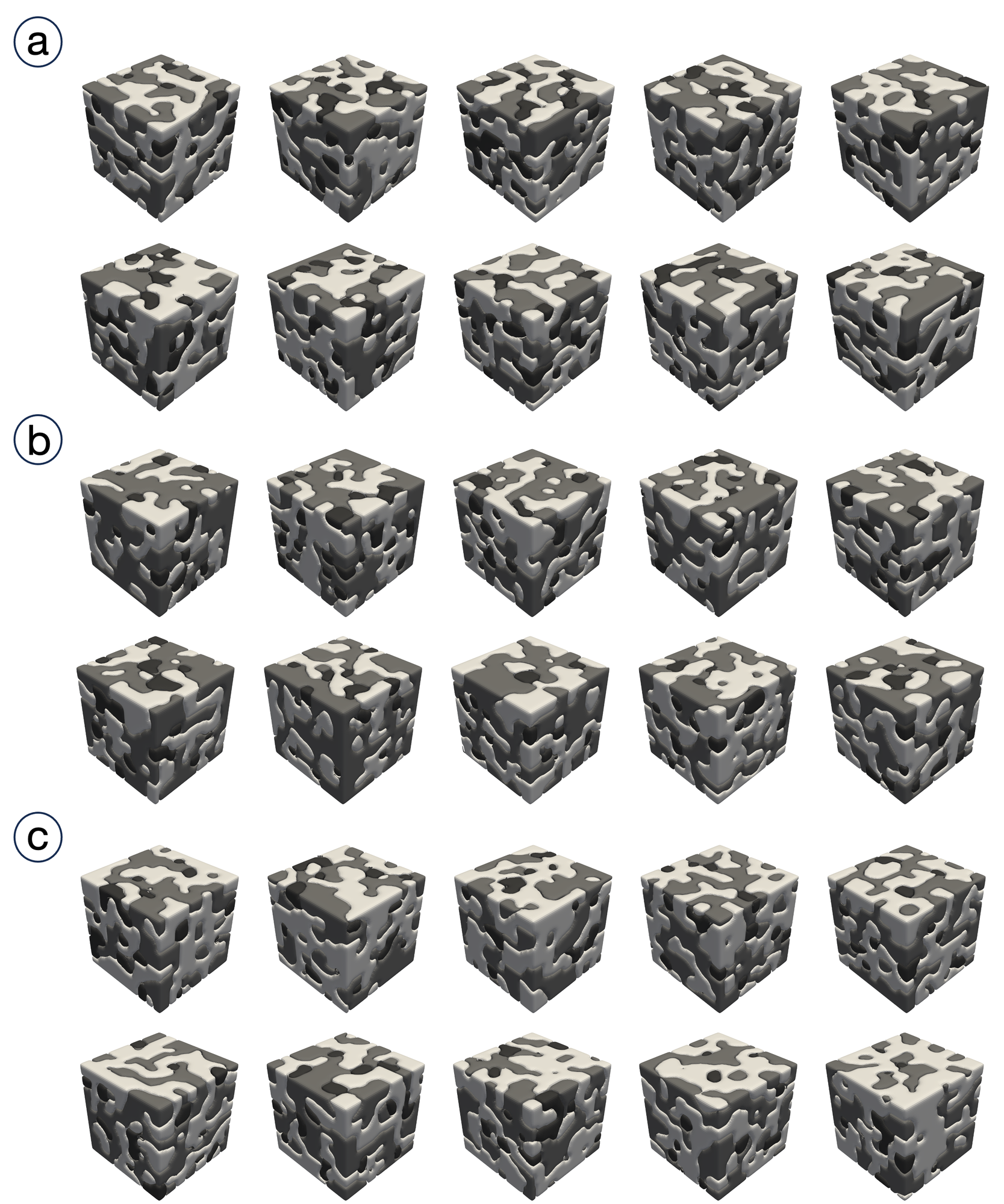}
    \caption{Visualization of the 2D-to-3D generated samples using MicroLad, along with their target objectives: (a) +$D_e^{\text{Pore}}$, (b) +$D_e^{\text{YSZ}}$ and (c) +$D_e^{\text{Ni}}$}
    \label{fig app: visualization De}
\end{figure}

\begin{figure}[H]
    \centering    \includegraphics[width=0.8\linewidth]{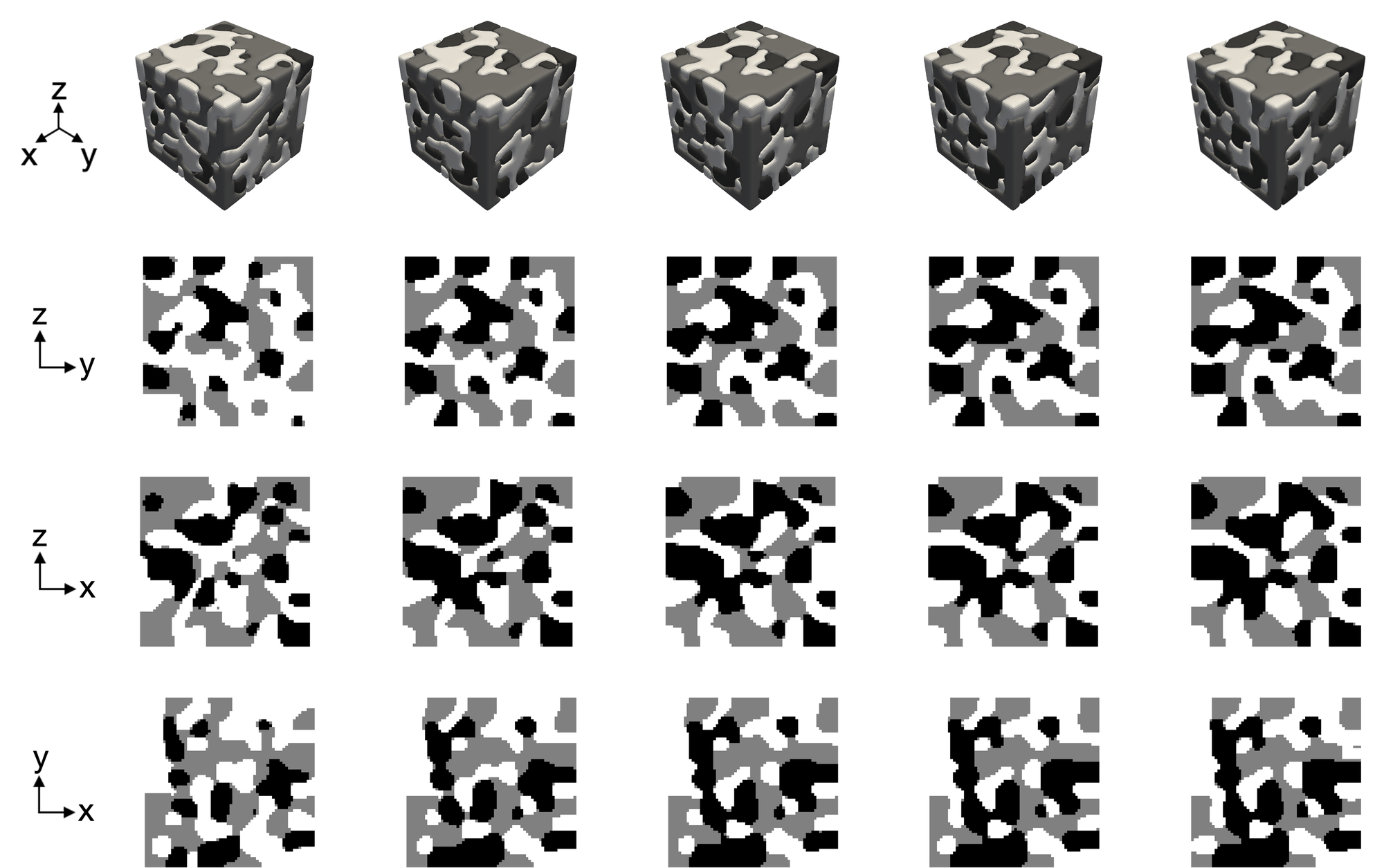}
    \caption{Progression of three orthogonal cross-sectional slices (through the center of each volume) during the 2D-to-3D microstructure generation process guided by a target objective +$V_f^{\text{Pore}}$.}
    \label{fig app: progression Vf}
\end{figure}

\begin{figure}[H]
    \centering    \includegraphics[width=0.8\linewidth]{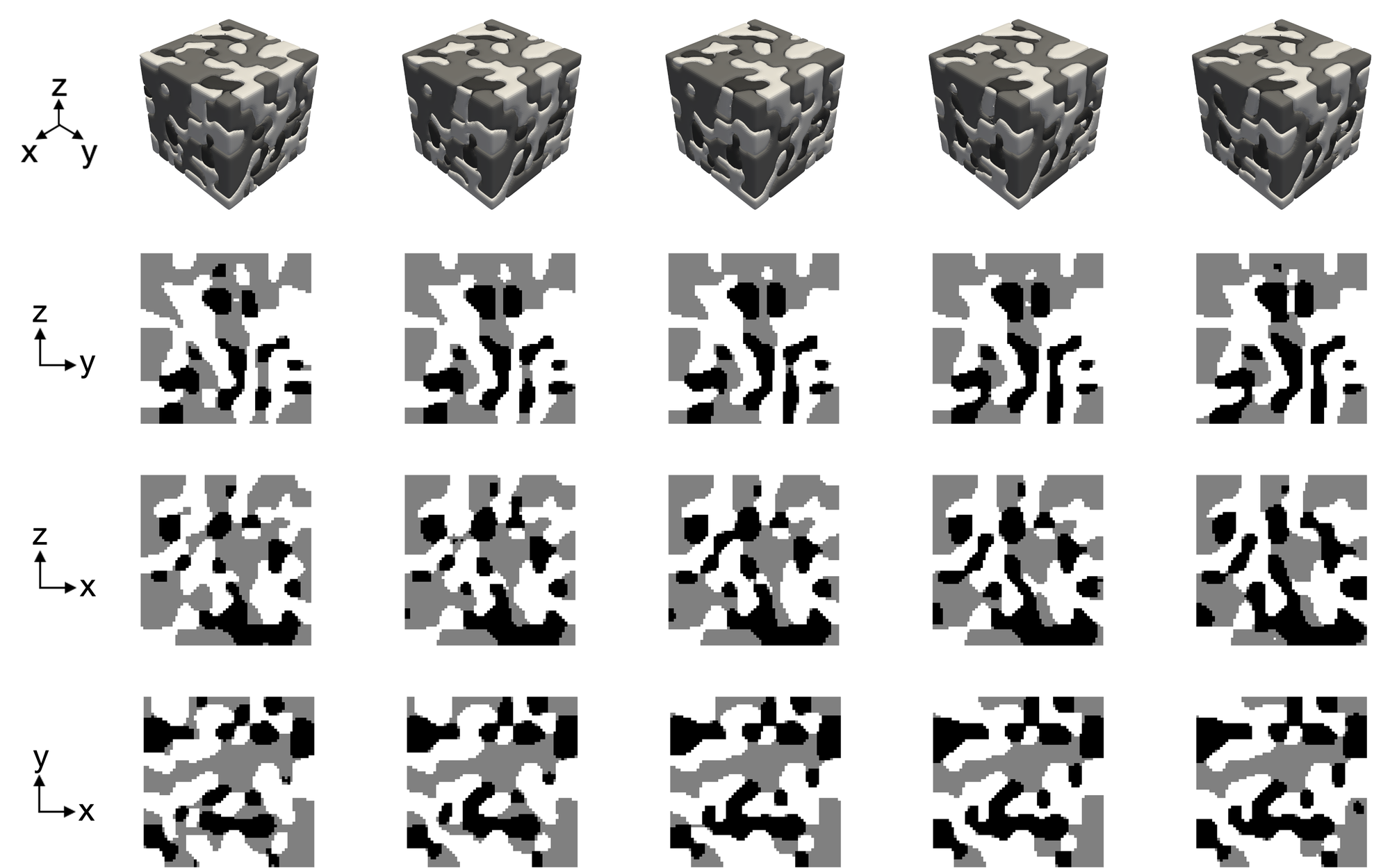}
    \caption{Progression of three orthogonal cross-sectional slices (through the center of each volume) during the 2D-to-3D microstructure generation process guided by a target objective +$SA^{\text{Pore}}$.}
    \label{fig app: progression SA}
\end{figure}

\begin{figure}[H]
    \centering    \includegraphics[width=0.8\linewidth]{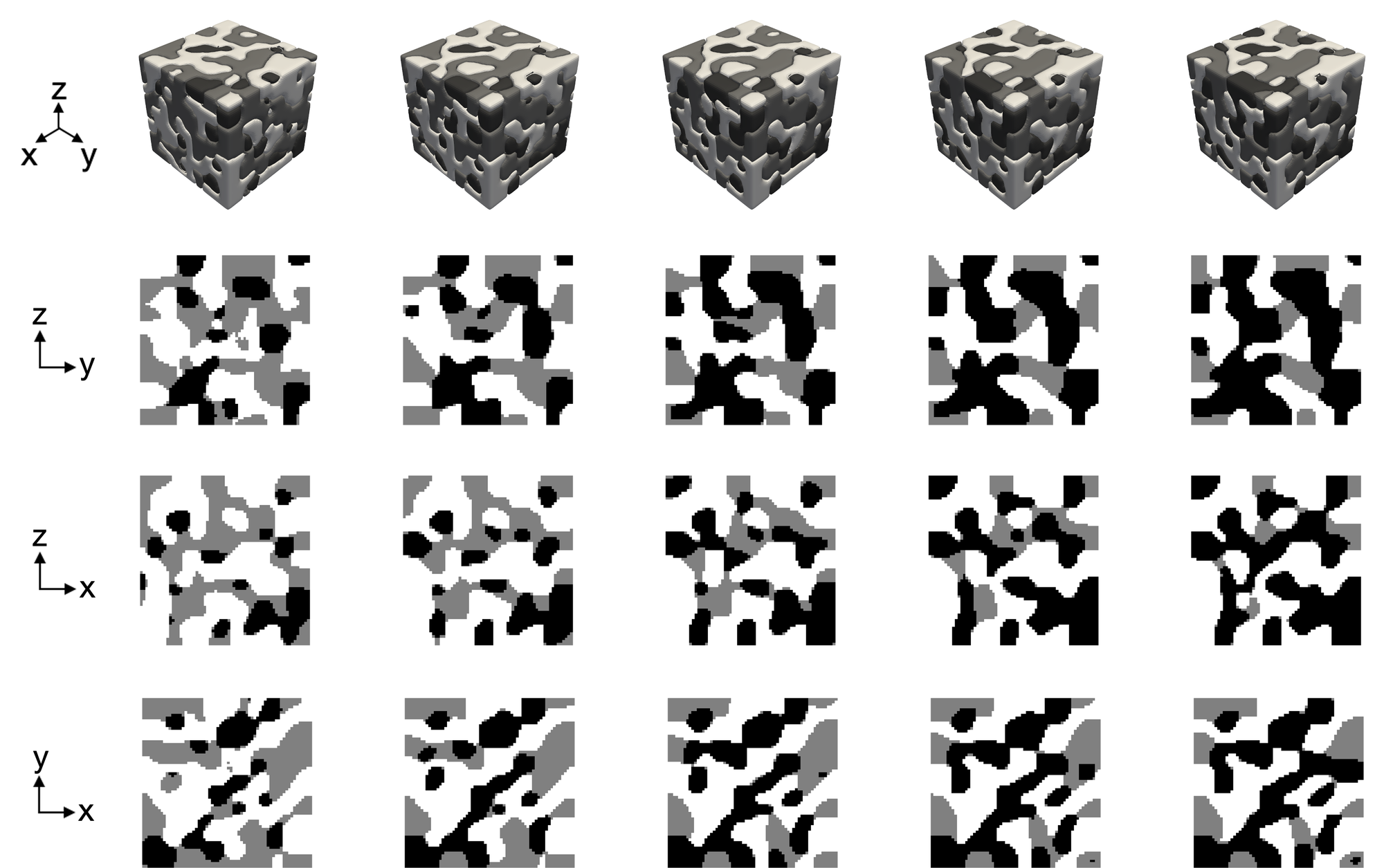}
    \caption{Progression of three orthogonal cross-sectional slices (through the center of each volume) during the 2D-to-3D microstructure generation process guided by the target objective ++\( D_e^{\text{Pore}} \), set to twice the maximum value in the training dataset.}
    \label{fig app: progression De}
\end{figure}

\begin{figure}[H]
    \centering    \includegraphics[width=1.0\linewidth]{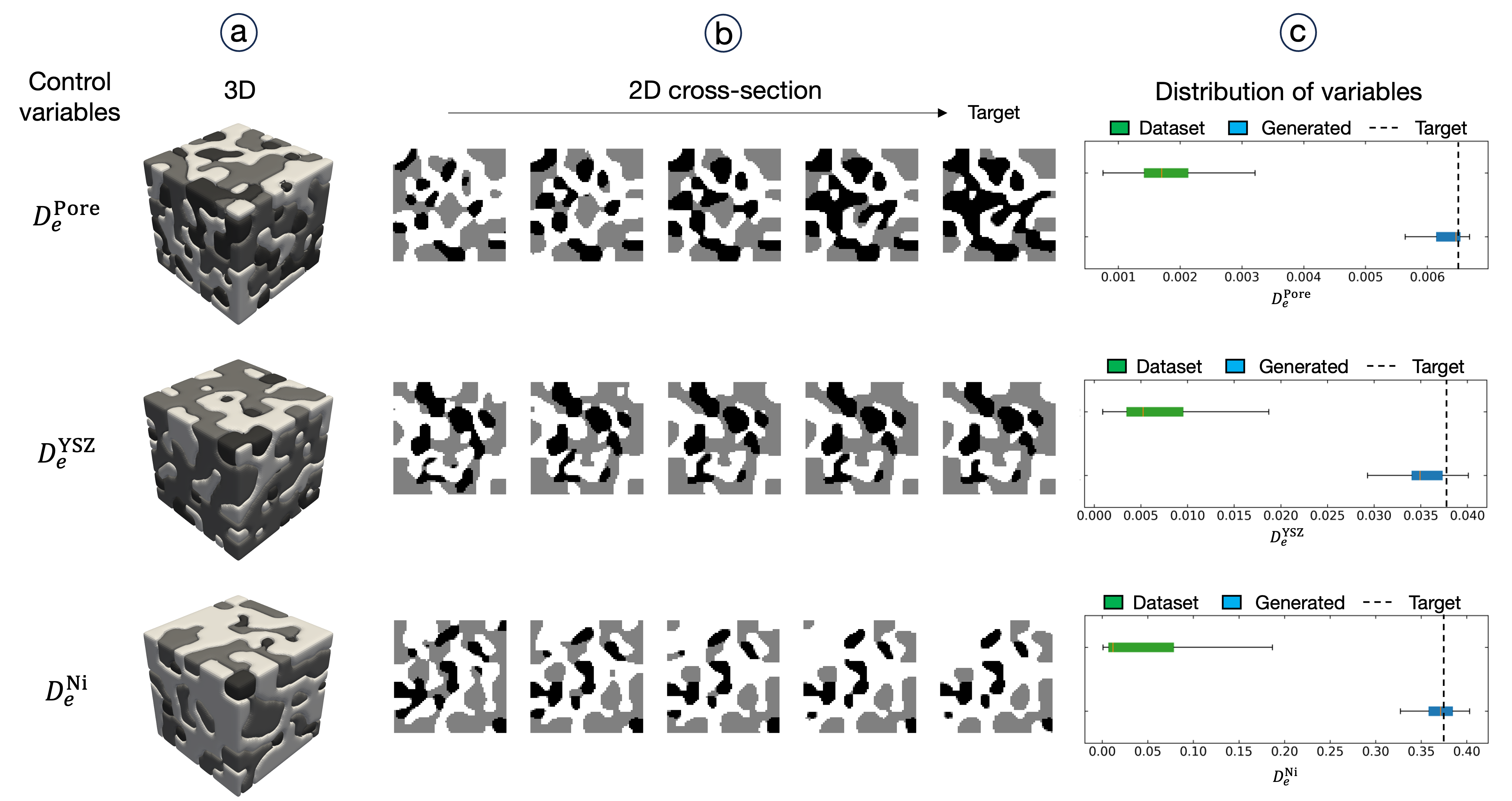}
    \caption{Inverse 2D-to-3D generation of three-phase microstructures with controlled relative diffusivity (twice the maximum in training dataset) for each material phase: (a) generated 3D volumes, (b) variation in 2D cross-sectional images (randomly selected slices) as SDS steps increase, and (c) distribution of variables in the training dataset compared to generated samples with target objectives.}
    \label{fig app: Inverse Deff x2}
\end{figure}

\section{Latent-space embedding and UMAP projection}
\label{sec:umap}

To visualize and analyze the structure of generated microstructures in latent space, we project a collection of 2D slices from 3D volumes onto a two-dimensional manifold using UMAP \cite{mcinnes2018umap}. The process begins by passing each microstructure slice \(\mathbf{x}\) through the pretrained encoder \(E_{\boldsymbol{\theta}}\), which produces spatial mean and log-variance maps:
\begin{equation}
    \left(\boldsymbol{\mu}(\mathbf{x}),\; \log \boldsymbol{\sigma}^2(\mathbf{x})\right) = E_{\boldsymbol{\theta}}(\mathbf{x}),
    \qquad \boldsymbol{\mu} \in \mathbb{R}^{C \times H \times W},\quad (C,H,W) = (4,16,16).
    \label{eq:umap_encoder}
\end{equation}

From the spatial mean tensor \(\boldsymbol{\mu}\), a compact descriptor is computed for each slice. For every channel \(c = 1,\dots,C\), we calculate the spatial mean and standard deviation:
\[
    \bar{\mu}_c = \frac{1}{HW} \sum_{h,w} \mu_{c,h,w},
    \qquad
    s_c = \sqrt{\frac{1}{HW} \sum_{h,w} \left(\mu_{c,h,w} - \bar{\mu}_c\right)^2}.
\]
These values are concatenated into a vector:
\begin{equation}
    \mathbf{d} = \left[ \bar{\mu}_1, \dots, \bar{\mu}_C,\; s_1, \dots, s_C \right]^{\top} \in \mathbb{R}^{2C}.
    \label{eq:umap_descriptor}
\end{equation}

Stacking the descriptors of all \(N\) slices yields the dataset \(\mathbf{D} \in \mathbb{R}^{N \times 2C}\), where each row encodes a single slice.

UMAP then constructs a fuzzy topological graph from the data matrix \(\mathbf{D}\). The pairwise Euclidean distances are computed as:
\[
    \delta_{ij} = \| \mathbf{d}_i - \mathbf{d}_j \|_2.
\]
For each point \(i\), the local connectivity radius is defined as \(\rho_i = \min_{j \ne i} \delta_{ij}\), and a local scale parameter \(\sigma_i\) is determined such that:
\[
    \sum_{j \ne i} \exp\left( - \frac{\max(0, \delta_{ij} - \rho_i)}{\sigma_i} \right) = \log_2 k_{\mathrm{NN}},
\]
where \(k_{\mathrm{NN}}\) is a user-defined number of nearest neighbors.

The local edge weights are then symmetrized to obtain:
\begin{equation}
    \omega_{ij} = \exp\left( - \frac{\max(0, \delta_{ij} - \rho_i)}{\sigma_i} \right),
    \qquad
    \tilde{\omega}_{ij} = \omega_{ij} + \omega_{ji} - \omega_{ij} \omega_{ji}.
    \label{eq:umap_weights}
\end{equation}

In the low-dimensional space, UMAP seeks a set of points \(\mathbf{z}_i \in \mathbb{R}^2\) that preserve the local topology, using a similarity function of the form:
\[
    p_{ij} = \left(1 + a \| \mathbf{z}_i - \mathbf{z}_j \|^{2b} \right)^{-1},
\]
with constants \(a\) and \(b\) set heuristically. The embedding is obtained by minimizing the cross-entropy loss between \(\tilde{\omega}_{ij}\) and \(p_{ij}\):
\begin{equation}
    \mathcal{L}(\{\mathbf{z}_i\}) = \sum_{i \ne j} \left[
        \tilde{\omega}_{ij} \log \frac{\tilde{\omega}_{ij}}{p_{ij}} +
        (1 - \tilde{\omega}_{ij}) \log \frac{1 - \tilde{\omega}_{ij}}{1 - p_{ij}}
    \right].
    \label{eq:umap_objective}
\end{equation}

Solving this optimization yields a low-dimensional representation \(\mathbf{Z} = [\mathbf{z}_1^{\top}, \dots, \mathbf{z}_N^{\top}]^{\top} \in \mathbb{R}^{N \times 2}\), which is used for visualization and analysis. The resulting scatter plots enable comparison of latent-space distributions across different objectives, datasets, or optimization steps, and are often annotated by volume fraction, surface area, or effective diffusivity.

\begin{bluecolorregion}
\section{Additional microstructure reconstruction results including anisotropic cases}\label{sec:Reconstruction of various types of microstructures}

To demonstrate the capability of the proposed framework in reconstructing diverse types of microstructures, several additional examples were considered, including spherical particle inclusion microstructures~\cite{lee2024multi} and anisotropic battery separators~\cite{lagadec2018microstructure}. As shown in Figure~\ref{fig app: recon particles}(a), the \(64\times64\) images of spherical particle inclusions were used to train the LDM, and the MicroLad framework was then applied to perform 2D-to-3D reconstruction. The reconstructed 3D microstructures are presented in Figure~\ref{fig app: recon particles}(b), and their corresponding 2D slices are shown in Figure~\ref{fig app: recon particles}(c). Although some reconstructed particles appear slightly deformed (i.e., not perfectly spherical) in the 2D slices, the overall 3D spherical morphology is well reproduced, as seen in Figure~\ref{fig app: recon particles}(b). Furthermore, the two-point correlation functions of the reconstructed samples show very good agreement with those of the original dataset (Figure~\ref{fig app: recon particles}(d)), with a relative error of only 0.7\% under \(S_2\) guidance.

In order to further demonstrate the capability of the proposed MicroLad framework for anisotropic microstructures, a conditional 2D LDM was trained to generate 2D microstructures corresponding to the xz, xy, and yz planes of a battery separator, similar to the anisotropic microstructure reconstruction version of MPDD presented in~\cite{lee2024denoising}. A single image was prepared for each plane: xz (\(650\times1300\)), xy (\(650\times499\)), and yz (\(499\times1300\)). The cropped \(64\times64\) regions from these images were used to train the conditional 2D LDM to generate samples based on the specified plane orientation. The MicroLad framework was then applied to reconstruct the 3D anisotropic microstructures by using the conditional LDM to denoise the 2D latent slices corresponding to each plane. Figure~\ref{fig app: recon not rotated}(a) shows the reconstructed 3D anisotropic microstructures, which were obtained using a 2D LDM trained on the corresponding 2D dataset (Figure~\ref{fig app: recon not rotated}(b)). The reconstructed 3D volumes and their 2D slices exhibit strong visual similarity across different planes, indicating successful reconstruction of anisotropic features. Moreover, the two-point correlation function comparisons for each plane (Figures~\ref{fig app: recon not rotated}(d)–(f)) demonstrate that the spatial correlations were well reproduced across the planes, with a maximum relative error of 2.6\% (on the yz plane). These results highlight the effectiveness of the proposed framework for 2D-to-3D reconstruction of anisotropic microstructures.

It is worth noting that, until now, the microstructures considered in this study are not perfectly isotropic or orthogonally anisotropic (i.e., they exhibit certain anisotropic features in non-orthogonal directions). Nevertheless, the visual similarities and spatial correlations were well preserved. To further demonstrate that the proposed MicroLad framework can reconstruct microstructures with pronounced non-orthogonal anisotropy, a battery separator microstructure rotated by \(45^\circ\) on the xz-plane was used for 2D-to-3D reconstruction, as shown in Figure~\ref{fig app: recon rotated}. As illustrated in Figure~\ref{fig app: recon rotated}(a), the 3D microstructures were reconstructed based on a dataset exhibiting diagonal anisotropy (Figure~\ref{fig app: recon rotated}(b)), and the 2D slices of the reconstructed 3D volumes (Figure~\ref{fig app: recon rotated}(c)) confirm that the non-orthogonal anisotropy was successfully reproduced. The two-point correlation function comparisons (Figures~\ref{fig app: recon rotated}(d)–(e)) show a maximum relative error of only 1.4\% on the yz-plane, further highlighting that the spatial correlations were well preserved. These results demonstrate the proposed framework’s capability to handle non-orthogonal anisotropy effectively. In addition, Bostanabad~\cite{Bostanabad2020Reconstruction} discussed that non-orthogonal anisotropy can be effectively captured in orthogonality-based microstructure reconstruction when the voxel field is updated simultaneously rather than through individual voxel updates. In this context, the performance of the proposed framework can be attributed to the MPDD mechanism, which performs multi-plane denoising processes across a large number of diffusion steps. This process effectively approximates a continuous stochastic differential equation, as discussed in~\cite{lee2024multi}, allowing the entire voxel field—rather than individual voxels—to be updated simultaneously from the prior to the data distribution across multiple planes.

\begin{figure}[H]
    \centering    \includegraphics[width=0.9\linewidth]{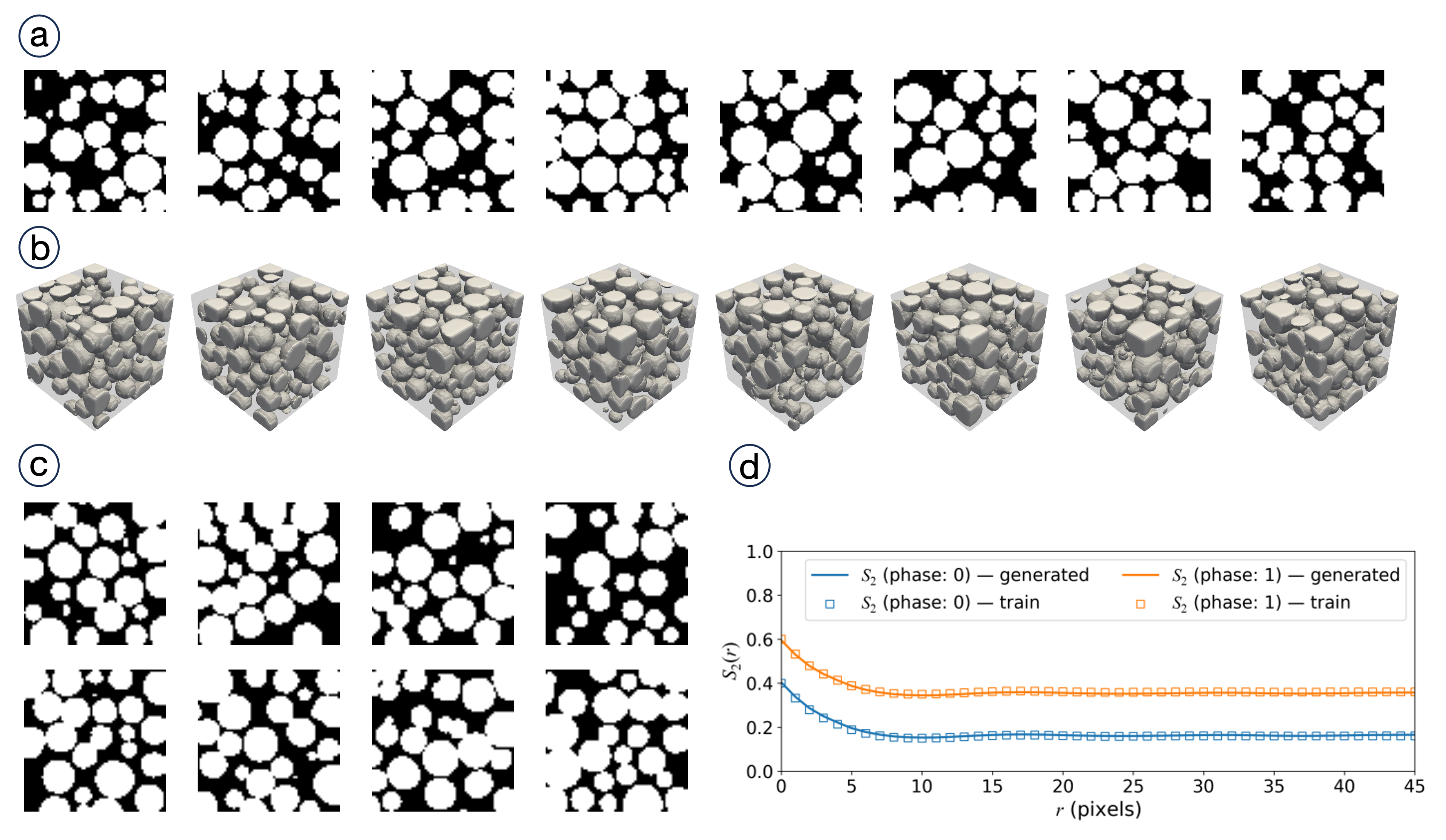}
    \caption{2D-to-3D microstructure reconstruction results for spherical particle inclusion microstructures: (a) original 2D microstructure images, (b) reconstructed 3D microstructure, (c) random cross-sectional slices of (b), (d) two-point correlation functions of 2D slices from the reconstructed 3D samples in (b) compared to those from the original 2D dataset in (a).}
    \label{fig app: recon particles}
\end{figure}

\begin{figure}[H]
    \centering    \includegraphics[width=0.9\linewidth]{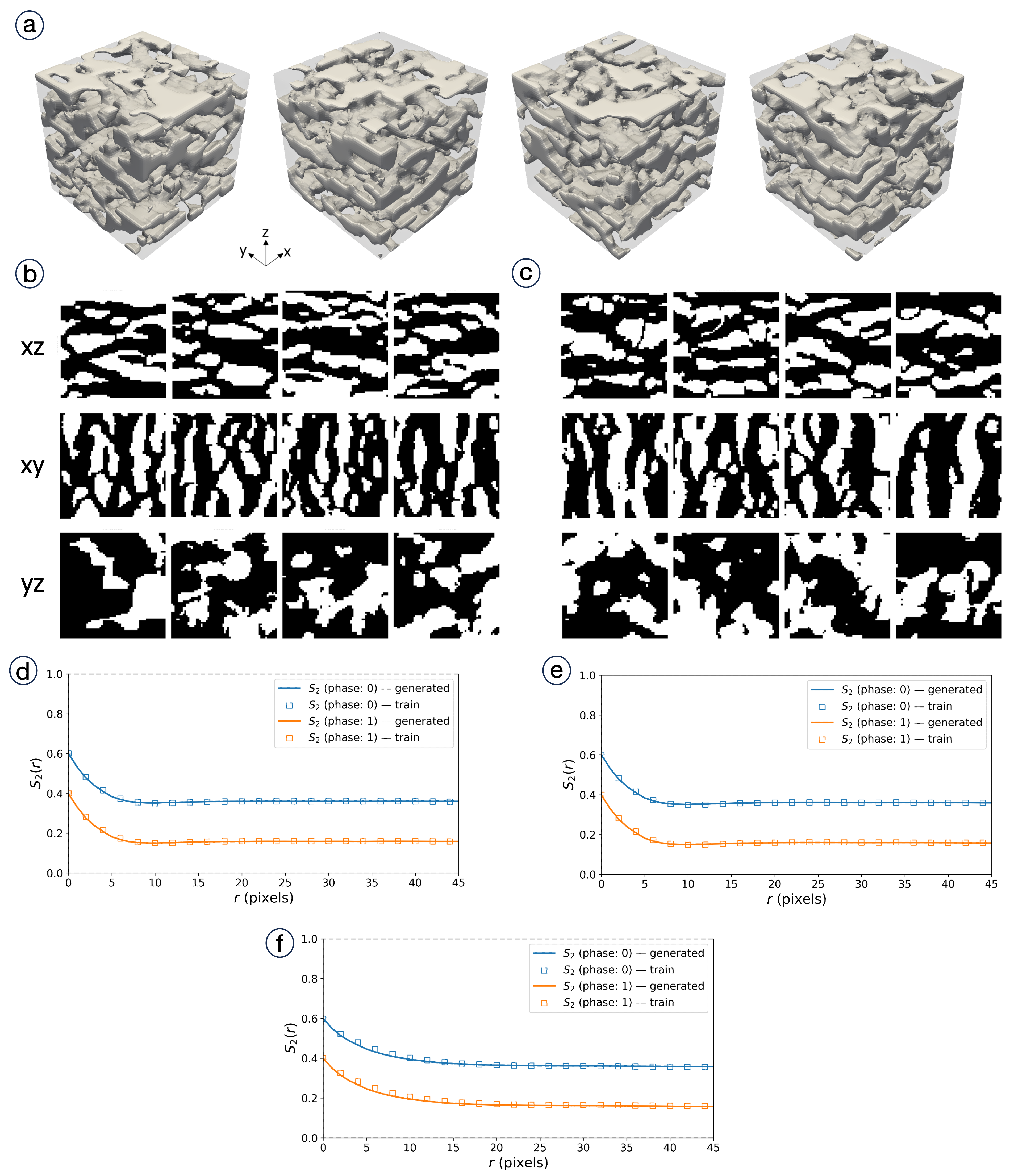}
    \caption{2D-to-3D microstructure reconstruction results for anisotropic microstructures (lithium-ion battery separator~\cite{lagadec2018microstructure}): (a) reconstructed 3D microstructure, (b) original 2D microstructure images, (c) random cross-sectional slices of (a), and (d)–(f) two-point correlation functions of 2D slices from the reconstructed 3D samples in (a) compared to those from the original 2D dataset in (b) for the xz, xy, and yz planes, respectively.}
    \label{fig app: recon not rotated}
\end{figure}

\begin{figure}[H]
    \centering    \includegraphics[width=0.9\linewidth]{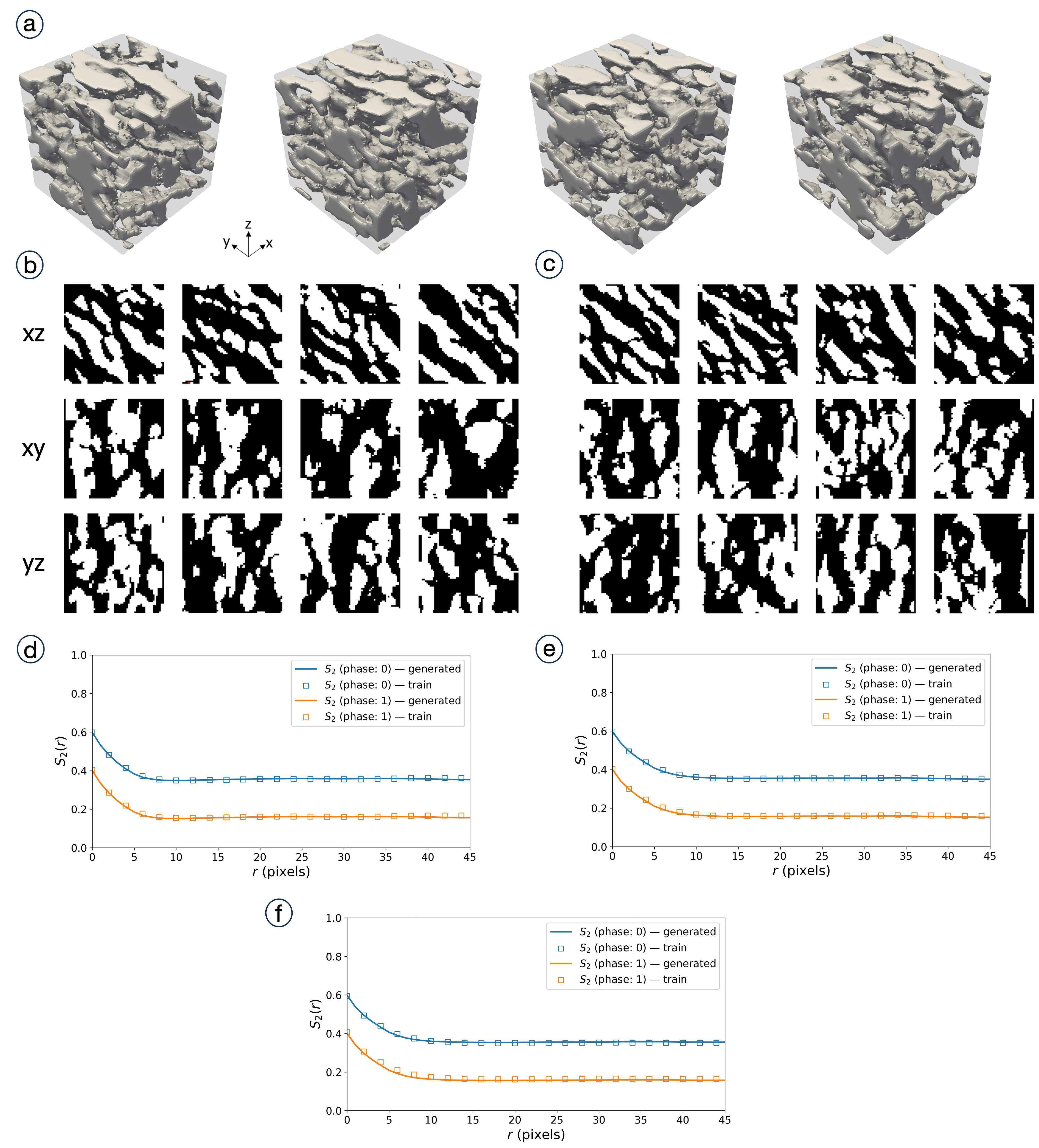}
    \caption{2D-to-3D microstructure reconstruction results for anisotropic battery separator microstructures rotated by \(45^\circ\) on the xz-plane: (a) reconstructed 3D microstructure, (b) original 2D microstructure images, (c) random cross-sectional slices of (a), and (d)–(f) two-point correlation functions of 2D slices from the reconstructed 3D samples in (a) compared to those from the original 2D dataset in (b) for the xz, xy, and yz planes, respectively.}
    \label{fig app: recon rotated}
\end{figure}

\end{bluecolorregion}

\section{Parametric study results for SDS}\label{sec:parameteric study for SDS}

To assess the sensitivity and effectiveness of SDS in guiding 3-D microstructure generation, we conduct a parametric study by varying three key hyperparameters: (i) the objective weight ($w$), which controls the strength of descriptor- or property-guided optimization, (ii) the SDS learning rate, and (iii) the number of SDS update steps.  The study is performed for three target objectives—volume fraction ($V_f$), surface area ($SA$), and effective diffusivity ($D_e$)—with all evaluations carried out on the pore phase using target objectives of \(+V_f^{\text{Pore}}\), \(+SA^{\text{Pore}}\), and \(+D_e^{\text{Pore}}\). For each configuration we report the objective error and the wall-clock time (s) required to generate a single 3-D volume.  Wall-clock times were measured on a single NVIDIA RTX 4090 GPU and include encoder/decoder I/O as well as back-propagation through the differentiable FEM solver.  The results, summarized in Tables~\ref{tab:vf_sweep}–\ref{tab:rd_sweep}, highlight the trade-offs between optimization accuracy and computational cost.

\begin{table}[H]
  \centering
  \caption{Parametric study of SDS parameters for $V_f$-guided generation.}
  \label{tab:vf_sweep}
  \begin{tabular}{llrrr}
    \toprule
  \(w_{V_f}\) & SDS learning rate & SDS steps &
  \(V_f\) error & Wall-clock time (s) \\
    \midrule
    {1}
      & 0.10 &  500 & 0.06262 &  61 \\
      &      & 1000 & 0.04244 & 106 \\
      &      & 2000 & 0.03507 & 184 \\
      &      & 3000 & 0.01118 & 270 \\[0.2em]
      & 0.05 &  500 & 0.09939 &  61 \\
      &      & 1000 & 0.09382 & 101 \\
      &      & 2000 & 0.06882 & 186 \\
      &      & 3000 & 0.03741 & 274 \\[0.2em]
      & 0.01 &  500 & 0.15433 &  62 \\
      &      & 1000 & 0.12856 & 103 \\
      &      & 2000 & 0.10210 & 187 \\
      &      & 3000 & 0.09602 & 277 \\
    \midrule
    {5}
      & 0.10 &  500 & 0.05681 &  60 \\
      &      & 1000 & 0.03538 & 101 \\
      &      & 2000 & 0.04064 & 184 \\
      &      & 3000 & 0.02920 & 268 \\[0.2em]
      & 0.01 &  500 & 0.14113 &  61 \\
      &      & 1000 & 0.10953 & 104 \\
      &      & 2000 & 0.08771 & 186 \\
      &      & 3000 & 0.06617 & 267 \\
    \bottomrule
  \end{tabular}
\end{table}

\begin{table}[H]
  \centering
  \caption{Parametric study of SDS parameters for $SA$-guided generation.}
  \label{tab:sa_sweep}
  \begin{tabular}{llrrr}
    \toprule
  \(w_{SA}\) & SDS learning rate & SDS steps &
  \(SA\) error & Wall-clock time (s) \\
    \midrule
    {1}
      & 0.10 &  500 & 0.02266 &  62 \\
      &      & 1000 & 0.02734 & 102 \\
      &      & 2000 & 0.01947 & 186 \\
      &      & 3000 & 0.01702 & 273 \\[0.2em]
      & 0.05 &  500 & 0.07357 &  59 \\
      &      & 1000 & 0.06704 & 102 \\
      &      & 2000 & 0.05207 & 186 \\
      &      & 3000 & 0.04819 & 271 \\[0.2em]
      & 0.01 &  500 & 0.07679 &  60 \\
      &      & 1000 & 0.07244 & 105 \\
      &      & 2000 & 0.07095 & 188 \\
      &      & 3000 & 0.06355 & 274 \\
    \midrule
    {5}
      & 0.10 &  500 & 0.02474 &  59 \\
      &      & 1000 & 0.02310 & 102 \\
      &      & 2000 & 0.02711 & 187 \\
      &      & 3000 & 0.02742 & 268 \\[0.2em]
      & 0.01 &  500 & 0.05948 &  58 \\
      &      & 1000 & 0.05301 & 103 \\
      &      & 2000 & 0.04693 & 195 \\
      &      & 3000 & 0.04815 & 288 \\
    \bottomrule
  \end{tabular}
\end{table}

\begin{table}[H]
  \centering
  \caption{Parametric study of SDS parameters for $D_e$-guided generation.}
  \label{tab:rd_sweep}
  \begin{tabular}{llrrr}
    \toprule
  \(w_{D_e}\) & SDS learning rate & SDS steps &
  \(D_e\) error & Wall-clock time (s) \\
    \midrule
    {1}
      & 0.10 &  500 & 0.00073 & 152 \\
      &      & 1000 & 0.00065 & 287 \\
      &      & 2000 & 0.00063 & 558 \\
      &      & 3000 & 0.00057 & 832 \\[0.2em]
      & 0.05 &  500 & 0.00093 & 152 \\
      &      & 1000 & 0.00081 & 290 \\
      &      & 2000 & 0.00073 & 557 \\
      &      & 3000 & 0.00074 & 824 \\[0.2em]
      & 0.01 &  500 & 0.00094 & 152 \\
      &      & 1000 & 0.00092 & 286 \\
      &      & 2000 & 0.00084 & 558 \\
      &      & 3000 & 0.00081 & 824 \\
    \midrule
    {10}
      & 0.10 &  500 & 0.00063 & 152 \\
      &      & 1000 & 0.00062 & 288 \\
      &      & 2000 & 0.00065 & 540 \\
      &      & 3000 & 0.00045 & 790 \\[0.2em]
      & 0.01 &  500 & 0.00085 & 146 \\
      &      & 1000 & 0.00080 & 276 \\
      &      & 2000 & 0.00075 & 536 \\
      &      & 3000 & 0.00076 & 797 \\
    \bottomrule
  \end{tabular}
\end{table}

\begin{bluecolorregion}

\section{Experiments with larger microstructure size}\label{sec:Experiments with larger microstructure size}

To evaluate the scalability of the proposed framework and confirm its applicability to larger microstructure sizes, additional experiments were conducted using \(256\times256\times256\) microstructures. As shown in Figure~\ref{fig app: visualization 256 recon}(a), 2D microstructure images of the SOFC anode with a size of \(256\times256\) were extracted to train a 2D LDM along with the corresponding VAE configured as described in Table~\ref{tab:latent_pipeline_256}. The trained MicroLad framework was then applied for 2D-to-3D reconstruction of the \(256\times256\times256\) microstructure. As shown in Figure~\ref{fig app: visualization 256 recon}(b), the proposed framework successfully generated 3D microstructures, and the 2D slices of the reconstructed 3D volumes (Figure~\ref{fig app: visualization 256 recon}(c)) exhibit visual characteristics closely resembling those of the original 2D slices. The two-point correlation functions in Figure~\ref{fig app: visualization 256 recon}(d) also show very good agreement between the generated samples and the original dataset. The quantitative evaluation of \(S_2\) accuracy for each phase, summarized in Table~\ref{tab:phase_recon_accuracy_256}, shows a maximum relative error of 5.23\%, confirming that the proposed MicroLad framework can be effectively extended to larger microstructure sizes while maintaining high reconstruction accuracy. Furthermore, it is noteworthy that the framework operates in a latent space of size \(32\times32\) using an LDM, which is lower-dimensional than the original pixel space (\(256\times256\)). The 2D-to-3D reconstruction of \(256\times256\times256\) microstructures required only 71\,s, which clearly demonstrates the computational efficiency of the proposed approach while maintaining high reconstruction accuracy.

Next, to validate the inverse-controlled 2D-to-3D generation performance for larger microstructure sizes, the inverse generation results targeting \(V_f\), \(SA\), and \(D_e\) are illustrated in Figures~\ref{fig app: visualization 256 Vf}, \ref{fig app: visualization 256 sa}, and \ref{fig app: visualization 256 deff}, respectively. As shown in these figures, the proposed MicroLad framework successfully generates microstructures that reach the maximum values of the dataset for each objective (i.e., the tail of the data distribution), consistent with the \(64\times64\times64\) cases presented in Section~\ref{Results: Inverse-controlled 2D-to-3D generated microstruture samples}. These results highlight the capability of the proposed framework for data expansion and inverse-controlled generation at larger microstructural scales. The inverse-controlled generation used 9000 SDS steps, requiring 21.6 minutes for volume-fraction control and 23.6 minutes for surface-area control to generate a single 3D volume. For \(D_e\)-guided generation, the process took 156.1 minutes because the FEM computations for \(256\times256\) microstructure images substantially increased the overall computational cost compared to the \(64\times64\) case, as expected. Nonetheless, this computational efficiency can be further improved by employing more efficient FEM solvers and surrogate modeling approaches~\cite{xue2023jax,kudela2022recent}, which enable differentiable and near-instantaneous evaluation of \(D_e\). Such improvements are suggested as promising directions for future work.

\begin{figure}[H]
    \centering    \includegraphics[width=0.9\linewidth]{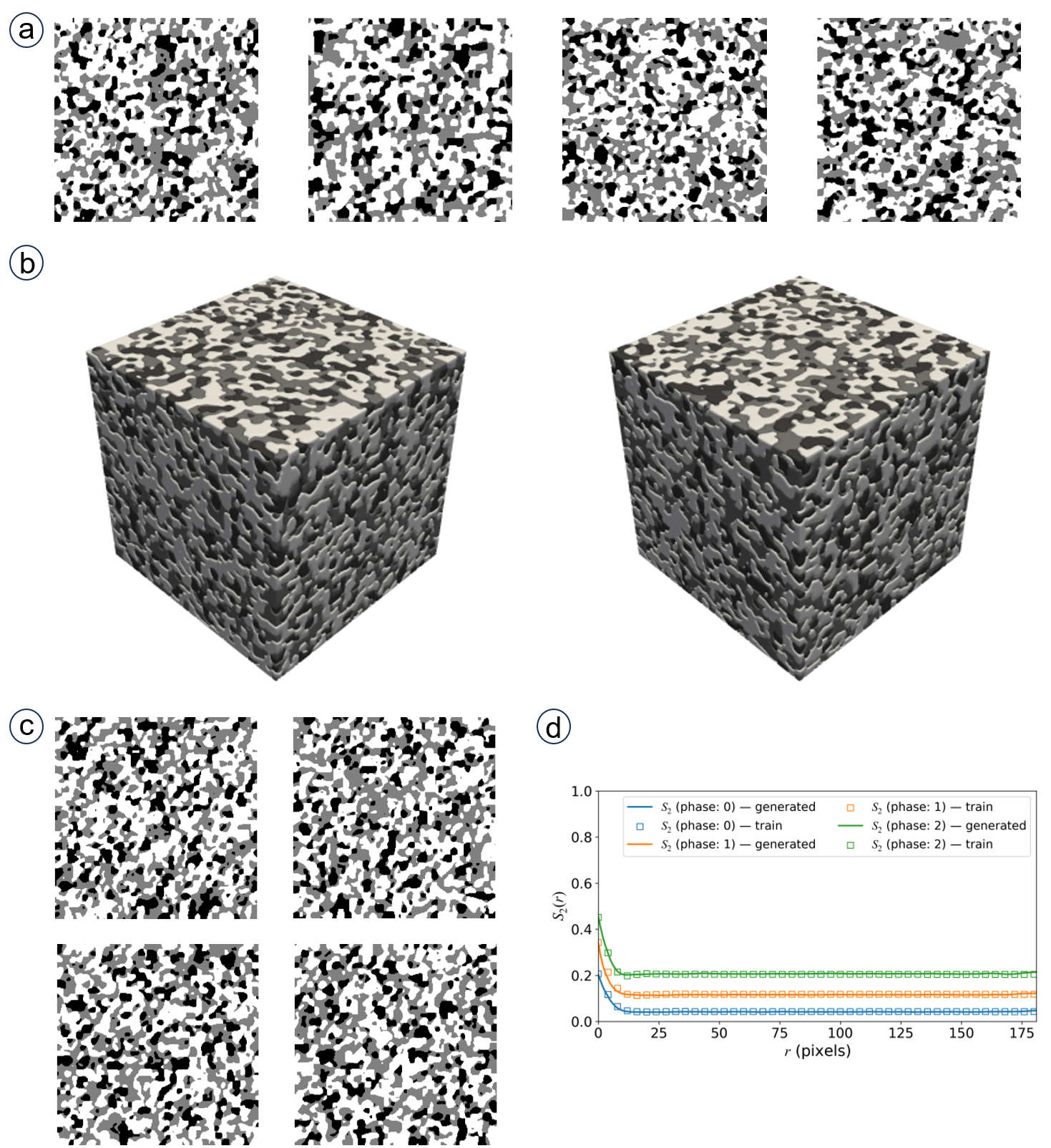}
    \caption{2D-to-3D microstructure reconstruction results for three-phase microstructure ($256\times256\times256$): (a) original 2D microstructure images, (b) reconstructed 3D microstructures, (c) random cross-sectional slices of (b), (d) two-point correlation functions of 2D slices from the reconstructed 3D samples in (b) compared to those from the original 2D dataset in (a)}
    \label{fig app: visualization 256 recon}
\end{figure}

\begin{figure}[H]
    \centering    \includegraphics[width=0.8\linewidth]{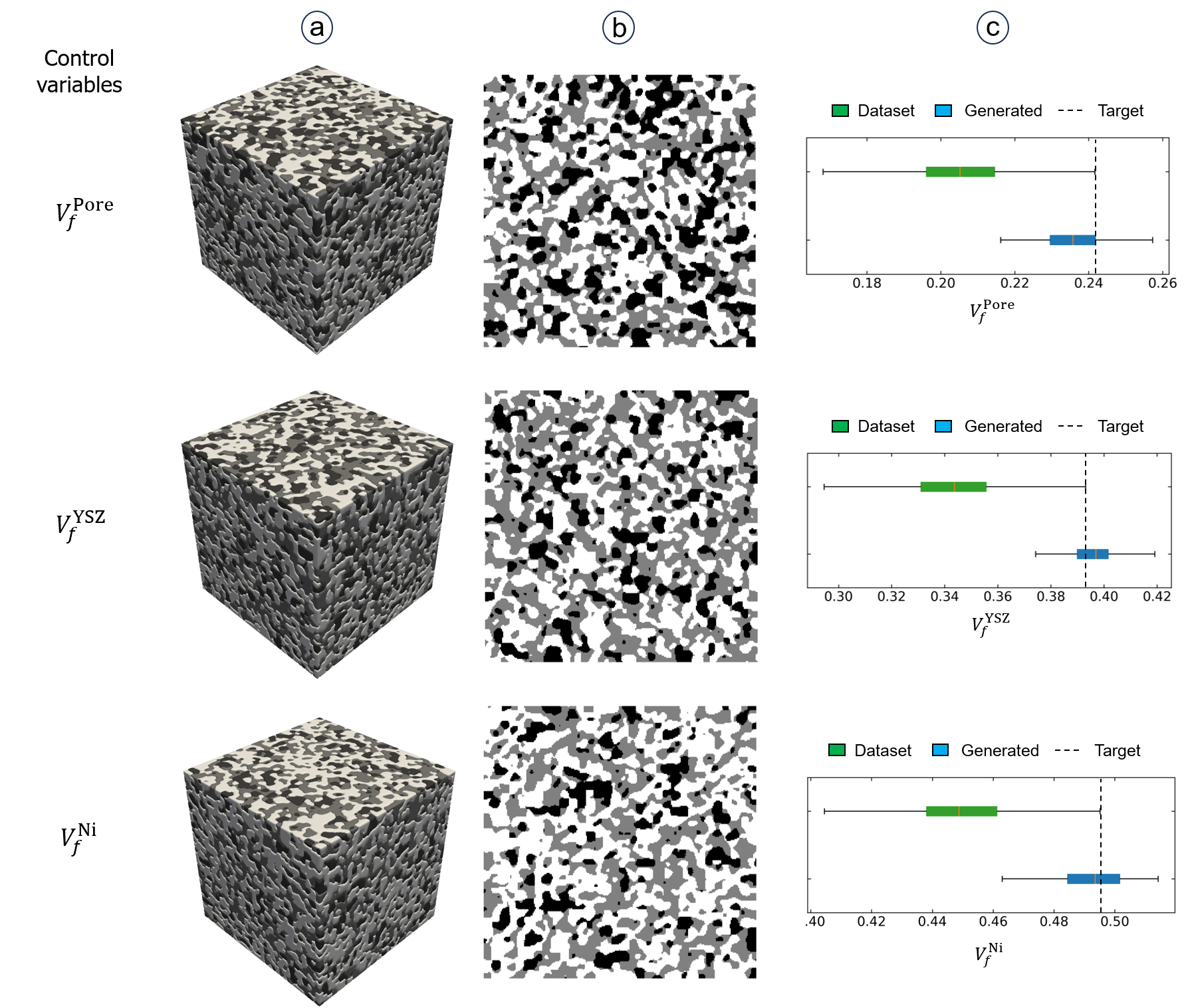}
    \caption{Inverse 2D-to-3D generation of three-phase microstructures ($256\times256\times256$) with controlled volume fractions for each material phase: (a) generated 3D volumes, (b) corresponding 2D cross-sectional slices, and (c) distribution of variables in the training dataset compared to generated samples with target objectives.}
    \label{fig app: visualization 256 Vf}
\end{figure}

\begin{figure}[H]
    \centering    \includegraphics[width=0.8\linewidth]{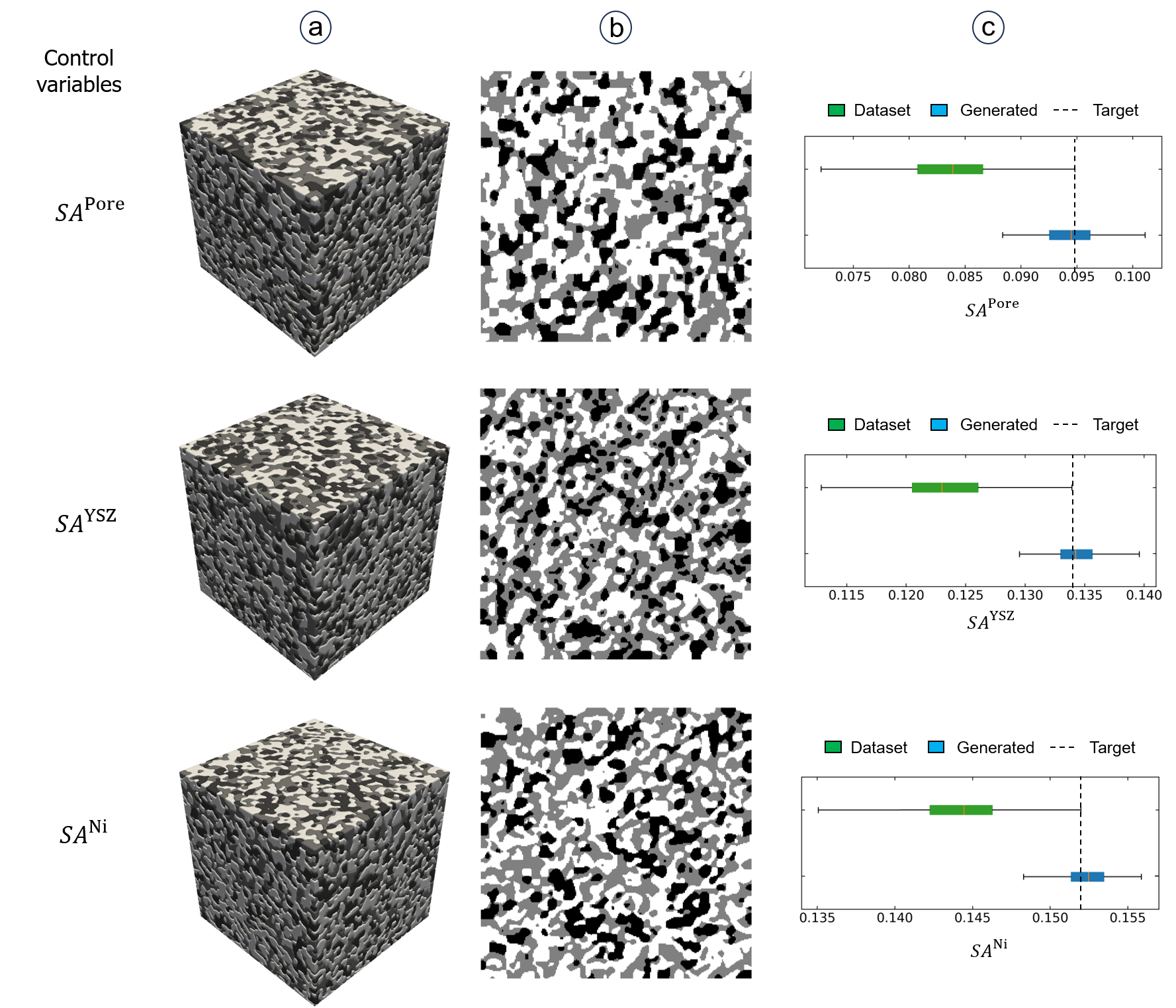}
    \caption{Inverse 2D-to-3D generation of three-phase microstructures ($256\times256\times256$) with controlled relative surface area for each material phase: (a) generated 3D volumes, (b) corresponding 2D cross-sectional slices, and (c) distribution of variables in the training dataset compared to generated samples with target objectives.}
    \label{fig app: visualization 256 sa}
\end{figure}

\begin{figure}[H]
    \centering    \includegraphics[width=0.8\linewidth]{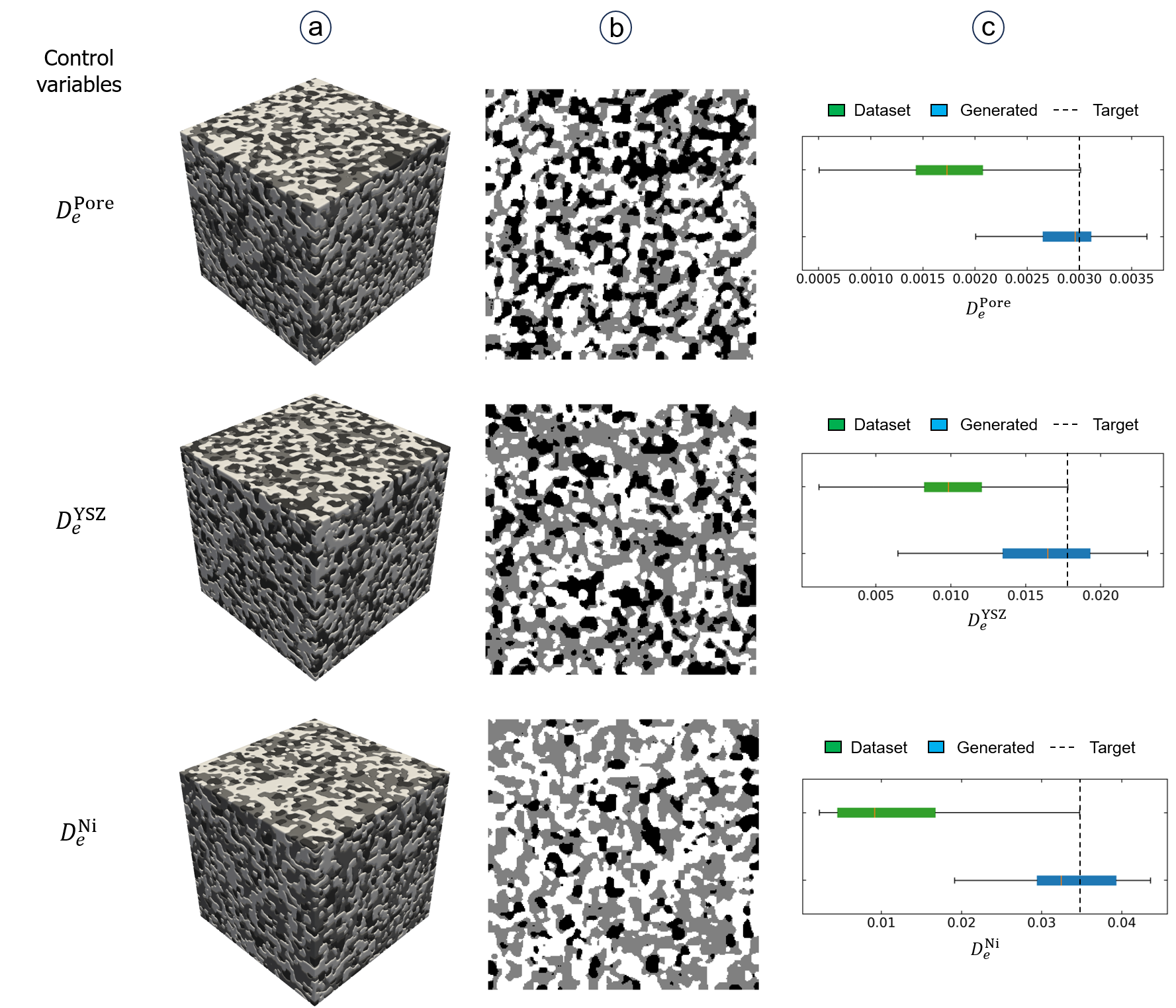}
    \caption{Inverse 2D-to-3D generation of three-phase microstructures ($256\times256\times256$) with controlled relative diffusivity for each material phase: (a) generated 3D volumes, (b) corresponding 2D cross-sectional slices, and (c) distribution of variables in the training dataset compared to generated samples with target objectives.}
    \label{fig app: visualization 256 deff}
\end{figure}

\begin{table}[H]
\centering
\caption{Encoder–latent-space U-Net–decoder pipeline for MicroLad with 256×256 input (latent map $4\times32\times32$).}
\label{tab:latent_pipeline_256}
\begin{tabular}{@{}lll@{}}
\toprule
\textbf{Layer} & \textbf{Operation} & \textbf{Output size} \\ \midrule
\multicolumn{3}{l}{\itshape VAE encoder}\\
Input             & $3\times3$ Conv                                  & $128\times256\times256$ \\
Downsample 1       & DownBlock ($\times2$ Res)                         & $128\times128\times128$ \\
Downsample 2       & DownBlock ($\times2$ Res)                         & $256\times64\times64$ \\
Downsample 3       & DownBlock ($\times2$ Res)                         & $512\times32\times32$ \\
Encoder head       & Res $+$ Attention $+$ Res                         & $512\times32\times32$ \\
Latent params      & $1\times1$ Conv ($\mu,\sigma$)                    & $4\times32\times32$ \\[4pt]

\multicolumn{3}{l}{\itshape U-Net in latent space}\\
Input              & $3\times3$ Conv                                   & $128\times32\times32$ \\
32×32 level        & $\times2$ Res (time-conditioned)                  & $128\times32\times32$ \\
Downsample to 16   & $3\times3$ Conv (stride 2)                        & $256\times16\times16$ \\
16×16 level        & $\times2$ Res $+$ Attention                       & $256\times16\times16$ \\
Downsample to 8    & $3\times3$ Conv (stride 2)                        & $512\times8\times8$ \\
8×8 level          & $\times2$ Res $+$ Attention                       & $512\times8\times8$ \\
Downsample to 4    & $3\times3$ Conv (stride 2)                        & $512\times4\times4$ \\
Bottleneck         & Res $+$ Attention $+$ Res                         & $1024\times4\times4$ \\
Upsample to 8      & ConvTranspose (factor 2) $+$ skip                 & $512\times8\times8$ \\
8×8 level          & $\times2$ Res $+$ Attention                       & $512\times8\times8$ \\
Upsample to 16     & ConvTranspose (factor 2) $+$ skip                 & $256\times16\times16$ \\
16×16 level        & $\times2$ Res $+$ Attention                       & $256\times16\times16$ \\
Upsample to 32     & ConvTranspose (factor 2) $+$ skip                 & $128\times32\times32$ \\
32×32 level        & $\times2$ Res                                     & $128\times32\times32$ \\
Output             & $3\times3$ Conv                                   & $4\times32\times32$ \\[4pt]

\multicolumn{3}{l}{\itshape VAE decoder}\\
Latent input       & $3\times3$ Conv                                   & $512\times32\times32$ \\
Decoder core       & Res $+$ Attention $+$ Res                         & $512\times32\times32$ \\
Upsample 1          & UpBlock (factor 2)                               & $256\times64\times64$ \\
Upsample 2          & UpBlock (factor 2)                               & $128\times128\times128$ \\
Upsample 3          & UpBlock (factor 2)                               & $64\times256\times256$ \\
Output             & $3\times3$ Conv $+$ Sigmoid                       & $1\times256\times256$ \\ \bottomrule
\end{tabular}
\end{table}

\begin{table}[ht]
  \centering
  \caption{Reconstruction accuracy for each phase based on $S_2$.}
  \renewcommand{\arraystretch}{1.2}
  \begin{tabular}{l c c}
    \toprule
    \textbf{Phase} & \textbf{MAE} & \makecell{\textbf{Relative error} (\%) \\ $\varepsilon_{\text{rel}}$} \\
    \midrule
    Pore & 0.0050 & 5.23 \\
    YSZ & 0.0023 & 0.55 \\
    Ni & 0.0025 & 0.41 \\
    \bottomrule
  \end{tabular}
  \label{tab:phase_recon_accuracy_256}
\end{table}

\end{bluecolorregion}

\begin{bluecolorregion}

\section{Evaluation of reconstruction in terms of 3D spatial correlation}\label{sec:Evaluation in terms of 3D spatial correlation}

It is noteworthy that the proposed framework focuses on the 2D-to-3D reconstruction and generation of microstructures, where the available reference data are inherently two-dimensional. However, to further demonstrate the reconstruction capability of the framework in terms of 3D spatial correlation, an example case was investigated. A random 3D microstructure (\(256\times256\times256\)) (Figure~\ref{fig app: 3D TPCF comparison}(a)) was first generated to serve as a reference 3D volume. Subsequently, a two-point-correlation-guided 2D-to-3D reconstruction was performed using the two-point correlation functions of the slices in the reference volume (as described in Section~\ref{sec:results_2dto3d_recon}) to reconstruct 3D microstructures (Figure~\ref{fig app: 3D TPCF comparison}(b)). The 3D two-point correlation functions of the reference and reconstructed volumes were then compared, as shown in Figure~\ref{fig app: 3D TPCF comparison}(c). The results indicate that the two-point correlation functions are in good agreement, demonstrating that the 3D statistical characteristics are well preserved. The relative errors for each microstructural phase (pore, YSZ and Ni) were 5.14\%, 1.12\%, and 1.01\%, respectively.

\begin{figure}[H]
    \centering    \includegraphics[width=0.8\linewidth]{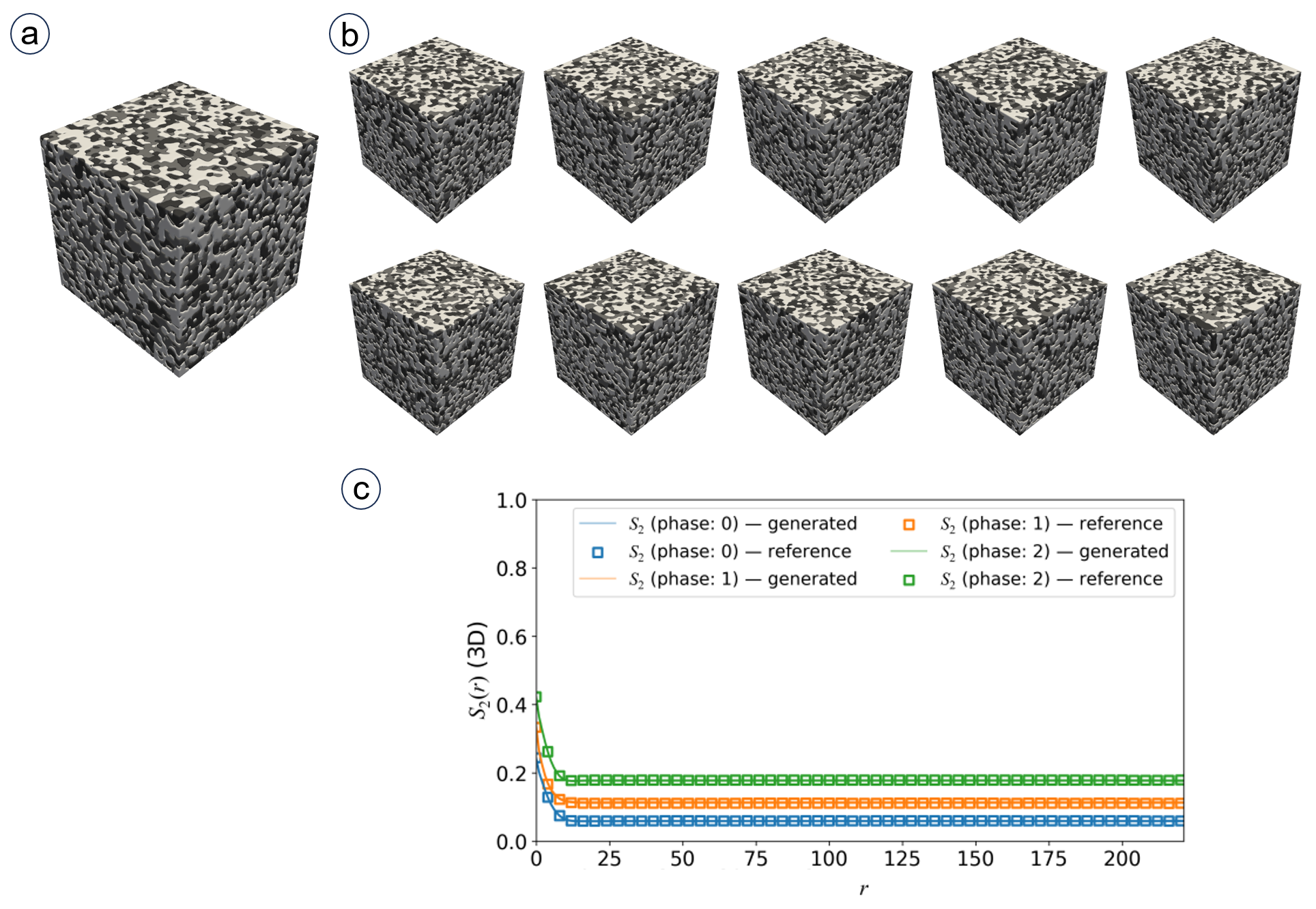}
    \caption{Evaluation of 3D spatial correlation: (a) random 3D microstructure, (b) reconstructed equivalent 3D microstructure, and (c) comparison of two-point correlation functions in 3D.}
    \label{fig app: 3D TPCF comparison}
\end{figure}

\end{bluecolorregion}

\section{Code availability}  
\label{sec:code}

Code for the MicroLad training and inference pipelines is available at https://github.com/KangHyunL/microlad.

\end{document}